\documentclass[aps,prc,reprint,superscriptaddress,floatfix,nofootinbib]{revtex4-2}
\usepackage[x11names]{xcolor}
\usepackage{bbm}
\usepackage{amssymb}
\usepackage[colorlinks=true, allcolors=blue]{hyperref}
\usepackage{graphicx}
\usepackage[caption=false]{subfig}
\usepackage{siunitx}
\usepackage{listings}
\usepackage{gensymb}
\usepackage{tikz}
\usepackage{tikz-3dplot}
\usepackage{circuitikz}
\usepackage{environ}
\usepackage{enumitem}

\usepackage[left]{lineno}
\usepackage{amsmath}  %% <- after lineno
\usepackage{cleveref}
\usepackage{etoolbox} %% <- for \cspreto, \csappto

\newcommand*\linenomathpatch[1]{%
  \cspreto{#1}{\linenomath}%
  \cspreto{#1*}{\linenomath}%
  \csappto{end#1}{\endlinenomath}%
  \csappto{end#1*}{\endlinenomath}%
}

\linenomathpatch{equation}
\linenomathpatch{gather}
\linenomathpatch{multline}
\linenomathpatch{align}
\linenomathpatch{alignat}
\linenomathpatch{flalign}
\crefname{section}{Sec.}{Secs.} % For sections
\crefname{appendix}{App.}{Apps.} % For appendices
\crefname{figure}{Fig.}{Figs.} % Figures
\crefname{equation}{Eq.}{Eqs.} % equations
\crefname{pluralequation}{Eqs.}{Eqs.} %subequations
\crefname{table}{Table}{Tables}
\usepackage{enumerate}
\usepackage{empheq} % For grouping equations in { with sub-labels
\usepackage{dsfont} % 
\hypersetup{
    colorlinks=True,
    citecolor=black,
    filecolor=black,
    linkcolor=blue,
    urlcolor=blue
}
\newcommand{\RE}{\mathrm{Re}}
\newcommand{\IM}{\mathrm{Im}}
\newcommand{\myspace}[1]{\qquad\!\!\!\!\hspace{#1em}} % Adds space in environ

\DeclareMathAlphabet\mathbfcal{OMS}{cmsy}{b}{n}

\NewEnviron{customalign}[1]{%
  \begingroup
  \makeatletter
  \fontsize{#1}{#1}\selectfont
  \def\maketag@@@#1{\hbox{\m@th\large\normalfont#1}}%
  \begin{align}
  \BODY
  \end{align}
  \endgroup
}

\NewEnviron{customsubeqn}[3]{%
  \begingroup
  \makeatletter
  \begin{subequations}
    \if\relax\detokenize{#3}\relax\else\label{#3}\fi
    \begingroup
    \def\customleftbrace{%
      \if\relax\detokenize{#2}\relax\empheqlbrace\else{#2}\empheqlbrace\fi
    }
    \begin{empheq}[left=\customleftbrace]{align}
      \BODY
    \end{empheq}
    \endgroup
  \end{subequations}
  \endgroup
}

\allowdisplaybreaks
\begin{document}
\title{Cyclotron Radiation Signal Characterization in Resonant Cavities for the Project 8 Neutrino Mass Experiment}
\newcommand{\Arlington}{\affiliation{Department of Physics, University of Texas at Arlington, Arlington, TX 76019, USA}}
\newcommand{\Berkeley}{\affiliation{Nuclear Science Division, Lawrence Berkeley National Laboratory, Berkeley, CA 94720, USA}}
\newcommand{\Mainz}{\affiliation{Institute for Physics, Johannes Gutenberg University Mainz, 55128 Mainz, Germany}}
\newcommand{\Mines}{\affiliation{Department of Physics, Colorado School of Mines, Golden, CO 80401, USA}}
\newcommand{\MIT}{\affiliation{Laboratory for Nuclear Science, Massachusetts Institute of Technology, Cambridge, MA 02139, USA}}
\newcommand{\PennStateARL}{\affiliation{Applied Research Laboratory, Pennsylvania State University, University Park, PA 16802, USA}}
\newcommand{\PennState}{\affiliation{Department of Physics, Pennsylvania State University, University Park, PA 16802, USA}}
\newcommand{\PNNL}{\affiliation{Pacific Northwest National Laboratory, Richland, WA 99354, USA}}
\newcommand{\Yale}{\affiliation{Wright Laboratory and Department of Physics, Yale University, New Haven, CT 06520, USA}}
\newcommand{\Livermore}{\affiliation{Lawrence Livermore National Laboratory, Livermore, CA 94550, USA}}
\newcommand{\Case}{\affiliation{Department of Physics, Case Western Reserve University, Cleveland, OH 44106, USA}}
\newcommand{\Heidelberg}{\affiliation{Institute for Theoretical Astrophysics, Heidelberg University, 69120 Heidelberg, Germany}}
\newcommand{\Illinois}{\affiliation{Department of Physics, University of Illinois Urbana-Champaign, Urbana, IL 61801, USA}}
\newcommand{\Indiana}{\affiliation{Center for Exploration of Energy and Matter and Department of Physics, Indiana University, Bloomington, IN, 47405, USA}}
\newcommand{\KIT}{\affiliation{Institute of Astroparticle Physics, Karlsruhe Institute of Technology, 76021 Karlsruhe, Germany}}
\newcommand{\Pitt}{\affiliation{Department of Physics \& Astronomy, University of Pittsburgh, Pittsburgh, PA 15260, USA}}
\newcommand{\Washington}{\affiliation{Center for Experimental Nuclear Physics and Astrophysics and Department of Physics, University of Washington, Seattle, WA 98195, USA}}
\newcommand{\Ghent}{\affiliation{Department of Physics and Astronomy, Ghent University, 9000 Ghent, Belgium}}

\author{A.~Ashtari~Esfahani}\altaffiliation{Present Address: TRIUMF, Vancouver, BC V6T 2A3, Canada}\Washington
\author{S.~Bhagvati}\PennState
\author{H.~P.~Binney}\MIT
\author{S.~B\"oser}\Mainz
\author{M.~J.~Brandsema}\PennStateARL
\author{N.~Buzinsky}\altaffiliation{Present Address: Center for Experimental Nuclear Physics and Astrophysics and Department of Physics, University of Washington, Seattle, WA 98195, USA}\MIT
\author{R.~Cabral}\Indiana
\author{M.~C.~Carmona-Benitez}\PennState
\author{C.~Claessens}\Washington
\author{L.~de~Viveiros}\PennState
\author{A. El Boustani}\Mainz
\author{M.~G.~Elliott}\Arlington
\author{S.~Enomoto}\Washington
\author{M.~Fertl}\Mainz
\author{J.~A.~Formaggio}\MIT
\author{B.~T.~Foust}\PNNL
\author{J.~K.~Gaison}\PNNL
\author{P.~Harmston}\Illinois
\author{K.~M.~Heeger}\Yale
\author{B.~J.~P.~Jones}\Arlington
\author{E.~Karim}\Pitt
\author{K.~Kazkaz}\Livermore
\author{P.~T.~Kolbeck}\Washington
\author{A.~Kurmus}\Washington
\author{M.~Li}\MIT
\author{A.~Lindman}\Mainz\Berkeley
\author{C.-Y.~Liu}\Illinois
\author{T.~Luo}\Berkeley
\author{C.~Matth\'e}\Mainz
\author{R.~Mohiuddin}\Case
\author{B.~Monreal}\Case
\author{B.~Mucogllava}\Mainz
\author{R.~Mueller}\email[Corresponding author: ]{rjm6826@psu.edu}\PennState 
\author{A.~Negi}\Arlington
\author{J.~A.~Nikkel}\altaffiliation{Present Address: Bamfield Marine Sciences Centre, Bamfield, British Columbia, Canada}\Yale
\author{E.~Novitski}\Washington
\author{N.~S.~Oblath}\PNNL
\author{M.~Oueslati}\Indiana
\author{J.~I.~Pe\~na}\MIT
\author{W.~Pettus}\Indiana
\author{A.~W.~P.~Poon}\Berkeley
\author{V.~S.~Ranatunga}\Case
\author{R.~Reimann}\Mainz
\author{A.~L.~Reine}\Indiana
\author{R.~G.~H.~Robertson}\Washington
\author{G.~Rybka}\Washington
\author{L.~Salda\~na}\Yale
\author{V.~Sharma}\Pitt
\author{P.~L.~Slocum}\Yale
\author{F.~Spanier}\Heidelberg
\author{J.~Stachurska}\Ghent
\author{Y.-H.~Sun}\Washington
\author{P.~T.~Surukuchi}\Pitt
\author{A.~B.~Telles}\Yale
\author{F.~Thomas}\Mainz
\author{L.~A.~Thorne}\Mainz
\author{T.~Th\"ummler}\KIT
\author{M.~Turqueti}\Berkeley
\author{W.~Van~De~Pontseele}\MIT\Mines
\author{B.~A.~VanDevender}\Washington\PNNL
\author{T.~E.~Weiss}\Yale
\author{M.~Wynne}\Washington
\author{A.~Ziegler}\altaffiliation{Present Address: HRL Laboratories LLC, Malibu, CA 90265, USA}\PennState
\collaboration{Project 8 Collaboration}\noaffiliation

%\author{Project 8 Collaboration}
%\affiliation{P8}
%-------------------------------------------------------------------------------------------------%
\begin{abstract}
    Many experimental methods in physics require understanding radiation from single particles into non-trivial electromagnetic mode structures. 
    Such characterization is critical for Cyclotron Radiation Emission Spectroscopy (CRES), an advancing new measurement technique that has the potential to greatly benefit fundamental physics measurements. 
    % The CRES technique uses the frequencies of the cyclotron radiation of charged particles to infer their energies.
    In CRES, charged particles emit cyclotron radiation at frequencies that provide their energy measurement. %
    As a notable example, the Project 8 experiment aims to kinematically infer the neutrino mass by measuring the energies of electrons emitted in tritium beta decay using CRES. %
    In near-term realizations of Project 8, resonant cylindrical cavities will be used for CRES readout, in a configuration with a magnetic field oriented along the symmetry axis, and electrons following helical cyclotron trajectories confined to the cavity interior. The physics of electromagnetic radiation in these environments is complicated, since it involves both the motion of the emitting particle and the mode structure imposed by the cavity. In this work, we derive and validate an analytic model for how an oscillating, trapped electron radiates into cavity modes, and the power and frequency content of the radiation that can be read out from these events. These results can be used to guide the design of cavities for future CRES and other experiments.
\end{abstract}

\maketitle

\tableofcontents
%-------------------------------------------------------------------------------------------------%
%%% Introduction
\section{Introduction}
The coupling of a single charged particle to spatially confined classical electromagnetic fields is a rich area of study with many applications in physics. In these cases, the spontaneous emission rate of the particle is modified because it couples to a different density of states compared to that of free space, a phenomenon known as the ``Purcell effect.'' The Purcell factor, which multiplies the free space emission rate when an emitter is strongly coupled to a cavity mode (indexed $\alpha$), was first estimated as~\cite{Purcell1995}
\begin{align}
    F_{\alpha} &= \frac{6 \pi c^3 Q_{\alpha}}{\omega^3 V} , \label{eqn:purcellfactor}
\end{align}
with the speed of light $c$, the cavity loaded quality factor $Q_{\alpha}$ of the mode, the angular frequency of emission $\omega$, and the cavity volume $V$. The Purcell factor can be more accurately determined if the particle's interaction with the environment's mode structure is modeled in detail. The ability to enhance or inhibit spontaneous radiation has been an important tool in precision physics, notably in the measurement of the electron magnetic dipole moment, wherein the lifetime of an electron confined in a Penning trap is extended beyond its free-space value by tuning the cyclotron frequency away from the mode structure of a weakly coupled cavity resonator~\cite{gabrielseObservationInhibitedSpontaneous1985, brownCyclotronMotionMicrowave1985,brownGeoniumTheoryPhysics1986,vandyckNewHighprecisionComparison1987,brownCyclotronMotionPenningtrap1988,odomFullyQuantumMeasurement2004,hannekeCavityControlSingleelectron2007,hannekeCavityControlSingleelectron2011,fanMeasurementElectronMagnetic2023}. The body of work regarding the lifetime and frequency shifts of an electron undergoing cyclotron motion within a cylindrical cavity~\cite{brownCyclotronMotionMicrowave1985, hannekeCavityControlSingleelectron2007, hannekeCavityControlSingleelectron2011} is an important foundation for our work here, where we investigate a similar interaction for the purpose of measuring electron energies using Cyclotron Radiation Emission Spectroscopy (CRES). 

The Project 8 collaboration seeks to measure the neutrino mass by performing CRES with electrons from a high-activity gaseous tritium source. CRES is a technique in which the frequency of radiation from a single charged particle in cyclotron motion provides a measurement of its kinetic energy~\cite{monrealRelativisticCyclotronRadiation2009,asnerSingleElectronDetection2015}. Energy measurements of tritium beta decay electrons near the endpoint are a well-known route to measuring the neutrino mass $m_\beta$~\cite{formaggioDirectMeasurementsNeutrino2021}. The most sensitive limits on this value to date have been achieved with this method \cite{akerDirectNeutrinomassMeasurement2025}. Project 8 has pioneered the tritium endpoint method with CRES with the goal of achieving a sensitivity $m_\beta < 40 \,\text{meV}/c^2$ ~\cite{ashtariesfahaniDeterminingNeutrinoMass2017,ashtariesfahaniBayesianAnalysisFuture2021,ashtariesfahaniProject8Neutrino2022}. 

The extreme rarity of signal events near the spectral endpoint ($\sim 10^{-13}$ of all decays fall in the last 1 eV) necessitates a detector with a large amount of source gas to achieve the required statistical sensitivity~\cite{ashtariesfahaniBayesianAnalysisFuture2021,ashtariesfahaniProject8Neutrino2022}. The precise frequency measurements required for CRES, however, depend on long observation times for each electron. Since the observation time is limited by scattering with the source gas, the gas density must be kept low, which in turn necessitates a large volume detector. Another important factor driving the design of CRES experiments is the optimization of the coupling of the cyclotron radiation to the detector. CRES has been demonstrated in waveguides~\cite{asnerSingleElectronDetection2015,ashtariesfahaniDeterminingNeutrinoMass2017,byronFirstObservationCyclotron2023,ashtariesfahaniTritiumBetaSpectrum2023,ashtariesfahaniCyclotronRadiationEmission2024} and described theoretically by Project 8 \cite{ashtariesfahaniElectronRadiatedPower2019} and He6-CRES \cite{buzinskyLarmorPowerLimit2024}. It has also been explored with free-space antennas~\cite{ashtariesfahaniDeterminingNeutrinoMass2017,ashtariesfahaniAntennaArraysNeutrino2025,ashtariesfahaniSYNCASyntheticCyclotron2023,ashtariesfahaniRealtimeSignalDetection2024,withingtonQuantumNoiseLimited2024}. However, scaling these concepts to the required size presents significant challenges. This paper focuses on a related approach: the use of resonant cavities for CRES.

\begin{figure}[ht] 
    \centering
    \includegraphics[width=.48\textwidth]{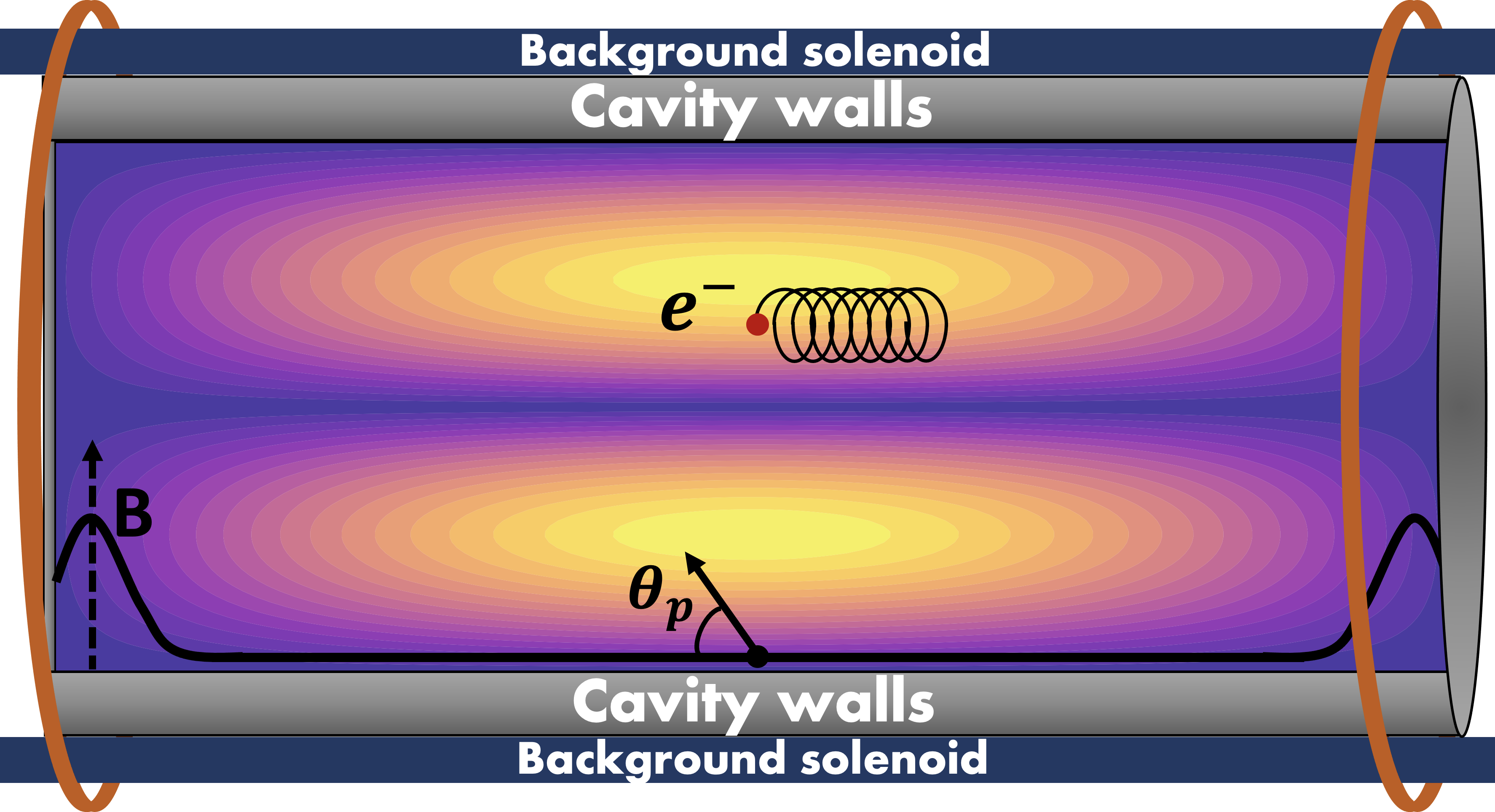}
    \caption{A cartoon of a cavity CRES experiment, showing a cross-section of the cavity's cylindrical symmetry axis. The electron is emitted with pitch angle $\theta_p$ relative to the magnetic field at the trap center and undergoes axial motion (parallel to the cavity's symmetry axis). The electron is bound by the magnetic bottle field produced by the two pinch trapping coils near the ends of the cavity. The heat map is the electric field magnitude of a cylindrical TE$_{011}$ mode.}
    \label{fig:cavityelectron}
\end{figure}

In CRES experiments, an electron's motion perpendicular to the local magnetic field $\mathbf{B}$ follows a helical path with cyclotron frequency
\begin{align}
    \omega_c &= \frac{eB}{\gamma m_e} = \frac{eB}{m_e + E_{kin}/c^2} , \label{eqn:omega_c}
\end{align}
where $B = |\mathbf{B}| $ is the magnetic field strength, $m_e$ is the electron mass, $e$ is the elementary electric charge ($e>0$), and , and $\gamma$ is the Lorentz factor for electron velocity $\mathbf{v}$. When the electron has non-zero velocity parallel to the background magnetic field, the electron follows a helical trajectory. The cyclotron radius is given by
\begin{align}
    \rho_c = \frac{m_e \gamma v_\perp }{e B} ,
\end{align}
where $v_\perp = |\mathbf{v}_\perp| = |\mathbf{v} \sin \theta_p|$ is the magnitude of the electron's velocity perpendicular to $\mathbf{B}$, and $\theta_p$ is the pitch angle between the electron's momentum and the magnetic field. Electrons must be confined axially to be observed for long enough to infer the cyclotron frequency. This is typically done with a magnetic bottle~\cite{ashtariesfahaniElectronRadiatedPower2019}, where magnetic trapping coils are placed at either end of the trapping region as in \cref{fig:cavityelectron}. The addition of two magnetic barriers localizes the electron's motion if its pitch angle at the trap center meets the criterion
\begin{align}
    \theta_p \geq \arcsin\left(\sqrt{\frac{B_0}{B_{\text{max}}}}\right) ,
\end{align}
with $B_{\text{max}}$ the maximum magnetic field of the trap and $B_0$ the magnetic field strength at the trap center. The addition of the trapping field introduces inhomogeneities in the magnetic field experienced by the charged particles, giving rise to slow grad-B drift motion with velocity $\mathbf{v}_{grad-B} = \frac{v_{\perp}^2}{2 B^2 \omega_{c}} \mathbf{B} \times \nabla_{\perp} B$ ~\cite{jacksonClassicalElectrodynamics2009}, where $\nabla_{\perp} B$ is the gradient of $B$ in the plane orthogonal to the dominant component of $\mathbf{B}$. The grad-B motion is analogous to magnetron motion in a Penning trap. The drift motion is quite slow compared to the cyclotron motion~\cite{ashtariesfahaniElectronRadiatedPower2019,ashtariesfahaniRealtimeSignalDetection2024}, such that the trapped helical trajectory described above is valid in the short-time approximation. The additional effects of simplified grad-B drift motion in the radiation spectrum will be discussed in \cref{sec:electronCoupling}.

The electron's rate of energy loss is tied to the rate of frequency change through \cref{eqn:omega_c} (an electron emitting very strongly will also change frequency quickly). The free-space power radiated by the electron is proportional to the square of frequency~\cite{ashtariesfahaniDeterminingNeutrinoMass2017}:
\begin{align}
    P(\gamma,\theta_p)&=\frac{e^2}{6\pi \epsilon_0 c}\omega_c^2 \gamma^2 (\gamma^2-1)\sin^2{\theta_p} , \label{eqn:freespacerate}
\end{align}
where $\epsilon_0$ is the permittivity of free space. Due to Purcell enhancement of the above (\cref{eqn:purcellfactor}), the rate of emission for electrons in CRES cavity experiments is partially tied to how many cavity modes are resonant near the cyclotron frequency. Operating the cavity at frequencies high above its fundamental resonant frequency increases the frequency mode density~\cite{hillElectromagneticFieldsCavities2009}. To exert more control of the electron energy loss rate, Project 8 is exploring operating the resonant cavity detector at frequencies near one of the cavity's fundamental modes. In these cases, the observed CRES signal power can scale more favorably to low frequencies via the Purcell factor in \cref{eqn:purcellfactor} than the free-space emission rate does. A cavity CRES experiment with a large detector volume ($\sim 10 \ \text{m}^3$) may then be realized with a single cavity operating in hundreds of MHz to tens of GHz range.

Cavity CRES presents additional theoretical challenges, relative to previous work on electron-cavity interactions in precision measurements, that require a distinct treatment. While existing frameworks for the Purcell enhancement/inhibition of cyclotron radiation in cavities~\cite{gabrielseObservationInhibitedSpontaneous1985, brownCyclotronMotionMicrowave1985,brownGeoniumTheoryPhysics1986,brownCyclotronMotionPenningtrap1988,hannekeCavityControlSingleelectron2007,odomFullyQuantumMeasurement2004, hannekeCavityControlSingleelectron2011,fanMeasurementElectronMagnetic2023} have been developed primarily for Penning trap experiments with electrons confined very close to the cavity center, CRES experiments detect an ensemble of electrons with allowed trajectories in a much larger fraction of the cavity volume. Additionally,  unlike in Penning trap experiments, in CRES the signal comes directly from collected cyclotron radiation. This makes it crucial to further develop a model of the signal and noise power extracted from one or more cavity readouts.
% CRES has been demonstrated only in waveguides~\cite{EsfahaniAshtari:2021idb,asnerSingleElectronDetection2015,byronFirstObservationCyclotron2023,buzinskyLarmorPowerLimit2024,ashtariesfahaniElectronRadiatedPower2019,esfahaniDeterminingNeutrinoMass2017,ashtariesfahaniCyclotronRadiationEmission2023}. 
% An alternative approach is a large, low-frequency cavity to contain the source. 
% The free-space power radiated by the electron is proportional to the square of frequency~\cite{esfahaniDeterminingNeutrinoMass2017}:
% \begin{eqnarray}
%     P(\gamma,\theta_p)&=&\frac{e^2}{6\pi \epsilon_0 c}\omega_c^2 \gamma^2 (\gamma^2-1)\sin^2{\theta_p} ,
% \end{eqnarray}
% where $\epsilon_0$ is the permittivity of free space. This equation implies a significant loss of signal strength at low frequencies. However, a number of countervailing factors emerge that outweigh the low frequency scaling and favor the use of large cavities. Firstly, the Purcell enhancement depends on cavity loaded $Q$ and on the total volume; in the cases of interest here, it is approximately $Q/80$ and can be an order of magnitude even before taking into account the low efficiency for an antenna array, which may be only tens of percent. At the same time, the noise power emitted from the cavity follows a similar frequency profile to the power emitted, but is suppressed by a factor of the cavity unloaded $Q$. The determination of cavity CRES signal and noise powers is the major theme of this paper.

We develop a broad theoretical framework that accounts for the three-dimensional motion of the electron and its coupling to the complete electromagnetic mode structure of resonant cavities.
In \cref{sec:responseModel}, we derive an analytic cavity response model for an arbitrary cavity mode. In \cref{sec:electronCoupling}, we evaluate the power spectrum of an electron in cyclotron motion in the presence of ideal cylindrical TE and TM modes. In \cref{sec:axialMotion}, we apply the formalism of \cref{sec:responseModel,,sec:electronCoupling} to the case of electron cyclotron motion that is periodically traversing the cavity's cylindrical symmetry axis. In \cref{sec:noiseModel}, we develop and characterize a model of how noise is generated and delivered to the cavity readout system.

\section{Signal Power in a Resonant Cavity} \label{sec:responseModel}
We first model the cavity response to a general emitter using the well-established modal decomposition technique ~\cite{kurokawaExpansionsElectromagneticFields1958,collinFoundationsMicrowaveEngineering2001,bladelElectromagneticFields2007}. The cavity response model detailed below accounts for the expected loss mechanisms in a cavity CRES experiment, and aims to translate the cavity response function to the Purcell enhancement factor of a general emitter. Later on, we apply the model to the special case of an electron undergoing cyclotron motion perpendicular to the cavity's symmetry axis. 

Any electromagnetic field can be Helmholtz-expanded in irrotational (curl-free) and solenoidal (divergence-free) modes as
\begin{align}
    \mathbfcal{E}(\mathbf{r},t) &= \sum_{\alpha} a_{\alpha}(t) \mathbf{f}_{\alpha}(\mathbf{r}) + \sum_\beta b_\beta(t) \mathbf{e}_\beta(\mathbf{r})  \label{eqn:EfieldExpansion}\\
    \mathbfcal{H}(\mathbf{r},t) &= \sum_{\alpha} c_{\alpha}(t) \mathbf{g}_{\alpha}(\mathbf{r}) + \sum_\beta d_\beta(t) \mathbf{h}_\beta(\mathbf{r}),  \label{eqn:HfieldExpansion}
\end{align}
where $\mathbf{f}_{\alpha}$ are electric irrotational (EI) with ${\nabla\times\mathbf{f}_{\alpha}=0}$, $\mathbf{g}_{\alpha}$ are magnetic irrotational (MI) with ${\nabla\times\mathbf{g}_{\alpha}=0}$, and ${\mathbf{e}_{\beta}}$ and ${\mathbf{h}_{\beta}}$ are solenoidal with $\nabla\cdot\mathbf{e}_{\beta}=\nabla\cdot\mathbf{h}_{\beta}=0$. 
The solenoidal modes are resonant within cavities, and the transverse electric (TE) and transverse magnetic (TM) modes are composed of solenoidal modes (or linear combinations thereof).
Each basis field is a solution to an eigenvalue equation associated with the eigenfrequencies $\omega_{\alpha}$, detailed in \cref{app:helmholtz}. 
We assume mutual orthogonality for both the electric field eigenmodes $\mathbf{e}_{\beta}, \mathbf{f}_{\alpha}$ and the magnetic field eigenmodes $\mathbf{h}_{\beta}, \mathbf{g}_{\alpha}$. 
The relationship between the resonating electric and magnetic fields is%
\begin{align}
    \nabla\times\mathbf{e}_{\alpha}(\mathbf{r}) &= -i \mu \omega_\alpha \mathbf{h}_{\alpha} \label{eqn:FaradaySourceFreeSoln}\\
    \nabla\times\mathbf{h}_{\alpha}(\mathbf{r}) &= i \epsilon \omega_\alpha \mathbf{e}_{\alpha} , \label{eqn:AmpereSourceFreeSoln}
\end{align}
where $\epsilon$ is the real part of permittivity of the cavity volume and $\mu$ is the cavity volume's permeability. This relation holds when the conductivity of the cavity volume, $\sigma(\mathbf{r})$ is small, such that it may be treated as a perturbation to the non-conductive cavity mode. We assume this is the case in the subsequent calculations.\footnote{The relationship between the fields of a cavity with conductive volume may be modeled directly by adding $\sigma(\mathbf{r}) \mathbf{e}_\alpha(\mathbf{r})$ to the RHS of \cref{eqn:AmpereSourceFreeSoln}.}

In practice, the cavity response has a significant dependence on radio frequency (RF) design choices, including single or multiple readouts, overcoupled operational modes (when more energy is extracted from the readout than is lost to heat, typically $Q_{\alpha}\sim\mathcal{O}(100-1000)$), Fano resonances, and heavily perturbed eigenmodes. 
We will next detail our cavity response model in order to specify the mechanisms of our expected signal loss and readout.
%------------------------------------------------------------------------%
\subsection{Modeling energy extraction} \label{ssec:cavityboundaries}
The energy stored in cavities can be lost through resistive heating of imperfectly conducting walls, resistive material within the cavity volume, and by emission through ports. The coupling of a cavity mode to these loss channels extends the relevant electromagnetic domain to include the cavity walls and RF components further down the readout chain. To avoid modeling these complex subsystems, we approximate the fields entering ports and conductors using impedance boundaries. These boundary conditions are further employed to model signal extraction by relating the electromagnetic field amplitudes in the cavity to the power flux through the boundaries. 

% The cavity volume in our model is distinguished from the port volume by the hallmark of a set of resonating eigenmodes (this distinction is necessary so that losses can propagate into a continuum). 
We define the cavity volume as the part of the apparatus describable by a set of resonant eigenmodes, while the port volume is the part of the apparatus with fields describable by the cavity eigenmodes and a set of non-resonant waveguide modes~\cite{kristensenTheoryCoupledModes2017,lobanovResonantstateExpansionThreedimensional2018}. At the interface of the cavity volume and port, we assume the electric and magnetic field coefficients obey the scalar impedance boundary condition
\begin{align}
    \mathbf{\hat{n}}(\mathbf{r})\times\mathbfcal{E}(\mathbf{r},t) = \mathbfcal{J}_s (\mathbf{r},t) = Z(\mathbf{r}) \mathbfcal{H}_{tan}(\mathbf{r},t) , \label{eqn:impedanceBdry}
\end{align}
where $Z(\mathbf{r})$ is the complex impedance on the cavity volume boundary $S = \bigcup S_{i}$ (with the unit vector $\mathbf{\hat{n}}(\mathbf{r})$ pointing out from the boundary). The tangential magnetic field and surface current at the boundary are $\mathbfcal{H}_{tan}(\mathbf{r},t)$ and $\mathbfcal{J}_s$, respectively. The unit normal vector $\hat{\mathbf{n}}(\mathbf{r})$ points outward. Each boundary segment $S_{i}$ is assumed to have a distinct impedance, and for simplicity, we will define the boundary segments such that $Z(\mathbf{r})$ is constant on each $S_{i}$.

\begin{figure}[ht] 
    \centering
    \includegraphics[width=.48\textwidth,trim=0 0 4 0,clip]{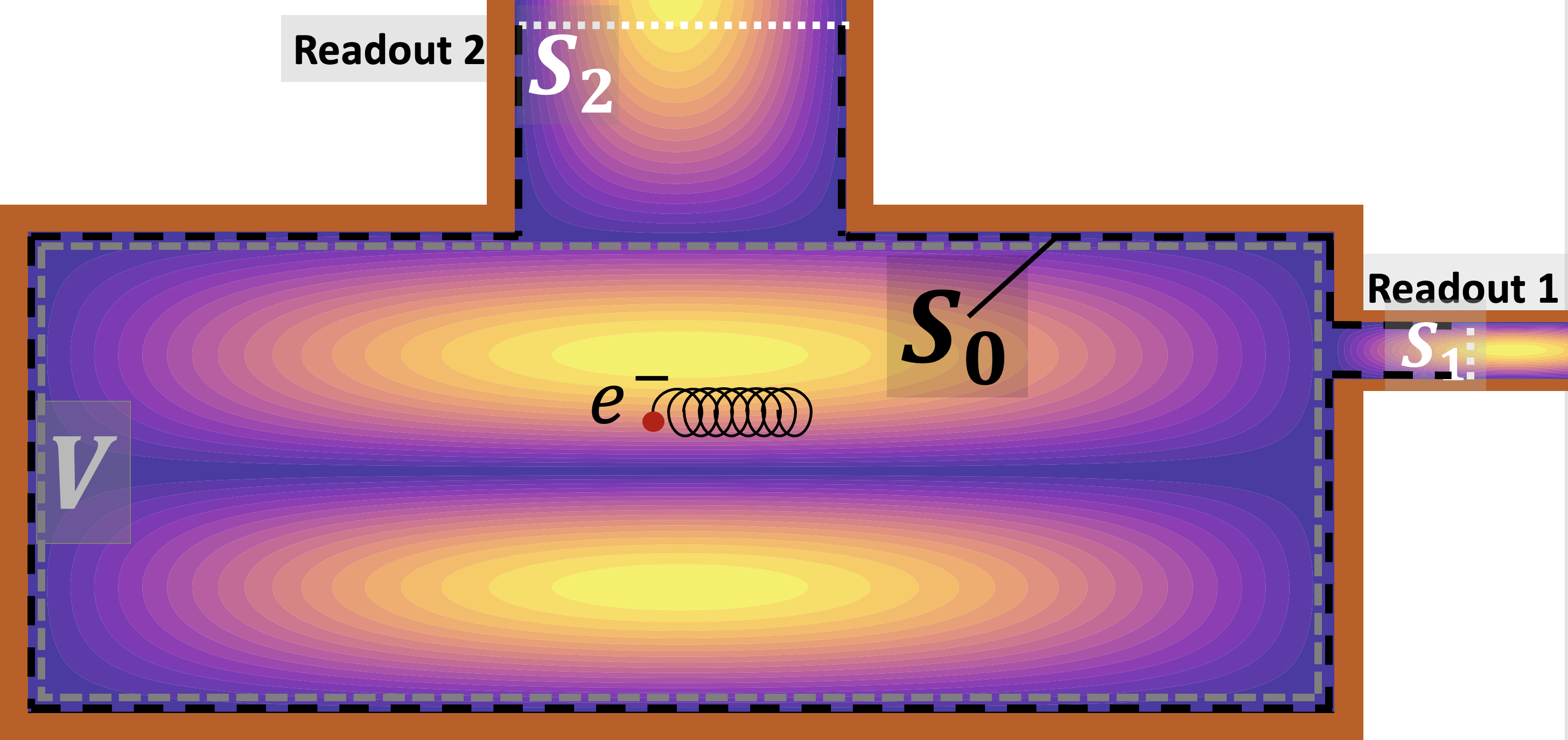}
    \caption{Sketch of the cavity volume and port boundaries. $S_0$ is the surface of the cavity walls (black outline), and $S_{i>0}$ are the cavity readout ports (each white dashed line). Each labelled section of the cavity boundary has a distinct impedance and quality factor. The cavity volume is enclosed in the gray dotted line. The background electric field pattern is representative of a cylindrical TE$_{011}$-like mode that is strongly coupled to multiple readouts.}
    \label{fig:cavitybdry}
\end{figure}

The cavity walls, denoted by the surface $S_0$ in \cref{fig:cavitybdry}, can be characterized by a skin depth into which the electromagnetic fields penetrate, leading to energy loss and resonant frequency shifts. 
This wall impedance perturbs the eigenmodes' frequencies, spatial structure, and quality factor. 
The port impedance boundaries are denoted $S_{i}$ with ${i>0}$ (with examples shown in \cref{fig:cavitybdry}), and are assumed to be located either at the interface of an RF component or at a set reference point to one of them, such that the input impedance is known from the RF specification and transmission line theory.
This assumption is justified since many RF components have standardized input impedance across their recommended operating frequency range.

Additionally, we will assume outgoing boundary conditions beyond the port interface, such that no radiation is reflected back into the cavity from past those boundaries. 
% As a result, the cavity electric fields no longer completely describes the radiation field at and beyond the boundary~\cite{kristensenTheoryCoupledModes2017, lobanovResonantstateExpansionThreedimensional2018}. 
% The mode electric field at these interfaces in our model is zero, while the non-zero outgoing field is described by the waveguide excitations dictated by the impedance boundary conditions. 
The loaded quality factor of the cavity for non-dispersive materials is defined as~\cite{lalanneLightInteractionPhotonic2018}
\begin{align}
    Q_{\alpha} &= \RE[\omega_{\alpha}] \frac{\text{Total stored energy}}{\text{Total power lost}} = -\frac{\RE[\omega_{\alpha}]}{2 \IM[\omega_{\alpha}]} \\
    &= \RE[\omega_{\alpha}] \frac{\frac{\mu}{4} \int_{V} |\mathbf{h}_{\alpha}(\mathbf{r})|^{2} dV + \frac{\epsilon}{4} \int_{V} |\mathbf{e}_{\alpha}(\mathbf{r})|^{2} dV}{\frac{\sigma}{2} \RE[\omega_{\alpha}] \int_V |\mathbf{e}_{\alpha}(\mathbf{r})|^{2} dV + \frac{1}{2} P_{\alpha}^{S} } ,
\end{align}
where $P_{\alpha}^{S} = \int_{S} \RE\left[ \mathbf{e}_{\alpha} \times \mathbf{h}^{*}_{\alpha}(\mathbf{r})\right] \cdot \mathbf{\hat{n}}(\mathbf{r}) dS$ is the total power radiated through the boundary.
The energy loss in the system is related to the complex eigenfrequency through its imaginary component, 
\begin{align}
    \omega_{\alpha} = \RE[\omega_{\alpha}]\left( 1 - i \frac{1}{2 Q_{\alpha}} \right) . \label{eqn:defnOmegaAlpha}
\end{align}
The resonant frequency and loaded quality factor also determine the cavity mode's bandwidth (BW), defined as the full width at half maximum of the mode's power spectral density: \begin{align}
    \text{BW}_{\alpha} = \RE[\omega_{\alpha}]/Q_{\alpha} . \label{eqn:BWdefn}
\end{align}%
Each lossy boundary segment has an associated quality factor that contributes to the overall loaded $Q$:
\begin{align}
    \int_{S} \RE\left[ \mathbf{e}_{\alpha} \times \mathbf{h}^{*}_{\alpha}(\mathbf{r})\right] \cdot \mathbf{\hat{n}}(\mathbf{r}) dS
    &= \RE[\omega_{\alpha}] \frac{2 W_{\alpha}}{Q_{\alpha,i}}, \label{eqn:Qdefn}
\end{align}
with $W_{\alpha}$ the total stored energy in the mode. The cavity electric and magnetic fields $\mathbf{e}_{\alpha}$, $\mathbf{h}_{\alpha}$ are normalized such that ${W_{\alpha} = \mu/2}$. %
The boundary quality factors $Q_{\alpha,i}$ are related to the loaded quality factor by
\begin{align}
\frac{1}{Q_{\alpha}} = \sum_i \frac{1}{Q_{\alpha,i}}, \label{eqn:impedanceAndQ}
\end{align}
when conductive losses within the cavity are negligible. We will later introduce the cavity bulk conductivity as a perturbative term in Maxwell's equations in order to better examine its role in non-resonant emission. %
We specifically denote the cavity walls with the index 0 (the surface $S_0$) and denote the corresponding unloaded quality factor $Q_{\alpha,0}$.
% When the cavity is excited close to resonance for a single mode $\alpha$ that is well separated from the other modes in frequency, 
% ${\mathbfcal{E}(\mathbf{r},t) \approx b_{m}(t) \mathbf{e}_{\alpha}(\mathbf{r})}$, ${\mathbfcal{H}(\mathbf{r},t) \approx d_{m}(t) \mathbf{h}_{\alpha}(\mathbf{r})}$,
% the power through a port can be approximated by~\cite{collinFoundationsMicrowaveEngineering2001}
% \begin{align}
%     \int_{S_{i}} \left(\mathbf{\hat{n}}(\mathbf{r})\times \mathbfcal{E}(\mathbf{r},t)\right)\cdot \mathbf{h}^{*}_{\alpha}(\mathbf{r}) dS 
%     &\approx Z_{i} d_{\alpha}(t) \frac{\mu \RE[\omega_{\alpha}]}{Q_{\alpha,i} \RE\left[Z_{i}\right]}, \label{eqn:Qdefn}
% \end{align}
% where $Q_{\alpha,i}$ is the quality factor of the mode with index $\alpha$ due to the presence of the boundary $S_{i}$, $\RE[\omega_{\alpha}]$ is the real part of the complex eigenfrequency, and $Z_{i}$ is the impedance of the boundary indexed $i$. 
% We define the loaded $Q$, $Q_{\alpha}$ in terms of the surface quality factors $Q_{\alpha,i}$,
% \begin{align}
% \frac{1}{Q_{\alpha}} = \frac{1}{Q_{\alpha, \sigma}} + \sum_i \frac{1}{Q_{\alpha,i}}. \label{eqn:impedanceAndQ}
% \end{align}
% \begin{align}
% \frac{1}{Q_{\alpha}} = \sum_i \frac{1}{Q_{\alpha,i}}. \label{eqn:impedanceAndQ}
% \end{align}
% We specifically denote the cavity walls with the index 0 (the surface $S_0$) and denote the corresponding unloaded quality factor $Q_{\alpha,0}$.
%------------------------------------------------------------------------%
\subsection{Cavity response model} \label{ssec:cavityresponse}
In this section, we derive expressions for the electric and magnetic field amplitudes ($a_{\alpha}, b_{\alpha}$ and $c_{\alpha}, d_{\alpha}$ in \cref{eqn:EfieldExpansion,eqn:HfieldExpansion}) by decomposing the source current into the eigenmodes of the cavity's electric field. This makes the calculation of the power emitted into each mode more tractable. 

Working in the frequency domain, we use field amplitudes with the coefficients 
\begin{align}
    a_{\alpha}(t) = \mathcal{F}^{-1} \{a_{\alpha}(\omega)\} = \frac{1}{2\pi} \int a_{\alpha}(\omega) e^{i\omega t} d\omega ,
\end{align}
where $\mathcal{F}^{-1}$ denotes the inverse Fourier transform. %
We use block-letter symbols to refer to the vector quantities of Maxwell's equations in the frequency domain.
The fields in Faraday's law ${\nabla\times\mathbf{E}(\mathbf{r},\omega) =  -i \omega\mu\mathbf{H}(\mathbf{r},\omega)}$ are expanded in cavity eigenmodes as in \cref{eqn:EfieldExpansion,eqn:HfieldExpansion}. Taking the inner product with the cavity electric fields and integrating over the cavity volume yields
\begin{align}
    0 &= -i\mu \omega c_{\alpha}(\omega)\\
    -i \omega_{\alpha} b_{\alpha}(\omega) &= -i\mu \omega d_{\alpha}(\omega) \label{eqn:b-cup2},
\end{align}
where the first equation is for the irrotational modes (where $\nabla\times\mathbf{f}_{\alpha}(\mathbf{r})=0$ eliminates the left-hand-side of Faraday's law), and the second equation is for resonating modes where \cref{eqn:FaradaySourceFreeSoln} has been used in the left-hand-side. Similarly, Ampere's law $\nabla\times\mathbf{H}(\mathbf{r},\omega) =  \sigma (\mathbf{r}) \mathbf{E}(\mathbf{r},\omega) + i \omega \epsilon\mathbf{E}(\mathbf{r},\omega) + \mathbf{J}(\mathbf{r},\omega) $ is used to produce the remaining coefficients
\begin{align}
    0 &= \left(\sigma + i\epsilon \omega\right)a_{\alpha}(\omega) + \int_V \mathbf{J}(\mathbf{r},\omega)\cdot \mathbf{f}_{\alpha}(\mathbf{r}) dV\\
    i \omega_{\alpha} \mu d_{\alpha}(\omega) &= \left(\sigma + i\epsilon \omega\right) b_{\alpha}(\omega) + \int_V \mathbf{J}(\mathbf{r},\omega)\cdot \mathbf{e}_{\alpha}(\mathbf{r}) dV \label{JFT},
\end{align}
where we have assumed a uniform conductivity throughout the cavity volume, and used \cref{eqn:FaradaySourceFreeSoln} in the left-hand-side of the second equation for resonating modes. Above, we have set $\int_V \left| \mathbf{e}_{\alpha} \right|^2 dV = 1$. The resonating field coefficients $b_{\alpha}$ and $d_{\alpha}$ are now linearly coupled. Solving this system yields
\begin{align}
    a_{\alpha}(\omega) &= \frac{-1}{\sigma+i\epsilon\omega}\int_V \mathbf{J}(\mathbf{r},\omega) \cdot \mathbf{f}_{\alpha}(\mathbf{r}) dV \label{eqn:am}\\
    b_{\alpha}(\omega) &= \frac{-1}{\sigma+i\epsilon\omega + \frac{k_{\alpha}^2}{i\mu\omega}}\int_V \mathbf{J}(\mathbf{r},\omega) \cdot \mathbf{e}_{\alpha}(\mathbf{r}) dV \label{eqn:bm} \\
    c_{\alpha}(\omega) &= 0 \label{eqn:cm} \\
    d_{\alpha}(\omega) &= \frac{1}{-i\mu\omega - \frac{k_{\alpha}^2}{\sigma+i\epsilon\omega}} \frac{-k_{\alpha}}{\sigma+i\epsilon\omega} \int_V \mathbf{J}(\mathbf{r},\omega)\cdot \mathbf{e}_{\alpha}(\mathbf{r}) dV  .
\end{align}

The irrotational mode amplitudes \cref{eqn:am,,eqn:cm} are in their final form. 
% As mentioned in the previous section, when driving near an isolated mode, we can relate the surface integrals to loss using \cref{eqn:Qdefn} in the frequency domain. The solenoidal magnetic field amplitude \cref{eqn:b-cup2} becomes
% \begin{align}
%     d_{\alpha} = \frac{k_{\alpha}}{-i\mu\omega - \frac{z'_{\alpha}\mu\omega_{\alpha}}{Q_{\alpha}}} b_{\alpha} \label{eqn:dm_bm}.
% \end{align}
% Solving for $b_{\alpha}$ and putting the denominator into resonant form (see \cref{app:bmderiv}), and using $k=\omega/c=\omega\sqrt{\mu\epsilon}$
% \begin{align}
%     b_{\alpha} &= \frac{-\frac{1}{\epsilon}\left(i\omega + \frac{z'_{\alpha} \omega_{\alpha}}{Q_{\alpha}}\right) \int_V \mathbf{J}\cdot \mathbf{e}_{\alpha} dV  }{\omega_{\alpha}^2 + i\omega \omega_{\alpha} \frac{z'_{\alpha}}{Q_{\alpha}} - \omega^2 + \frac{\sigma}{\epsilon} \left( i\omega + \frac{z'_{\alpha} \omega_{\alpha}}{Q_{\alpha}} \right) } \label{eqn:bmfinal}.
% \end{align}
% The second term in the numerator represents a small correction to the standard cavity electric field mode amplitude. We define the internal quality factor $1/Q_0 = \sigma / \epsilon \omega_m $, and use the wall impedance \cref{eqn:impedanceBdry} to find,
% \begin{align}
%     b_m &= \frac{-\frac{i \omega}{\epsilon}\left(1 + \frac{\omega_m(\text{sgn}(\omega)-i)}{\omega Q_m}\right) \int_V \mathbf{J}\cdot \mathbf{e}_m dV  }{\omega_m^2\left(1 + \frac{(1+i\text{sgn}(\omega))}{Q_m Q_{0,m}}\right) + i \omega \omega_m \left(\frac{(1 + i\text{sgn}(\omega))}{Q_m} + \frac{1}{Q_{0,m}} \right) - \omega^2} \label{eqn:bmfinal}
% \end{align}
Defining 
\begin{align}
L_{\alpha}(\omega) &= \frac{1}{\omega_{\alpha}^2 + i\omega \sigma / \epsilon - \omega^2} \label{eqn:cavitylorentzian} ,
\end{align}
for convenience, where $\omega_{\alpha}$ is related to the loaded quality factor though \cref{eqn:defnOmegaAlpha}, we can write
\begin{align}
    b_{\alpha}(\omega) &= \frac{- i \omega}{\epsilon} L_{\alpha}(\omega) \int_V \mathbf{J}(\mathbf{r},\omega) \cdot \mathbf{e}_{\alpha}(\mathbf{r}) dV \label{eqn:shortbm} \\
    d_{\alpha}(\omega) &= c \omega_{\alpha} L_{\alpha}(\omega ) \int_V \mathbf{J}(\mathbf{r},\omega)\cdot \mathbf{e}_{\alpha}(\mathbf{r}) dV . \label{eqn:shortdm}
\end{align}

These coefficients are the main result of this section, and we will next see how they can be used to calculate the power emitted into the mode. Before we turn our attention to that topic, we note an important limitation of our model. In this derivation we have assumed the eigenmode normalization of high-$Q$ cavities, $\int_V \left| \mathbf{e}_{\alpha} \right|^2 dV = 1$. For extremely low-Q cavities, where much of the stored energy leaves the cavity each cycle, additional complications arise. Foremost, the free space radiated field partly constitutes the mode volume. 
% For the present topic, this means that increasing the readout cable length arbitrarily will decrease the normalization factor similarly, decreasing the field amplitude and power emitted as a result. Since this framework assumes outgoing boundary conditions past the boundary, such that the reflection characteristics of the resonator do not change with increasing the length of the signal line, this feature is unphysical. 
As a consequence, the electric field normalization $\int_V \left| \mathbf{e}_{\alpha} \right|^2 dV$ tends to diverge, and the power emitted in our model incorrectly tends to zero.
We leave to future work this analysis using the framework of quasinormal modes applied to this case of low-$Q$ cavities \cite{sauvanTheorySpontaneousOptical2013,muljarovExactModeVolume2016,lalanneLightInteractionPhotonic2018}.
%------------------------------------------------------------------------%
\subsection{Calculating the power emitted} \label{ssec:poweremitted}
Here we provide the expression for the power transferred from a source current into the cavity modes. The calculation of the power transferred from a source current to the output ports is detailed in \cref{ssec:signalpower}. The time-averaged electromagnetic power delivered from a current $\mathcal{J}(\mathbf{r},t)$ to an electric field $\mathcal{E}(\mathbf{r},t)$ in a single cycle with period $T$ is
\begin{align}
    \bar{P}(t) &= \frac{1}{T} \int^{T+t}_{t} \int_V \mathbfcal{E}(\mathbf{r},t') \cdot \mathbfcal{J}(\mathbf{r}, t') dV dt' . \label{eqn:power-def}
\end{align}
We now focus on the case of an electric current oscillating at a single frequency $\omega_0$. Physical electric currents and fields that are oscillating at a single frequency can be written as
\begin{align}
    \mathbfcal{J}(\mathbf{r},t) &= \frac{1}{2\pi}\left(\mathbf{J}(\mathbf{r},\omega_0) e^{i\omega_0 t} + \mathbf{J}^*(\mathbf{r},\omega_0) e^{-i\omega_0 t}\right) \\
    \mathbfcal{E}(\mathbf{r},t) &= \frac{1}{2\pi}\left(\mathbf{E}(\mathbf{r},\omega_0) e^{i\omega_0 t} + \mathbf{E}^*(\mathbf{r},\omega_0) e^{-i\omega_0 t}\right).
\end{align}
The time-averaged electromagnetic power delivered is then
\begin{align}
    \bar{P}(t) &= \frac{1}{2\pi^2} \RE \left[ \int_V \mathbf{E}(\mathbf{r},\omega_0)\cdot \mathbf{J}^* (\mathbf{r},\omega_0) dV \right] .  \label{eqn:power-def-freq}
\end{align}

The above volumetric integral lends itself to an expansion in terms of cavity modes. We can expand the electric current in terms of cavity eigenmode electric fields:
\begin{align}
    \mathbf{J}(\mathbf{r},\omega) &= \sum_{\alpha} q_{\alpha}(\omega) \mathbf{f}_{\alpha}(\mathbf{r}) + \sum_\beta s_\beta(\omega) \mathbf{e}_\beta(\mathbf{r})  \label{eqn:currentExpansion} ,
\end{align}
which relates to the time domain electric current through
\begin{subequations} \label{eqns:jalpha-exp}
\begin{align}
    q_{\alpha}(\omega) &= \mathcal{F} \left\{ \int_V \mathbfcal{J}(\mathbf{r},t) \cdot \mathbf{f}_{\alpha}(\mathbf{r}) dV \right\} \label{eqn:qnml-def}  \\
    s_{\alpha}(\omega) &= \mathcal{F} \left\{ \int_V \mathbfcal{J}(\mathbf{r},t) \cdot \mathbf{e}_{\alpha}(\mathbf{r}) dV \right\} . \label{eqn:snml-def}
\end{align}
\end{subequations}
The spectral coefficients $s_{\alpha}$ of the current corresponding to the cavity's resonating modes are crucial in calculating the resonant radiated power. The analogous spectral coefficients $q_{\alpha}$ of the cavity's irrotational mode amplitudes are necessary to fully describe the electric current but do not contribute to resonant emission. 

In the cavity, under the assumption that only one mode $\alpha$ is significantly excited, the electric field is $\mathbf{E}(\mathbf{r},\omega) \approx b_{\alpha}(\omega) \mathbf{e}_{\alpha}(\mathbf{r}) $. Using the orthogonality of the eigenmodes and \cref{eqn:bm}, the rate of energy exchange is
\begin{align}
    P_{\alpha}(\omega) &= \frac{-1}{2 \pi^2}\RE\left[ \int_V \left(b_{\alpha}(\omega)\mathbf{e}_{\alpha}(\mathbf{r})\right) \cdot \left(s^*_{\alpha}(\omega) \mathbf{e}_{\alpha}(\mathbf{r}) \right) dV \right] \\
    &= \frac{-1}{2 \pi^2}\RE\left[ b_{\alpha}(\omega) s^*_{\alpha}(\omega) \right] \label{eqn:powerformula} \\
    &= \frac{-1}{2 \pi^2}\RE\left[ L_{\alpha} (\omega) \left|s_{\alpha}(\omega) \right|^2 \right] \\
    &= \frac{1}{2 \pi^2}\RE\left[ \frac{\frac{1}{\epsilon} i \omega  }{\omega_{\alpha}^2 + i\omega \sigma/\epsilon - \omega^2 } \right] \left|s_{\alpha}(\omega) \right|^2 \label{eqn:powerformula-exp} .
\end{align}

For the special case when the cavity is empty ($\sigma=0$), using the definition \cref{eqn:snml-def} in $b_{\alpha}$ (\cref{eqn:shortbm}) one finds that (with $\Omega_{\alpha}=\RE[\omega_{\alpha}]$)
\begin{align}
    P_{\alpha}(\omega) &= \frac{1}{2 \pi^2 \epsilon}\left[ \frac{Q_{\alpha} \frac{\omega}{\Omega_{\alpha}^{2}} \left|s_{\alpha}(\omega) \right|^2 }{1 + \frac{Q_{\alpha}^2}{\Omega_{\alpha}^4} \left(\Omega_{\alpha}^2-\omega ^2 - \frac{\Omega_{\alpha}^2}{(2 Q_{\alpha})^{2}}\right)^2} \right] \label{eqn:powercondensed} .
\end{align}
The formula above represents a common case with a simple closed form, and is useful for illustration. The Purcell signal enhancement from the loaded cavity quality factor is clearly demonstrated, and the signal-diminishing effect of the Purcell mode volume is implicit in the electric field normalization in $s_{\alpha}$. More complicated contributions to the signal amplitude result from the modal decomposition of the electron's current into the cavity mode's electric field. We will return to this expression when we model the noise coming from the cavity in \cref{sec:noiseModel}.

The same calculation may be done for the irrotational modes:
\begin{align}
    P^{EI}_{\alpha}(\omega) &= \frac{1}{2 \pi^2}\frac{\sigma}{\sigma^2 + \epsilon^2 \omega^2}\left|q_{\alpha}(\omega) \right|^2 \label{eqn:powerformula-exp-irr} .
\end{align}
The power emitted in these modes is only non-zero if the cavity volume is conductive. The Project 8 neutrino mass experiment plans to observe many decays within the cavity detector, leading to charged particle pileup. This pileup can be treated effectively as a plasma with non-zero conductivity. Consequently, an electron may lose energy to the cavity's conducting bulk through the Coulomb interaction. However, the plasma density in Project 8 experiments is expected to be small and the power lost into irrotational modes is expected to be negligible.

For modes degenerate in frequency ($\textit{e.g.}$ cylindrical solenoidal modes with radial mode number $n>0$), or even for modes that are simply close in frequency, the power lost to the modes will add. For nearby modes with indices $\alpha_{1}$ and $\alpha_{2}$, $\mathbf{E}(\mathbf{r}, \omega) \approx b_{\alpha_{1}}(\omega) \mathbf{e}_{\alpha_{1}}(\omega) + b_{\alpha_{2}}(\omega) \mathbf{e}_{\alpha_{2}}(\mathbf{r})$. The total power emitted is then the sum
\begin{align}
    P_{\alpha_{1} + \alpha_{2}}\left(\omega\right) 
    &= \frac{1}{2\pi^2} \left( \RE\left[b_{\alpha_{1}}(\omega) s^*_{\alpha_{1}}(\omega)\right] \right. \nonumber\\
    &\myspace{4} \left. + \RE\left[b_{\alpha_{2}}(\omega) s^*_{\alpha_{2}}(\omega)\right] \right) \\
    &= P_{\alpha_{1}}(\omega) + P_{\alpha_{2}}(\omega) .
\end{align}
This holds when the modes are orthogonal on the boundary, which means that their losses won't couple. Specifically for the case of degenerate $n>0$ modes with identical quality factors, the total power is proportional to $\left|s_{\alpha_{1}}(\omega) \right|^2 + \left|s_{\alpha_{2}}(\omega) \right|^2$.
\section{CRES Phenomenology in a Cylindrical Cavity} \label{sec:electronCoupling}
To apply the modal decomposition method, the current of an electron undergoing cyclotron motion must be expressed in the frequency domain in terms of the cavity modes. Here we decompose the current into harmonics of the cyclotron frequency using the Graf Addition Theorem \cite{abramowitzHandbookMathematicalFunctions2013}, a technique first employed for CRES by the He6-CRES collaboration in \cite{buzinskyLarmorPowerLimit2024}. We note that the coupling integral \cref{eqn:snml-def} is the same in both the cylindrical waveguide and cylindrical cavity cases up to the $z$ normalization of the cavity mode. This section details the calculation for all electric field modes within the cylindrical cavity. The same steps for the resonating TE/TM modes will be used to calculate the result for the electric irrotational modes, which contain information about the electron's near field. This calculation has potential applications to the near-field dynamics due to the electron's image charge fully within this mathematical framework. 

\subsection{Eigenmodes of the cylindrical cavity} \label{ssec:cylindricalModes}
\begin{figure}[ht]
    \centering
    \begin{tikzpicture}[scale=1.2,>=Latex,font=\Large]
\usetikzlibrary{arrows.meta}
%
% Cylinder parameters
\def\radius{2.4}
\def\length{4}
\def\smallrad{0.5}
\def\zsmall{\length/2} % z position of small oval
\def\rsmall{-\radius/2 * 1.1} % r position of small oval
\def\rolabelz{\zsmall + 0.}
\def\perspective{-0.4} % perspective compression factor (same as ellipse)
%
% Helical trajectory parameters
\def\helixradius{-\rsmall} % radius of the helix
\def\phirevolutions{1} % number of revolutions in phi (azimuthal angle)
\def\zoscillations{5} % number of oscillations in z direction
\def\totalangle{360*\phirevolutions} % total angle range for one complete cycle
%
% Draw cylinder
\draw[ultra thick] (0,0) ellipse (\radius cm and \perspective cm); % front ellipse (bottom)
\draw[ultra thick] (0,\length) ellipse (\radius cm and \perspective cm); % back ellipse (top)
\draw[ultra thick] (-\radius,0) -- (-\radius,\length);
\draw[ultra thick] (\radius,0) -- (\radius,\length);
%
% Back half of helical trajectory (drawn first so it appears behind)
\begin{scope}
  \clip (-\radius, \perspective) rectangle (\radius, \length-\perspective);
  % Create blur effect with multiple overlapping lines
  \foreach \w/\op/\col in {6pt/0.15/pink, 4pt/0.2/pink, 2pt/0.25/pink, 1pt/0.6/gray} {
    \draw[\col, line width=\w, opacity=\op]
      plot[domain=180:180+\totalangle, samples=100, smooth, variable=\t]
      ({\helixradius * cos(\t)}, 
       {(\length/2) + (\length/2) * sin(\zoscillations*\t) + \perspective/2 * \helixradius * sin(\t)});
  }
  \foreach \t in {0,35,70,110,150,175} {
  \draw[->, gray, thick] 
    ({\helixradius * cos(\t)}, {(\length/2) + (\length/2) * sin(\zoscillations*\t) + \perspective/2 * \helixradius * sin(\t)}) --
    ({\helixradius * cos(\t+1)}, {(\length/2) + (\length/2) * sin(\zoscillations*(\t+1)) + \perspective/2 * \helixradius * sin(\t+1)});
}
\end{scope}
%
%
% Smaller oval in the middle
\draw[ultra thick,dashed] (\rsmall,\zsmall) ellipse (\smallrad cm and 0.2cm);
%
% Axes
\draw[ultra thick, ->] (0,-0.01) -- (0,\length+1) node[above] {$\mathbf{z}$};
\draw[ultra thick, ->] (0,0) -- (\radius+1,0) node[right] {$\mathbf{\rho}$};
%
% Radius label at top
\draw[thick, <->] (0,\length) -- (\radius,\length)
    node[midway,above] {$a$};
%
% Length label
\draw[thick, <->] (\radius+0.4,0) -- (\radius+0.4,\length)
    node[midway,right] {$L$};
%
% Radius label for small oval
\draw[<->] (\rsmall,\zsmall+0.3) -- (\rsmall - \smallrad,\zsmall+0.3)
    node[midway,above] {$\rho_c$};
%
% Mark a point on the small ellipse and label it
\filldraw[black] (\rsmall - \smallrad,\zsmall) circle (2pt) node[left] {$e^-$};
%
% Radial position of guiding center
\draw[line width=1pt,->] (0,\rolabelz) -- (\rsmall,\rolabelz) node[midway,above] {$r_0$};
%
% Add solid arrow from bottom of z axis to z location of smaller circle, labeled z_c
\draw[thick,<->] (\rsmall-\smallrad,0) -- (\rsmall-\smallrad,\zsmall - .1) node[midway,left] {$z_c$};
%
% Front half of helical trajectory (drawn last so it appears in front)
\begin{scope}
  \clip (-\radius, \perspective) rectangle (\radius, \length - \perspective);
  \draw[gray, line width=1pt, opacity=.5]
    plot[domain=0:180, samples=50, smooth, variable=\t]
    ({\helixradius * cos(\t)}, 
     {(\length/2) + (\length/2) * sin(\zoscillations*\t) + \perspective/2 * \helixradius * sin(\t)});
  \draw[gray, line width=1pt, opacity=.5]
    plot[domain=360:360+180, samples=50, smooth, variable=\t]
    ({\helixradius * cos(\t)}, 
     {(\length/2) + (\length/2) * sin(\zoscillations*\t) + \perspective/2 * \helixradius * sin(\t)});
\end{scope}
%
% Add arrows to show direction of motion on visible parts
% \foreach \t in {90, 450} {
%   \draw[->, red, thick] 
%     ({\helixradius * cos(\t)}, {(\length/2) + (\length/2) * sin(4*\t) + \perspective * \helixradius * sin(\t)}) --
%     ({\helixradius * cos(\t+30)}, {(\length/2) + (\length/2) * sin(4*(\t+30)) + \perspective * \helixradius * sin(\t+30)});
% }
%
% Add a legend for the helical path
% \draw[red, thick] 
%   (-\radius-0.3,\length+0.7) -- (-\radius+0.3,\length+0.7) 
%   node[right] {Electron trajectory};
%
\end{tikzpicture} 
    \caption{Cylindrical cavity of radius $a$ and length $L$. The electron's cyclotron radius is denoted by $\rho_c$ and its guiding center position by $r_0$. The axial motion (along $\hat{z}$) and grad-B drift motion (along $\hat{\phi}$) are shown by the gray line. The depicted period of drift motion is for illustrative purposes and is much less than what is expected in Project 8 experiments.}
    \label{fig:cylindrical-cavity}
\end{figure}
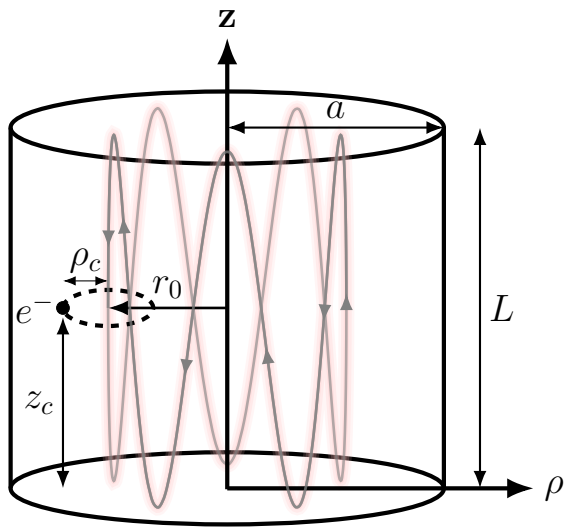

The full computation of the cylindrical cavity's electric field requires knowledge of the TE, TM, and EI modes. For a cylindrical cavity of radius $a$ and length $L$ as in \cref{fig:cylindrical-cavity}, the electric fields of these modes can be written in terms of their scalar potentials (adapted from~\cite[Ch.~5]{collinFieldTheoryGuided1991}),
% \begin{align}
%     \psi^M_{nml}(\mathbf{r}(t)) &= A^M_{nml} J_n(k^M_{n m} \rho) \cos(n \phi) S^M (k_{l} z) ,
% \end{align}
\begin{subequations}
\begin{align} \label{eqns:scalarpots}
    \psi^{TE}_{nml}(\mathbf{r}) &= A^{TE}_{nml} J_n(k^{TE}_{n m} \rho) \cos(n \phi) \sin(k_{l} z) \\
    \psi^{TM}_{nml}(\mathbf{r}) &= -A^{TM}_{nml} J_n(k^{TM}_{n m} \rho) \sin(n \phi) \cos(k_{l} z) \\
    \psi^{EI}_{nml}(\mathbf{r}) &= A^{EI}_{nml} J_n(k^{EI}_{n m} \rho) \sin(n \phi) \sin(k_{l} z) ,
\end{align}
\end{subequations}
where $n$ is the azimuthal $\hat{\phi}$ index, $m$ is the radial $\hat{\rho}$ index, and $l$ is the symmetry axis $\hat{z}$ index with corresponding wavenumber $k_{l} = \pi l / L$. Each wavenumber corresponds to the number of nodes along its corresponding axis. The symbol $J$ denotes the cylindrical Bessel functions. For TE modes, the radial wavenumber is $k^{TE}_{n m} = \chi'_{nm} = p'_{nm}/a$, and for TM modes, $k^{TM}_{n m} = \chi_{nm} = p_{nm}/a$, where $p_{nm}$ ($p'_{nm}$) are the $m^{th}$ zeros of the $n^{th}$ Bessel functions (Bessel function derivatives), respectively. The electric irrotational modes have $k^{EI}_{n m} = p_{nm}/a$. The $n>0$ modes have two orientations: the substitution ${n \phi \rightarrow n \phi + \pi/2} $ converts them. The pre-factors $A^M_{nml}$ in our convention normalize the electric fields to unity. They are
\begin{subequations} \label{eqn:eigennorms}
\begin{align}
    A^{TE}_{nml} &= \frac{2}{ \chi'_{nm} \sqrt{2^{\delta_{n0}} V {J_n}(\chi'_{nm} a) {J''}_n(\chi'_{nm} a)}} \label{eqn:TE_norm} \\
    A^{TM}_{nml} &= \frac{2}{{\chi}_{nm} {J'}_n(\chi_{nm} a) \sqrt{2^{\delta_{n0} + \delta_{l0}} V } }\label{eqn:TM_norm} \\
    A^{EI}_{nml} &= \frac{2}{k_{nml} {J'}_n(\chi_{nm} a) \sqrt{2^{\delta_{n0}} V }} \label{eqn:EI_norm} ,
\end{align}
\end{subequations}
where $V=\pi a^2 L$ is the volume of the cavity and $\delta_{n,0}=1$ if $n=0$ and zero otherwise. The resonant wavenumbers $k_{nml}$ (indexed as $k_{\alpha}$ in \cref{sec:responseModel}) are defined as
\begin{align}
    f^{M}_{nml} &= \frac{c}{2\pi} k^{M}_{nml} \\
    k^{M}_{nml} &= \sqrt{\left(k^M_{nm}\right)^2 + k_{l}^2} .
\end{align}
Above, $f^M_{nml}$ are the corresponding resonant frequencies when the mode label $M$ is in reference TE or TM. The electric fields are defined by 
\begin{subequations}\label{eqns:efieldsbymode}
    \begin{align}
    \mathbf{e}^{\,TE}_{nml} &= \mathbf{\nabla} \times \left(\mathbf{\hat{z}} \psi^{TE}_{nml}\right) \\
    \mathbf{e}^{\,TM}_{nml} &= \frac{1}{k_{nml}} \mathbf{\nabla} \times \mathbf{\nabla} \times \left(\mathbf{\hat{z}} \psi^{TM}_{nml}\right) \label{eqn:tmmode} \\
    \mathbf{f}^{\,EI}_{nml} &= \mathbf{\nabla} \psi^{EI}_{nml} \label{eqn:eimode}.
\end{align}
\end{subequations}
%---------------------------------------------------------------------------------------%
\subsection{Electron cyclotron-cavity coupling} \label{ssec:couplingIntegrals}
We derive explicit expressions for the decomposition of the current generated by a single electron into the cavity modes, resulting in the spectral coefficients $s_{\alpha}(\omega)$ and $q_{\alpha}(\omega)$ in \cref{eqns:jalpha-exp}, which allow for direct numerical evaluation of the power emitted into the modes (\cref{eqn:powerformula-exp,eqn:powerformula-exp-irr}). We first write these coefficients in a generalized form in order to identify similarities between modes that allow for faster calculation:
% \begin{align}
%     s^M_{\alpha}(\omega) &= -e \mathcal{F} \left\{ \int_V \delta\left(\mathbf{r} - \mathbf{r}_c(t)\right) \mathbf{v}_c(t) \cdot\mathbf{e}^M_{\alpha}(\mathbf{r}) dV \right\} , \label{eqn:snml-expanded}
% \end{align}
\begin{align}
    j^M_{\alpha}(\omega) &= -e \mathcal{F} \left\{ \int_V \delta\left(\mathbf{r} - \mathbf{r}_c(t)\right) \mathbf{v}_c(t) \cdot\mathbf{E}^M_{\alpha}(\mathbf{r}) dV \right\} , \label{eqn:snml-expanded}
\end{align}
where the electron's velocity is $\mathbf{v}_{c}(t)$. %
Here, $j^M_{\alpha}$ can be either $s^M_{\alpha}$ (with $\mathbf{E}^M_{\alpha}=\mathbf{e}^M_{\alpha}$) or $q^M_{\alpha}$ (with $\mathbf{E}^M_{\alpha}=\mathbf{f}^M_{\alpha}$). %
In this section we assume that the electron undergoes ideal cyclotron motion with no $z$ momentum and no losses, and so take $\rho_c, \omega_c, z_c$ to be constant. %
Later, in \cref{sec:axialMotion}, we will allow these quantities to vary in ways that are representative of periodic motion in a highly uniform magnetic bottles. %
In the electron's guiding center coordinate system, the electron's momentum is only in $\hat{\phi}$, and so re-casting the above integral in this coordinate system simplifies the evaluation of $j^M_{\alpha}(\omega)$. % 

In casting \cref{eqn:snml-expanded} into the new coordinate system, we note that the delta function transforms under a change of coordinates as $ \delta(\mathbf{r}) \rightarrow \frac{\delta(\mathbf{r}_c)}{|J|} $, where $|J|$ is the determinant of the Jacobian of the transformation. 
Similarly, the volume element transforms as $ dV \rightarrow |J| dV_c$, and the two $|J|$'s cancel out in $j^M_{\alpha}(\omega)$. 

\begin{figure}[ht]
    \centering
    % File: graf_figure.tikz
% \usetikzlibrary{angles,quotes}
% \begin{tikzpicture}
%     \coordinate (o) at (0,0);
%     \coordinate (o1) at (1.5,1.5);
%     \coordinate (p) at (1.5,.75);
%     \coordinate (x) at (1,0);
%     % Draw the main coordinate system
%     \draw[->] (-.2,0) -- (3,0) node[anchor=north west] {$x$};
%     \draw[->] (0,-.2) -- (0,3) node[anchor=south east] {$y$};
%     % Origin points
%     \filldraw[black] (o) circle (1.5pt) node[anchor=south east] {$O$};
%     \filldraw[black] (o1) circle (1.5pt) node[anchor=south west] {$O_1$};
%     % Dashed circle for new origin
%     \draw[thick, dashed] (o1) circle (.75);
%     % Coordinates in old system
%     \draw[->, thick] (o) -- (1.5, .75) node[midway, anchor=south] {$\rho$};
%     \pic [draw, ->, "$\phi_0$", angle radius = 1.25cm, angle eccentricity=1.5] {angle = x--o--p};
% \end{tikzpicture}
\usetikzlibrary{arrows.meta, calc, angles, quotes}
\begin{tikzpicture}[
    scale=1.5, % Adjust scale of the entire figure
    font=\large, % Default font size for text
    every node/.style={scale=1.2}, % Scale down node text slightly
    angle_font/.style={scale=1.2} % Smaller font for angle labels
]
    % --- Parameters ---
    % You can change these values to alter the geometry
    \def\R{2.4}        % Magnitude of vector R (O to O1)
    \def\alphaDeg{30}   % Angle of vector R with x-axis, in degrees
    \def\rhoVal{2.}    % Magnitude of vector rho (O to P)
    \def\phiZeroDeg{50} % Angle of vector rho with x-axis, in degrees
    % --- Coordinates ---
    % Origin O
    \coordinate (O) at (0,0);
    % Origin O1, defined by R and alpha relative to O
    \coordinate (O1) at (\alphaDeg:\R);
    % Point P, defined by rhoVal and phiZeroDeg relative to O
    \coordinate (P) at (\phiZeroDeg:\rhoVal);
    % Helper coordinate for defining angles from the main x-axis (O->X_main_ref is along +x)
    \coordinate (X_main_ref) at ($(O) + (1,0)$);
    % Helper coordinate for defining angles from the shifted x1-axis (O1->X_shifted_ref is along +x1)
    \coordinate (X_shifted_ref) at ($(O1) + (1,0)$);
    % --- Main Coordinate System (O, x, y) ---
    \draw[->, >=Latex] (-0.5,0) -- (3.8,0) node[anchor=north west] {$x$};
    \draw[->, >=Latex] (0,-0.5) -- (0,2.5) node[anchor=south east] {$y$};
    \fill (O) circle (1.2pt) node[anchor=north east] {$O$};
    % --- Shifted Coordinate System (O1, x1, y1) ---
    % Axes are parallel to original axes, drawn dashed and gray
    % \draw[->, >=Latex, gray, dashed] ($(O1) + (-0.8,0)$) -- ($(O1) + (1.2,0)$) node[anchor=north west, black] {$x_1$};
    % \draw[->, >=Latex, gray, dashed] ($(O1) + (0,-0.8)$) -- ($(O1) + (0,1.2)$) node[anchor=south east, black] {$y_1$};
    \draw[->, >=Latex, gray, dashed] ($(O1) + (-0.8,0)$) -- ($(O1) + (1.2,0)$) node[anchor=north west, black] {};
    \draw[->, >=Latex, gray, dashed] ($(O1) + (0,-0.8)$) -- ($(O1) + (0,1.2)$) node[anchor=south east, black] {};
    \fill (O1) circle (1.2pt) node[anchor=north west] {$O_1$};
    % --- Point P ---
    \fill (P) circle (1.2pt) node[anchor=south east] {$P$};
    % --- Vectors ---
    % Vector R (from O to O1)
    \draw[->, thick, blue, -{Stealth[length=2.5mm, width=2mm]}] (O) -- (O1)
        node[pos=.75, sloped, above, font=\large] {$r_0$};
    % Vector rho (from O to P)
    \draw[->, thick, red, -{Stealth[length=2.5mm, width=2mm]}] (O) -- (P)
        node[midway, sloped, above, xshift=-1mm, font=\large] {$\rho$};
    % Vector rho1 (from O1 to P)
    \draw[->, thick, green!60!black, -{Stealth[length=2.5mm, width=2mm]}] (O1) -- (P)
        node[midway, sloped, above, font=\large] {$\rho_1$};
    % --- Angles ---
    % Angle alpha (for vector R w.r.t. x-axis)
    \pic [draw, ->, blue, angle_font, "$\phi_0$", angle radius=0.6cm, angle eccentricity=1.5]
        {angle = X_main_ref--O--O1};
    % Angle phi0 (for vector rho w.r.t. x-axis)
    % Drawn with a slightly larger radius if alpha is small, to avoid overlap
    \ifdim \phiZeroDeg pt > \alphaDeg pt
        \ifdim \alphaDeg pt > 0pt % Check if alpha is not zero to avoid issues with angle order
            \coordinate (AngleStartNodePhi0) at (O1); % Start phi0 from R if phi0 > alpha
        \else
            \coordinate (AngleStartNodePhi0) at (X_main_ref); % Default start from x-axis
        \fi
    \else
         \coordinate (AngleStartNodePhi0) at (X_main_ref); % Default start from x-axis
    \fi
     % The above logic for AngleStartNodePhi0 is to draw phi0 as an arc from R to rho if phi0>alpha for some representations.
     % For angles relative to x-axis always, use X_main_ref.
    \pic [draw, ->, red, angle_font, "$\phi$", angle radius=1.1cm, angle eccentricity=1.35, pic text options={yshift=-2mm, xshift=-1mm}]
        {angle = X_main_ref--O--P};
    % Angle phi1 (for vector rho1 w.r.t. x1-axis)
    \pic [draw, ->, green!60!black, angle_font, "$\phi_1$", angle radius=0.2cm, angle eccentricity=1.5, pic text options={xshift=2mm}]
        {angle = X_shifted_ref--O1--P};
    % Circle around O1
    \draw[thick, dashed] (O1) circle (.85);
\end{tikzpicture} 
    \caption{Illustration of the parameters used in the Graf Addition Theorem. Here, $O$ is the cavity-centered coordinate system, and $O_1$ is the electron's guiding center. The phase in the guiding center coordinate system is defined relative to the phase in the cavity-centered coordinate system.}
    \label{fig:graf-addition}
\end{figure}
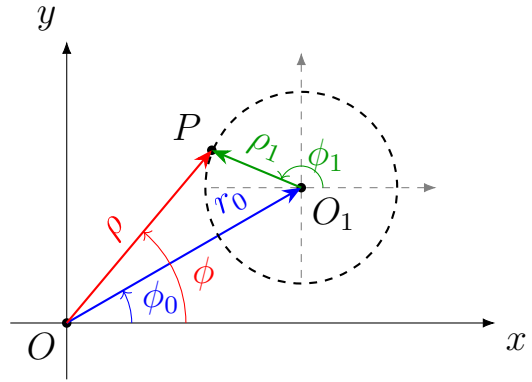

We transform polar coordinates $\mathbf{r}=(\rho,\phi)$ in the cavity-centered coordinate system $O$ to polar coordinates $\mathbf{r}_1=(\rho_1, \phi_1)$ in the electron's guiding center coordinate system $O_1$, with origin located at $\mathbf{r}_0=(r_0, \phi_0)$ in $O$ (see \cref{fig:graf-addition}). The prescription for this transformation is
\begin{align}
    \rho &= \sqrt{\rho_1^2 + r_0^2 - 2 \rho_1 r_0 \cos(\phi_1 - \phi_0)} \\
    \phi_1 + \phi_0 &= \arcsin\left(\frac{\rho}{\rho_1} \sin(\phi - \phi_0)\right) .
\end{align}
This transformation allows for the use of the Graf Addition Theorem \cite{abramowitzHandbookMathematicalFunctions2013},
\begin{align}
    J_n\left(\rho\right) &= e^{- i n \Delta \phi} \sum^{\infty}_{u=-\infty} J_{n+u}(r_0) J_{u}(\rho_1) e^{iu(\pi + \phi_0 - \phi_1)}, \label{eqn:graf}
\end{align}
where $\Delta \phi = \phi - \phi_0$. The above identity will be substituted into \cref{eqns:scalarpots} in order to write the eigenmode scalar potentials in the guided center coordinate system. As differential operations are independent of coordinate system, the electric fields will be calculated from \cref{eqns:efieldsbymode} in $O_1$.
% When evaluating \cref{eqn:snml-def} in the guiding center coordinate system, only the $\hat{\phi}$ component contributes for TE modes. 

% It is advantageous to re-write the integral \cref{eqn:snml-def} in a cylindrical coordinate system centered at the electron's guiding center. 
% \subsubsection{The power spectrum in cylindrical eigenmodes} \label{ssec:powerspectrum}
In the electron's guiding center coordinates $O_1$, the electron's velocity is in one direction $\mathbf{v}_c = \omega_c \rho_c \hat{\phi}_1$ only, which allows us to focus only on the $\hat{\phi}_1$ component of the electric field, $E^M_{\phi_1,\alpha}(\mathbf{r}_1)$. The electron's phase (and its $\phi_1$ position) is $\phi_c(t) = \int \omega_c(t') dt' = \omega_c t + \phi'_0$, where $\phi'_0$ is the starting phase of its orbit in the guiding center. Via \cref{eqn:graf}, it is apparent that when $\phi_1 = \phi_c(t)$ the quantity $u\phi_0$ gets replaced by $u(\phi_0-\phi'_0)$. This replacement can be made whenever the impact of the cyclotron's starting phase is of interest.

A similar replacement allows for investigation of the effects of the grad-B motion of the electron when the local magnetic field is inhomogeneous. If the magnetic field inhomogeneities are centered around the cavity's $\hat{z}$ axis, the position of the electron's guiding center moves in a circle with radius $r_{0}$ at frequency $\omega_{g}$. If the motion is further simplified by assuming the electron does not lose energy, and that variations in its cyclotron radius can be ignored, it is straightforward to apply this to \cref{eqn:graf} by the substitution $\phi_0 \rightarrow \phi_0 + \omega_{g} t$.%
\footnote{The drift motion replacement rule results in a modification of~\cref{eqn:phinu_FT},%
\begin{align}
    \mathcal{F}\left\{ \Phi_{n,u} \right\} &= \pi \omega_c \left(e^{i(n-u)\phi_0} \delta(\omega - u \omega_c - (n - u) \omega_{g}) \right. \nonumber\\
    &\myspace{3} \left. + e^{-i(n-u)\phi_0}\delta(\omega + u \omega_c + (n-u) \omega_{g}) \right) . \label{eqn:phinu_GradB_FT}
\end{align}%
The radiated frequency is split in two and the new frequencies $\omega=p\omega_{c} + {(n-p)\omega_{g}}$, $\omega=p\omega_{c} - {(n+p)\omega_{g}}$ (with $p=|u|$ when both $(p \pm n)\omega_{g} < p \omega_{c}$) are shifted from the cyclotron frequency by integer multiples of the grad-B motion's frequency. The amplitudes of the shifted spectrum peaks are not in general equal, and can be found by evaluating the coefficients in \cref{eqn:sm_integrand}. CRES experiments may require such magnetic field homogeneity that the frequency of drift motion is well below the target frequency resolution, in which case this splitting cannot be observed in the power spectrum.
}

Continuing without the drift motion, \cref{eqn:snml-expanded} condenses because the radial position $\rho_c$ and $\hat{z}$ position $z_c$ are constant, leading to
\begin{align}
    j^M_{\alpha}(\omega)
    &= -e \mathcal{F} \left\{ \int\limits^{2\pi}_0 d\phi_1 \ \delta\left(\phi_1 - \omega_c t\right) \left. \omega_c \rho_c E^{M}_{\phi_1,\alpha}\right|_{\rho_c,z_c} \right\}. \label{eqn:sm_formal}
\end{align}
% \begin{align}
%     s^M_{\alpha}(\omega)
%     &= -e \mathcal{F} \left\{ \int\limits^{2\pi}_0 d\phi_c \ \delta\left(\phi_c - \omega_c t\right) \omega_c \rho_c e^{M}_{\phi,\alpha}(\phi_1,\rho_c,z_c) \right\}.
% \end{align}
% \begin{widetext}
% \begin{align}
%     s^M_{nml}(\omega)
%     &= -e \omega_c \rho_c \int \left[\int^{2\pi}_0 \delta\left(\phi - \omega_c t\right) e^{\,M}_{\phi,nml} d\phi \right]_{\rho_c,z_c} e^{-i\omega t} dt  \\
%     &= -e B_{nml}^M Z_l(z_c) \sum^{\infty}_{u=-\infty} \mathrm{P}^M_{nm}(u,\rho_c) \int^{\infty}_{-\infty} \Phi^M_{nm}(u,\omega_c) e^{-i\omega t} dt \\
%     &= -e \pi B_{nml}^M Z(z_c) \sum^{\infty}_{u=-\infty} \mathrm{P}^M_{nm}(u,\rho_c) \left(\delta(u\omega_c - \omega) e^{i (u (\phi_0 + \phi'_0) - n\phi'_0)} \right. \label{eqn:snml-preFT} \\
%     &\qquad\qquad\qquad \left. + \delta(u\omega_c + \omega) e^{-i (u (\phi_0 + \phi'_0) - n\phi'_0)}\right) \nonumber
% \end{align}
% \end{widetext}
The explicit form of the guiding center electric fields are found in \cref{app:efields}. For the present, we will anticipate that the integrand quantities $\omega_c \rho_c e^{M}_{\phi,\alpha}$ share a common structure across the different mode types, and define
\begin{align}
    \left. \omega_c \rho_c E^{M}_{\phi,\alpha}\right|_{\rho_c,z_c} 
    &= B_{nml}^M Z_l \sum^{\infty}_{u=-\infty} \mathrm{P}^M_{nm,u} \Phi_{n,u} \label{eqn:sm_integrand} ,
\end{align}
with the functions $B_{nml}^M, Z_l, \mathrm{P}^M_{nm,u},$ and $\Phi_{n,u}$ given in \cref{tab:coupling_coefficients}.
% \begin{align}
%     \left. \omega_c \rho_c e^{M}_{\phi,\alpha}\right|_{\rho_c,z_c} 
%     &= B_{nml}^M Z_l \sum^{\infty}_{u=-\infty} (-1)^u \mathrm{P}^M_{nm,u} \Phi_{n,u} \label{eqn:sm_formal} \\
%     \Phi_{n,u} &= \omega_c \cos(u(\phi_0 - \phi_c(t)) - n\phi_0) \\
%     Z_l &= \sin(k_{l} z_c) ,
% \end{align}
% and 
% \begin{align}
%     B_{nml}^M &= \chi'_{nm} A_{nml}^{TE} \ \ (TE), \ \frac{k_{l}}{k_{nml}} A_{nml}^{TM} \ (TM) \label{eqn:Bnorms}\\
%     \mathrm{P}^{TE}_{nm,u} &= \rho_c J_{n+u}(\chi'_{nm} r_0) J'_{u}(\chi'_{nm} \rho_c) \\
%     \mathrm{P}^{TM/EI}_{nm,u} &= u J_{n+u}(\chi_{nm} r_0) J_{u}(\chi_{nm} \rho_c) .
% \end{align}
\begin{table}[ht]
\centering
\caption{Function definitions for \cref{eqn:sm_integrand}, categorized by mode type.}
\label{tab:coupling_coefficients}
\begin{tabular}{ll}
\hline
\multicolumn{2}{l}{\textbf{Common among all mode types}} \\
\hline\\[-.9em]
\multicolumn{2}{l}{$\Phi_{n,u} = (-1)^u\omega_c \cos(u(\phi_0 - \phi_c(t)) - n\phi_0)$} \\[0.5em]
\multicolumn{2}{l}{$Z_l = \sin(k_{l} z_c(t))$} \\[0.5em]
\hline
\multicolumn{2}{l}{\textbf{Dependent on mode type}} \\
\hline\\[-.9em]
TE modes \myspace{2} & $B_{nml}^{TE} = \chi'_{nm} A_{nml}^{TE}$ \\[0.5em]
& $\mathrm{P}^{TE}_{nm,u} = \rho_c(t) J_{n+u}(\chi'_{nm} r_0) J'_{u}(\chi'_{nm} \rho_c(t))$ \\[0.5em]
TM modes & $B_{nml}^{TM} = \frac{k_{l}}{k_{nml}} A_{nml}^{TM}$ \\[0.5em]
& $\mathrm{P}^{TM/EI}_{nm,u} = u J_{n+u}(\chi_{nm} r_0) J_{u}(\chi_{nm} \rho_c(t))$ \\[0.5em]
EI modes & $B_{nml}^{EI} = A_{nml}^{EI}$ \\
\hline
\end{tabular}
\end{table}
% \begin{align}
%     \int_V \omega_c \rho_c \delta\left(\mathbf{r} - \mathbf{r}_c(t)\right) e^{\,M}_{\phi,nml} dV &:= B_{nml}^M Z^M_l(z_c) \sum^{\infty}_{u=-\infty} \mathrm{P}^M_{nm}(u,\rho_c) \Phi_{n}(u,\omega_c)
% \end{align}

In this section we assume the cyclotron radius and frequency are constant, and as a result only the $\Phi_{n,u}$ in \cref{eqn:sm_integrand} are time-dependent. They Fourier-transform as
\begin{align}
    \mathcal{F}\left\{ \Phi_{n,u} \right\} &= \pi \omega_c \left(e^{i(n-u)\phi_0} \delta(\omega - u \omega_c) \right. \nonumber\\
    &\myspace{3} \left. + e^{-i(n-u)\phi_0}\delta(\omega + u \omega_c) \right) . \label{eqn:phinu_FT}
\end{align}
It is apparent now that the frequency response is decomposed into positive and negative harmonics of the cyclotron frequency. In summing over $u$ in \cref{eqn:sm_integrand}, there are two terms for every harmonic $p$ that contribute to \cref{eqn:sm_formal}:
% As in \cref{ssec:poweremitted}, we will consider the 
% For the $p^{th}$ harmonic of the cyclotron frequency, the sum over $u$ both terms in the above 
% The $p^{th}$ harmonic of the cyclotron frequency has the expansion coefficient
\begin{align}
    j^M_{nml}(p \omega_c) 
    &= -e \pi B_{nml}^M Z_l e^{-i p \phi_0} \nonumber\\
    & \myspace{1} \cdot \left(\mathrm{P}^M_{nm,p} e^{i n \phi_0} + \mathrm{P}^M_{nm,-p} e^{-i n \phi_0} \right) .
\end{align}
To use this result in the calculation of the power emitted (\cref{eqn:powerformula}), we need to evaluate the magnitude squared:
\begin{align}
    \left| j^M_{nml}(p \omega_c) \right|^2 
    &= e^2 \pi^2 (B_{nml}^M)^2 Z_l^2
    \left[\left(\mathrm{P}^M_{nm,p}\right)^2 + \left(\mathrm{P}^M_{nm,-p} \right)^2 \right. \nonumber \\
    &\myspace{2} \left. + 2 \mathrm{P}^M_{nm,p} \mathrm{P}^M_{nm,-p} \cos(2 n \phi_0) \right] . \label{eqn:sm_sq_ab}
\end{align}

The modes of the ideal cylindrical cavity indexed by $nml$ with $n>0$ have two independent orientations relative to the cavity, resulting in two degenerate modes. As mentioned above, the transformation between them can be made by the substitution $n\phi_0 \rightarrow n\phi_0 + \pi/2$. In this section, we will consider the total power emitted into the mode indexed $nml$ to be the combined power emitted into both mode orientations (when $n>0$). The result of $\left| s^M_{nml}(p \omega_c) \right|^2 $ for the second mode orientation can be found simply by the substitution $\cos(2n\phi_0) \rightarrow \cos(2n(\phi_0 + \pi/2)) = -\cos(2n\phi_0)$ in \cref{eqn:sm_sq_ab}. In adding the two to get the total power (see \cref{ssec:poweremitted}), the angular dependence drops out. 

We define $|j^{M \pm}_{nml}(\omega)|^2$ as this mode orientation sum of $|j^{M}_{nml}(\omega)|^2$. For the edge case of $n=0$, where there is only one mode orientation, one can use the fact that $\mathrm{P}^M_{0m,p} = \mathrm{P}^M_{0m,-p}$ to write the result in a form for all $nml$,
\begin{align}
    \left| j^{M\pm}_{nml}(p \omega_c) \right|^2 
    &= 2 e^2 \pi^2 (B_{nml}^M)^2 Z_l^2
    \sum\limits_{u\in \{p,-p\}}(\mathrm{P}^M_{nm,u})^2 .
\end{align}
Taking the normalizations from \cref{eqn:TE_norm,eqn:TM_norm}, we find
\begin{align}
    \left| s^{TE\pm}_{nml}(p \omega_c) \right|^2 &= 8 \pi^2 \frac{e^2 \omega_c^2}{ 2^{\delta_{n0}} V } \frac{\rho_c^2 \sin^2(k_{l} z_c)}{J''_n(\chi'_{nm} a) J_n(\chi'_{nm} a)} \nonumber \\
    & \myspace{2} \cdot \sum\limits_{u\in \{p,-p\}} J^2_{n+u}(\chi'_{nm} r_0) {J'}^2_{u}(\chi'_{nm} \rho_c), \label{eqn:TEpower}
\end{align}
and for the TM modes,
\begin{align}
    \left| s^{TM\pm}_{nml}(p \omega_c) \right|^2 &= 8 \pi^2 \frac{e^2 \omega_c^2}{ 2^{\delta_{n0}}V } \frac{p^2 k^2_{l} \sin^2(k_{l} z_c) }{k^2_{nml} \chi^2_{nm} {J'_n}^2(\chi_{nm} a)} \nonumber\\
    & \myspace{2} \cdot \sum\limits_{u\in \{p,-p\}} J^2_{n+u}(\chi_{nm} r_0) {J}^2_{u}(\chi_{nm} \rho_c) . \label{eqn:TMpower}
\end{align}
Above, the $l=0$ normalization factor has been omitted because the cyclotron motion does not couple to those modes ($k_l=0$). The power can be found via \cref{eqn:powerformula}. The calculations in \cref{fig:phaseIItheorycomp} and the next section show how the electron's emitted power depends on radial position within the cavity with comparison to prior Project 8 calculations. The above results agree with the analogous treatment of the electron's current for waveguides from He6-CRES~\cite{buzinskyLarmorPowerLimit2024} and approximate agreement with prior cavity treatments from Penning trap experiments~\cite{brownCyclotronMotionMicrowave1985,hannekeCavityControlSingleelectron2007,hannekeCavityControlSingleelectron2011}.

Finally, for the non-resonant modes,
\begin{align}
    \left| q^{EI\pm}_{nml}(p \omega_c) \right|^2 &= 8 \pi^2 \frac{e^2 \omega_c^2}{ 2^{\delta_{n0}} V } \frac{p^2 \sin^2(k_{l} z_c) }{k^2_{nml} {J'}^2_n(\chi_{nm} a)} \nonumber\\
    & \myspace{2} \cdot \sum\limits_{u\in \{p,-p\}} J^2_{n+u}(\chi_{nm} r_0) {J}^2_{u}(\chi_{nm} \rho_c) . \label{eqn:EIpower}
\end{align}
The power emitted into these modes can be calculated with \cref{eqn:powerformula-exp-irr}.
It is important to note that real-world cavities may have imperfections or readout geometries that break the degeneracy of the $n>0$ modes, and in those cases the power spectra must be examined with \cref{eqn:sm_sq_ab}. We defer the discussion of choice of operating mode to \cref{ssec:modechoice}. 
%-------------------------------------------------------------------------------------------------%
\subsection{Comparison to CRES in waveguides} \label{ssec:comparisionwaveguides}
Previous work by Project 8 in Phase II set a limit on neutrino mass through CRES in waveguides~\cite{ashtariesfahaniTritiumBetaSpectrum2023,ashtariesfahaniCyclotronRadiationEmission2024}. CRES in waveguides and CRES in cavities are related by the coupling integral, or the field decomposition of the electron's current. In this section, we compare the predicted power with the above decomposition of the current with the predicted power in Phase II~\cite{ashtariesfahaniElectronRadiatedPower2019}. In that work, as in the electron magnetic dipole moment calculations~\cite{brownCyclotronMotionMicrowave1985,hannekeCavityControlSingleelectron2007,hannekeCavityControlSingleelectron2011}, the electron-cyclotron current was dipole approximated rather than treated in full as was done in the preceding section. The resulting comparison is then also relevant to the cavity-electron coupling calculations that serve as the bedrock for this work.
\begin{figure}[ht] 
    \centering%
    \subfloat{%
        \includegraphics[width=.98\columnwidth]{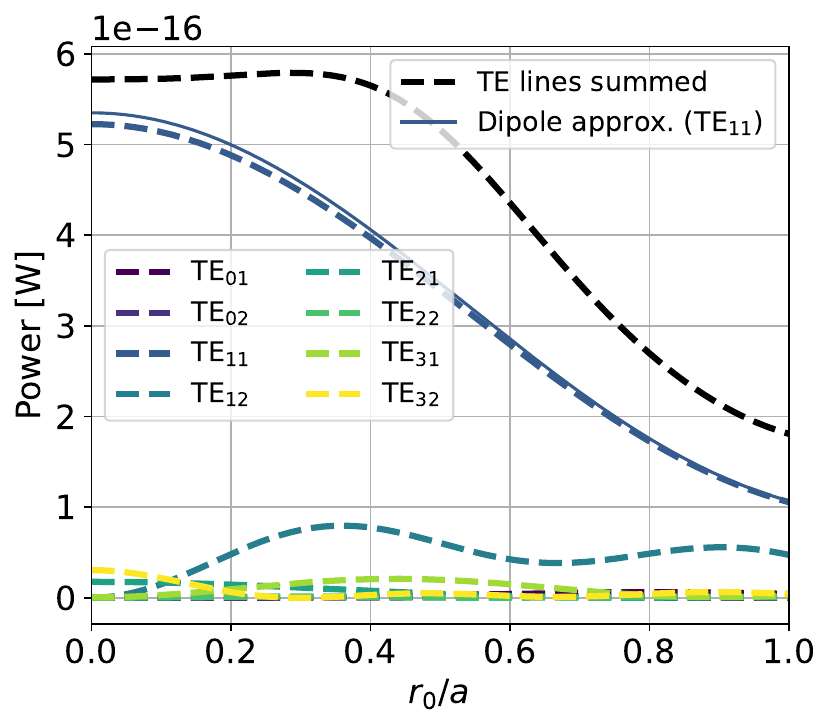}%
        \put(-240,205){\color{black}\textbf{(a)}}%
        \label{fig:TEnm}%
    }%
    \\[-1ex]%
    \subfloat{%
        \includegraphics[width=.98\columnwidth]{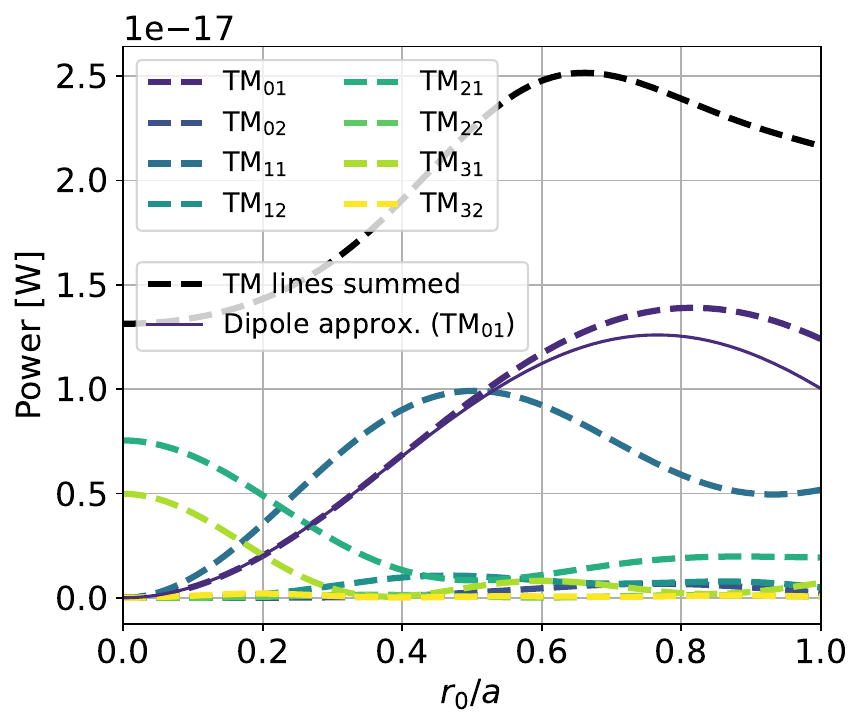}%
        \put(-240,205){\color{black}\textbf{(b)}}%
        \label{fig:TMnm}%
    }%
    \caption{Power radiated into forward/backward propagating \textbf{(a)} TE modes and \textbf{(b)} TM modes in a cylindrical waveguide with radius $a=$ 5.3 mm, up to the sixth harmonic ($p$ in the previous section) of the cyclotron frequency. The convergence properties of the power emitted are further explored in \cite{buzinskyLarmorPowerLimit2024}. The dashed lines represent the sum of the power lost to possible harmonics of the cyclotron frequency for the first eight TE/TM propagating modes. The solid lines represent the point dipole approximation of \cite{ashtariesfahaniElectronRadiatedPower2019}, which accounts for the first cyclotron harmonic only.}
    \label{fig:phaseIItheorycomp}
\end{figure}

The modal decomposition of the current $j^M_{\alpha}(\omega)$ is shared between ideal cylindrical cavities and waveguides (with the proper normalization adjustments), so it is of interest to compare our current work to previous waveguide results, particularly the TE/TM modes. Importing the standard waveguide coupling techniques used in \cite{jacksonClassicalElectrodynamics2009, ashtariesfahaniElectronRadiatedPower2019, buzinskyLarmorPowerLimit2024}, the power emitted into forward/backward modes in a waveguide is given by
\begin{align}
    P^\pm_{\alpha}(\omega) &= \left\{
        \begin{array}{c}
            \frac{\mu \omega}{8 k_z}_{(TE)} \\
            \frac{k_z}{8\epsilon \omega}_{(TM)}
        \end{array} 
    \right\} \left| s^M_{\alpha}(\omega) \right|^2 .
\end{align}
% To find the total power radiated into the mode, we must multiply again by 2 to account for the forward/backward propagating modes ($\pm$ above). %
The calculation of the power spectrum $P(\omega_c)$ for the cylindrical waveguide TE and TM modes using the results of \cref{ssec:couplingIntegrals} is shown in \cref{fig:phaseIItheorycomp}, where the $\text{TE}_{11}$ and $\text{TM}_{01}$ modes specifically are compared to previous work \cite{ashtariesfahaniElectronRadiatedPower2019}. %
The $\text{TE}_{11}$ and $\text{TM}_{01}$ modes are highlighted because they are propagating at the cyclotron frequency: the difference in power emitted into the propagating modes is marginal for most radii, but the total power emitted into the cyclotron frequency harmonics creates an excess of emitted power at all radii~\cite{buzinskyLarmorPowerLimit2024}.

In the absence of the waveguide context, \cref{fig:phaseIItheorycomp} also serves as a comparison between the electron-cavity coupling derived in~\cite{gabrielseObservationInhibitedSpontaneous1985,hannekeCavityControlSingleelectron2007,hannekeCavityControlSingleelectron2011} and our work. %
The cavity response and electron emission rate are derived differently in those prior cavity works, and the coupling factor as a function of radius plotted here seems to be the most direct comparison. %
Our results indicate good agreement with and may be used to further refine the electron magnetic dipole moment calculations.%
%-------------------------------------------------------------------------------------------------%
% -------------%
\begin{figure*}[t]
    \centering 

    % --- Top Row ---
    \subfloat{%
        \includegraphics[width=0.335\textwidth]{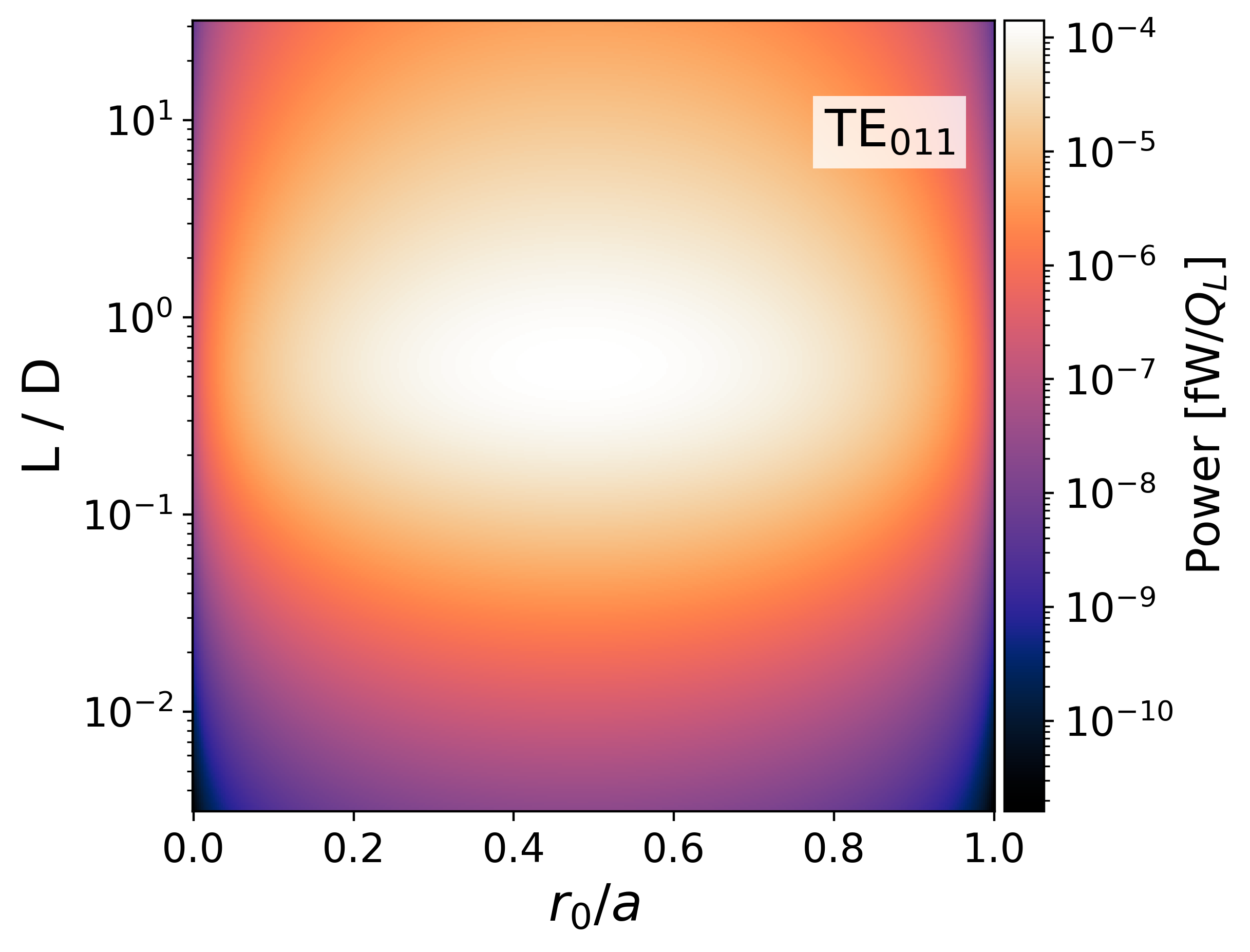}%
        \put(-140,120){\color{black}\scriptsize \textbf{(a)}}%
        \label{fig:powerheatmap_a}%
    }\hfill
    \subfloat{%
        \includegraphics[trim=1 1 0 0, clip, width=0.325\textwidth]{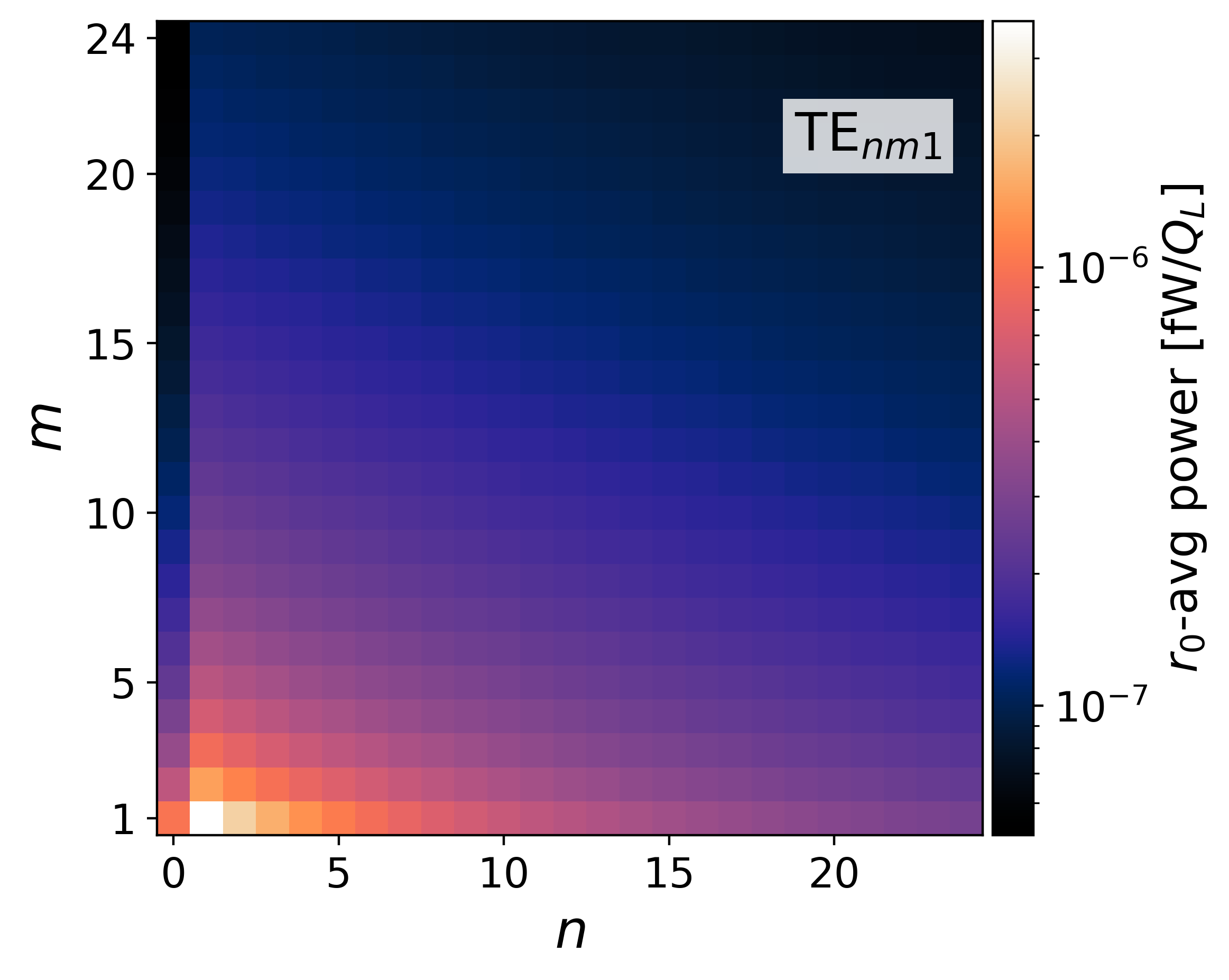}%
        \put(-145,120){\color{white}\scriptsize \textbf{(c)}}%
        \label{fig:powerheatmap_c}%
    }\hfill
    \subfloat{%
        \includegraphics[trim=40 0 0 88, clip, width=0.33\textwidth]{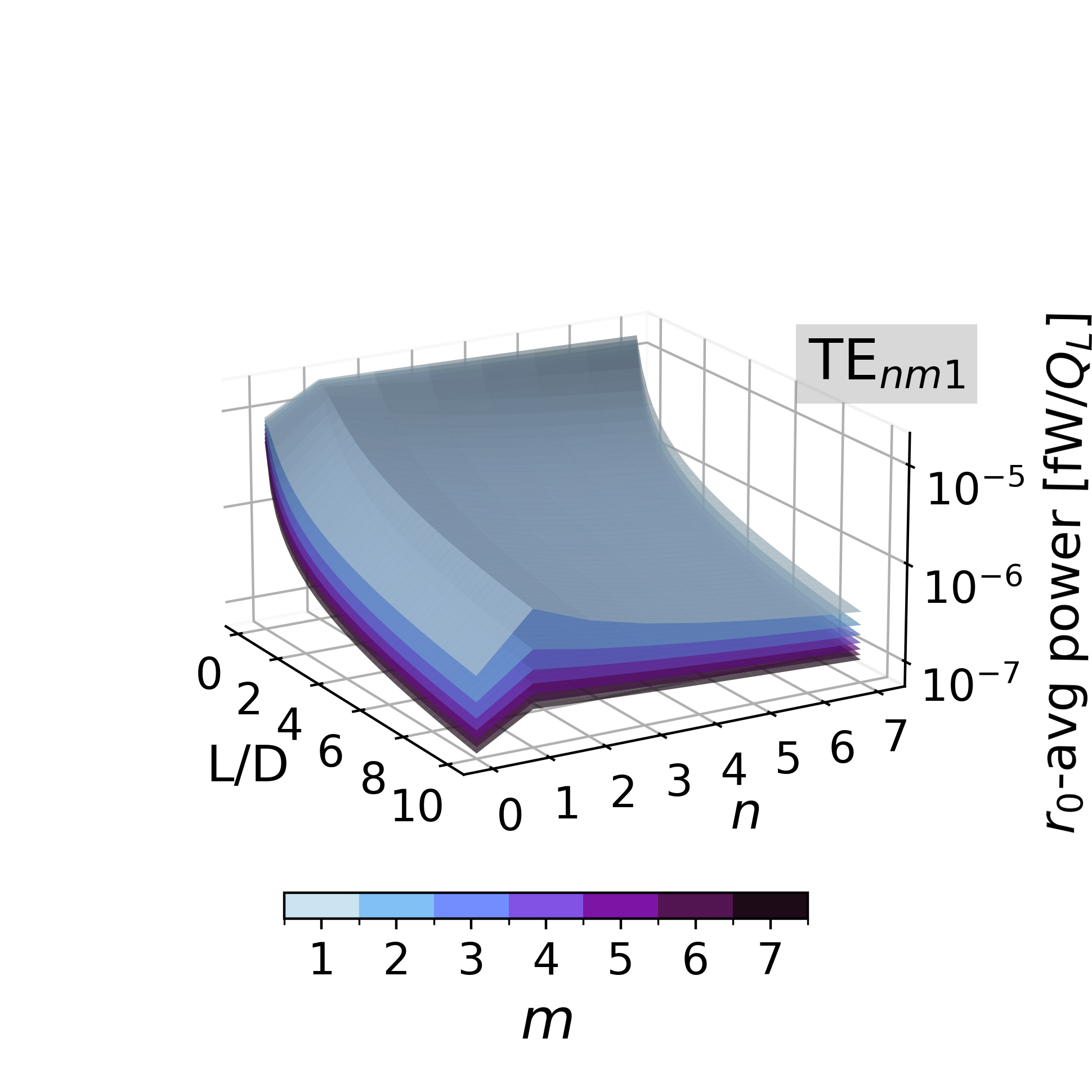}%
        \put(-158,120){\scriptsize \textbf{(e)}}%
        \label{fig:powerheatmap_e}%
    }
    \\[-3ex]
    % --- Bottom Row ---
    \subfloat{%
        \includegraphics[width=0.335\textwidth]{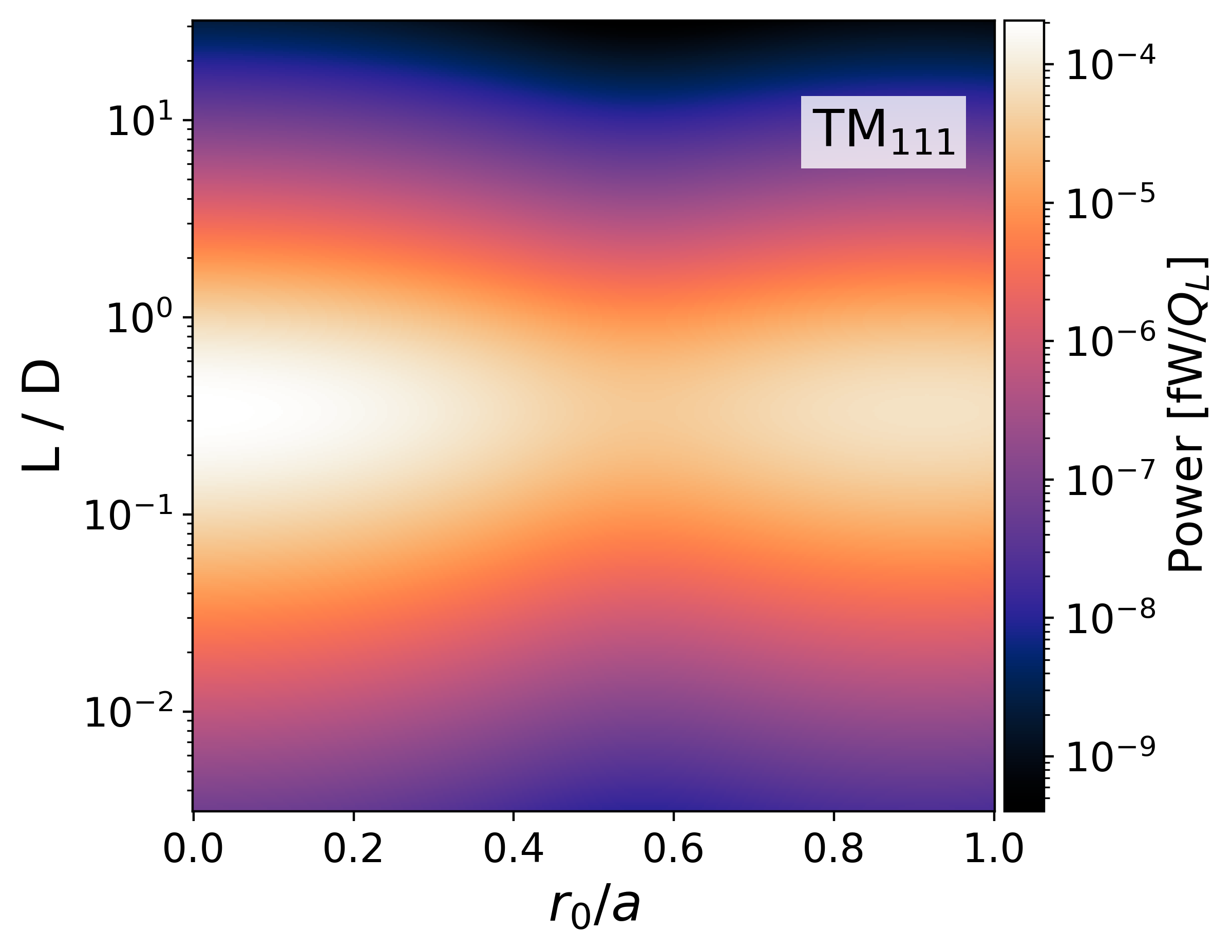}%
        \put(-140,120){\color{white}\scriptsize \textbf{(b)}}%
        \label{fig:powerheatmap_b}%
    }\hfill
    \subfloat{%
        \includegraphics[width=0.33\textwidth]{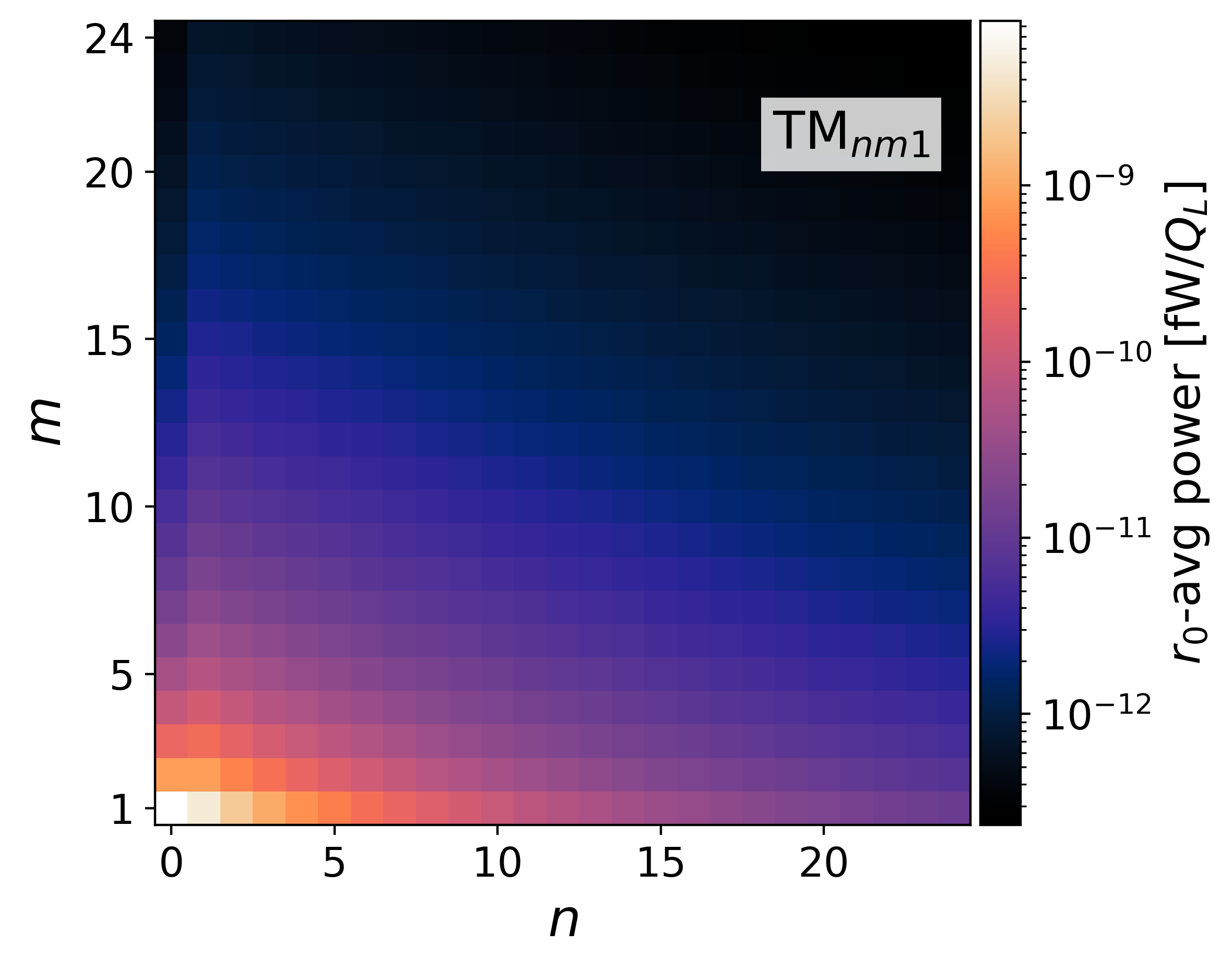}%
        \put(-145,120){\color{white}\scriptsize \textbf{(d)}}%
        \label{fig:powerheatmap_d}%
    }\hfill
    \subfloat{%
        \includegraphics[trim=40 0 0 88, clip, width=0.33\textwidth]{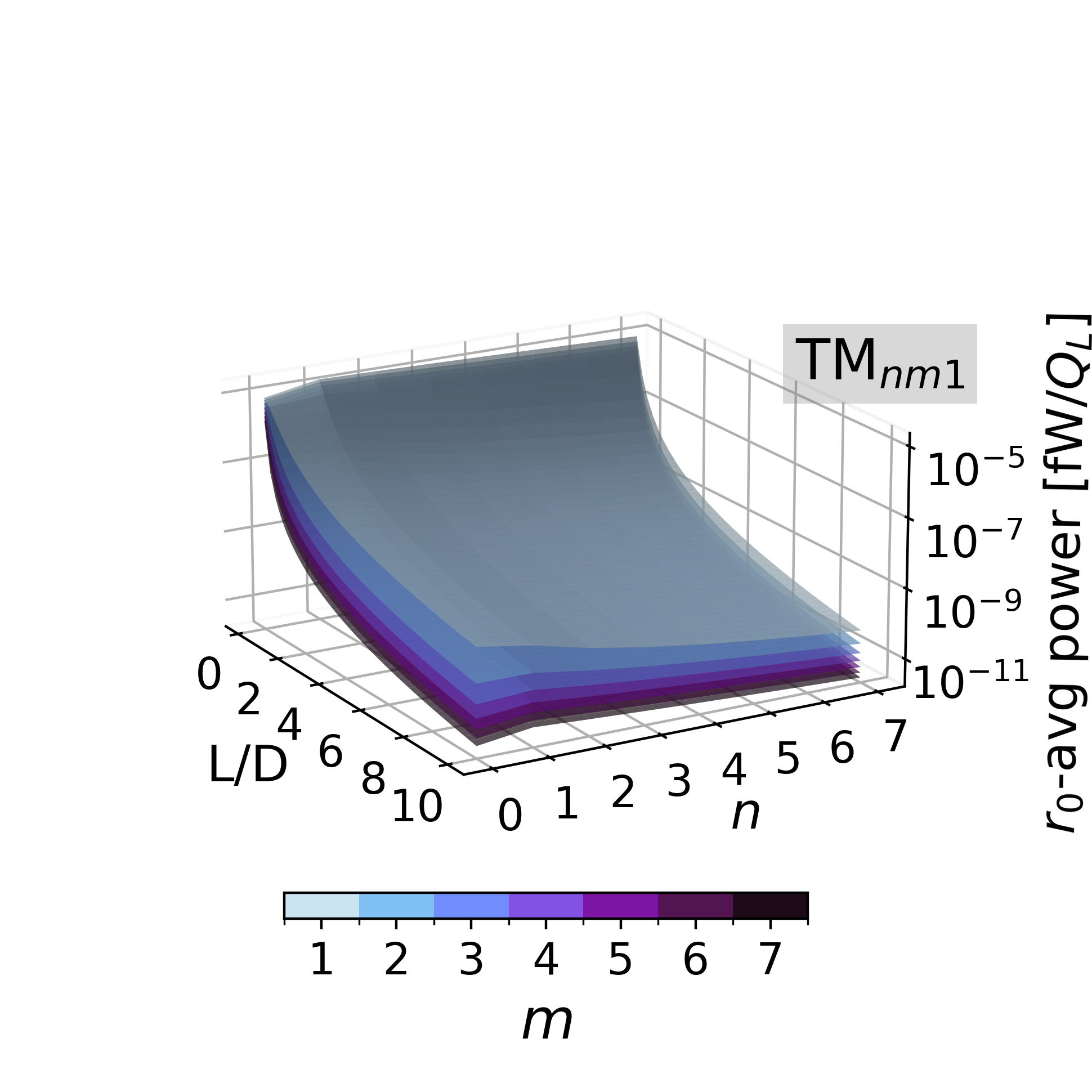}%
        \put(-158,120){\scriptsize \textbf{(f)}}%
        \label{fig:powerheatmap_f}%
    }

    % --- Main Caption ---
    \caption{Mode orientation summed power emitted in cavity CRES experiments for a choice of experimental constraint. All plots are for electrons emitted with a 90$^\circ$ pitch angle and 18.6 keV kinetic energy within ideal cylindrical cavities, and with cyclotron frequency exactly resonant with the mode being calculated (ignoring cavity $Q$ induced frequency shifts). 
    \textbf{(a,b)} Fixed cavity frequency: The resonant power emitted by the same electron into the (TE$_{011}$,TM$_{111}$) mode, both resonant at 1 GHz. Plotted for values of cavity's length-to-diameter ratio, L/D, and guiding center radial position $r_0$ within a cavity of radius $a$. 
    \textbf{(c,d)} Fixed cavity volume and shape: The average power emitted into the (TE$_{nm1}$,TM$_{nm1}$) modes by an electron in a 1 m$^3$ cavity with an L/D of 8. The $n=0$ column appears disjoint because it has one less mode (orientation) to emit into.
    \textbf{(e,f)} Fixed cavity volume: The average power emitted within a 1 m$^3$ cavity, normalized by the cavity loaded quality factor, and calculated as a function of L/D and indexed by the mode indices $n$ and $m$. 
    For plots \textbf{c–f}, powers are averaged over all radii in the volume assuming that electrons are uniformly distributed throughout the cavity.
    }
    \label{fig:powerheatmap} 

\end{figure*}
    % Frequency scaling of the power emitted into the TE$_{011}$ mode with length-to-diameter ratio L/D=8.19 and $Q_{TE_{011}}$=1000. The cyclotron frequency is chosen to be exactly resonant with the TE$_{011}$ mode frequency. The power is plotted for many values of the radial position of the electron's guiding center of cyclotron motion $r$ as a fraction of cavity radius $a$. 
    % (b) The power emitted into the TE$_{nm1}$ mode by an electron in a 1 m$^3$ cavity with a L/D of 8. Calculated with the cyclotron frequency exactly on resonance for each case. 
    % (c) The average power emitted assuming a uniform density of electrons within a 1 m$^3$ cavity, normalized by the cavity loaded quality factor. Calculated for many L/D's and indexed by the mode indices n and m. Lower L/D's favor higher average signal power due to the expanded physics volume.
    % (d) The resonant power emitted by the same electron into the TE$_{011}$ mode (resonant at 1 GHz), with $Q_{TE_{011}}$=1000. Plotted for many values of cavity's L/D. 
    % (e) Same as (b), but for TM modes. 
    % (f) Same as (c), but for TM modes. Note the wider range and logarithmic scaling.
    % }
%-------------------------------------------------------------------------------------------------%
\subsection{Choice of cavity configuration for CRES} \label{ssec:modechoice}
The choice of operating eigenmode(s), cavity relative dimensions, and cavity volume all affect the power spectrum of cavity CRES experiments. 
In \cref{app:convergence}, we detail the convergence properties of the power emitted as a function of the cylindrical mode indices.
In \cref{fig:powerheatmap}, we provide a sample of the possible parameter space that may serve to inform future experiments. 
Whether operating in the frequency space of lower order modes, where the power emitted is maximized, or in the higher order modes, one must consider the mode density in the operating frequency interval. 
For example, TE$_{011}$ and TM$_{111}$ are degenerate in ideal cavities (\cref{fig:powerheatmap_a,fig:powerheatmap_c}). 
Reading out only one of these modes may create a situation where some notable power is emitted but not measured. 

As an example, for an operational mode with a loaded $Q$ of 1,000 and parasitic degenerate mode with $Q$ of 100,000, the electrons in the hundredth of the operational bandwidth that overlaps with the parasitic mode can lose energy at a rate of $\mathcal{O}(100)$ times that which is measured.
The cavity's length-to-diameter ratio ($L/D$) may be adjusted to skew the relative power emission to one mode or the other, as shown in \cref{fig:powerheatmap}.
Other strategies to ameliorate this issue that may allow for a simpler signal analysis could include shifting the resonant frequency, suppressing the $Q$, or reading the signal out of nearby modes.
% The Project 8 neutrino mass experiment plans to observe many decays within the cavity detector, leading to charged particle pileup. This pileup can be treated effectively as a plasma with finite conductivity. As such, an electron can lose energy to the cavity's conducting bulk through the Coulomb interaction. Here we investigate the electron power lost to the medium through irrotational cavity modes (those which do not have a corresponding magnetic field).
% The power spectrum of the electric irrotational modes can be found using the same methodology as before. Using the IE electric field normalization from \cref{ie_norm}, the power spectrum for the $p^{th}$ cyclotron harmonic:
% \begin{widetext}
% \begin{align}
%     P(\pm p\omega_c) &= \RE\left[\frac{-c^2}{\sigma + i\epsilon p \omega_c}\right] \frac{2 e^2 p^2 \cos^2(k_{z,l} z_c)}{(\pi L a^2) {J'_n}^2(\chi_{nm} a)}  \\
%     &\qquad\qquad\qquad *
%     \left\{
%         \begin{array}{c}
%             \frac{1}{2} \left(J_{p}(\chi_{0m} r_0) {J}_{p}(\chi_{0m} \rho_c) + J_{-p}(\chi_{0m} r_0) {J}_{-p}(\chi_{0m} \rho_c) \right)^2 \quad (n=0)\\
%             2\left(J^2_{n+p}(\chi_{nm} r_0) {J}^2_{p}(\chi_{nm} \rho_c) + J^2_{n-p}(\chi_{nm} r_0) {J}^2_{-p}(\chi_{nm} \rho_c)\right) \quad (n \neq 0)
%         \end{array}
%     \right. \label{eqn:ei-power},
% \end{align}
% \end{widetext}
% which is zero unless the cavity bulk is conductive. Project 8 expects a plasma density of --, corresponding to a bulk conductivity of --.
\section{Cavity CRES with Axial Motion} \label{sec:axialMotion}
In CRES experiments, electrons are confined to a magnetic bottle where they undergo axial motion along the trap's symmetry axis as illustrated in \cref{fig:cylindrical-cavity}
% \footnote{That is, unless they are born with 90\degree\ pitch angle in a local minimum of the trap.}
. 
This motion modulates the CRES signal into a comb-like structure. The frequency and amplitude of these sideband components are the focus of this section.%
\footnote{%
For TM modes, electron motion parallel to the cylinder's symmetry axis can also contribute to resonant energy through the coupling to the electric field's $\hat{z}$ component. These spectral features are different from those discussed in this section. See \cref{sec:AxialTM} for more details.} %
This investigation builds on prior work regarding axial motion in waveguides \cite{ashtariesfahaniElectronRadiatedPower2019} and extends the cavity research \cite{hannekeCavityControlSingleelectron2011} to apply to periodic axial motion spanning the cavity's length. %
As in \cref{sec:electronCoupling}, we focus on cylindrical cavities, though the methodology can in principle be applied to arbitrary cavity eigenmodes. %

We first examine the case where an electron undergoing cyclotron motion enters and leaves the trap and cavity unimpeded, akin to the spectral contribution of a single half-period of the orbit. %
The result serves to build physical intuition for the periodic case and is applicable to CRES experiments with axial periods much longer than the observation window. % 
Next, we provide a general framework for calculating the spectral amplitudes of periodic axial motion. %
This framework is then applied to the case of an electron trapped in an idealized magnetic ``box'' trap coupled to a single resonating cavity mode. %
The resulting sideband distribution provides a first-order estimate of cavity CRES spectral structure arising from the complex dynamics of a charged particle in real-world magnetic bottles. %
% We will then provide estimates of the number of peaks in the frequency comb in realistic magnetic traps, and investigate the impact of coupling directly to the $\hat{z}$ component of the TM electric fields.

\subsection{Radiation spectrum of a throughgoing electron} \label{sssec:streakinge}
We first examine the simple case of an experiment without a magnetic trap, in which an electron with a helical trajectory is injected into the cavity at $z=0$ and exits $z=L$. This limiting case establishes a baseline for the signal modulation due to axial motion within cavities, and is comparable to the Doppler shift. The cavity in this case is imagined to have small inlets or open-ended mode geometries that allow for electron injection and/or gas flow, such as in \cite{wengerResonantFrequencyOpenEnded1967}. We assume a constant axial velocity $v_z$ along the $\hat{z}$-axis such that $z_{c}(t) = v_z t$. Additionally, we assume a uniform magnetic field in the cavity region such that $\rho_c(t)$ and $\omega_c(t)$ are constant. For any mode, the spectral amplitudes $j^M_{nml}$ (\cref{eqn:snml-expanded}) take the form
\begin{align}
    j^M_{nml}(\omega) &= -e \omega_c \rho_c \mathcal{F} \left\{ \int\delta\left(\mathbf{r} - \mathbf{r}_c(t)\right) E^{\,M}_{\phi,nml} dV \right\} \nonumber\\
    &= -e A_{nml}^M \sum\limits^{\infty}_{u=-\infty} \mathrm{P}^M_{n m,u} \mathcal{F} \left\{ \text{rect} \left(\frac{t}{v_z / L} \right) \right. \nonumber \\
    & \myspace{7} \cdot \left. \Phi_{n,u}(t) Z_l(t) \vphantom{\left(\frac{t}{v_z / L} \right)} \right\}, \label{eqn:axial-snml}
\end{align}
in the guiding center coordinate system, where $\text{rect}()$ is the rectangle function defined as
\begin{align}
    \text{rect}\left(\frac{t}{d}\right) = 
    \begin{cases}
        0,  & \text{if } |t|>\frac{d}{2} \\
        \frac{1}{2},  & \text{if } |t| = \frac{d}{2} \\
        1, & \text{if } |t| < \frac{d}{2},
    \end{cases}
\end{align}
which is used to limit the interaction to the spatial region of the cavity mode volume. 

The Fourier transform in \cref{eqn:axial-snml} is
\begin{align}
    \mathcal{F} \left\{ \text{rect} \left(\frac{t}{v_{z} / L} \right) \omega_c \right. & \cos(u(\phi_0 - \omega_c t) - n\phi_0) \nonumber\\
    &\left. \cdot \sin(k_l v_{z} t) \vphantom{\left(\frac{t}{v_{z} / L} \right)}\right\} .
\end{align}
After an application of the Fourier shift theorem, this becomes a Fourier transform of a rect function, giving
\begin{align}
    & \frac{i \omega_c T_a}{4}\sum_\pm \pm \left[ e^{i (u-n) \phi_0} \text{sinc} \left(\frac{T_a}{2} (\omega - (u \omega_c \mp k_{l} v_{z}))\right)
    \right. \nonumber\\
    &\myspace{1} \left. + e^{-i (u-n) \phi_0} \text{sinc}\left(\frac{T_a}{2} (\omega + (u \omega_c \pm k_{l} v_{z}))\right) \right] . \label{eqn:streakingspectra}
\end{align}
When the electron traverses the full cavity length, the time the electron spends in the cavity as $T_a = L / v_{z} $. In the limit where the throughgoing electron undergoes many cycles of cyclotron motion while within the cavity, the radiated power concentrates centered around two frequencies--a blue and red shift of the cyclotron frequency. The inherent spectrum resolution is dependent on the ratio of time spent within the cavity to cyclotron period. %
The above method is applicable to general electron motion with constant axial speed: By (i) adding an initial phase to match the starting $\hat{z}$-position, $z_{1}$, through ${Z_{l}(t)\rightarrow Z_{l}(t+z_{1}/v_{z})}$ and (ii) changing the observation period $T_{a}$ to match the observation time, one may use the above method to calculate the short-time windowed spectrum without any conditions on entering/exiting the resonator.
% \begin{align}
%     s^M_{nml}(\omega,u) &= -e A_{nml}^M \mathrm{P}^M_{nm,u} \int\limits^\infty_{-\infty}\cos\left(u(\omega_c t + \phi'_0 + \phi_0) - n\phi_0'\right) \nonumber\\
%     &\myspace{7} \times \left[ rect \left(\frac{t}{v_{z} / L} \right) \cos\left(k_{l} v_{z} t\right) \right] e^{-i \omega t}  dt ,
% \end{align}

\subsection{Radiation spectrum of an electron with periodic axial motion} \label{sssec:axialgeneral}
A trapped electron undergoes periodic motion at its frequency of axial oscillation, $\omega_{z}$. Realistic magnetic traps, where the field smoothly increases at the trap edges, will cause variations in the cyclotron radius $\rho_c(t)$ and frequency $\omega_c(t)$ that are also periodic in $T_{a}=2\pi/\omega_{z}$. The expression for the electron's current decomposition \cref{eqn:snml-def} now treats $\rho_c$, $\omega_c$, and $z_c$ as functions of time:
\begin{align}
    j^M_{nml}(\omega)
    &= -e B^M_{nml} \sum^{\infty}_{u=-\infty} \mathcal{F} \left\{ \mathrm{P}^M_{nm,u}(t) \Phi_{n,u}(t) Z_l(t) \right\} \label{eqn:snml-axial} .
\end{align}
We can Fourier decompose these functions into spectral components harmonic in $\omega_{z}$,
\begin{subequations} \label{eqns:axialexpansions}
\begin{align}
    % \rho_c(t) &= \bar{\rho}_c + \sum_{\nu = 1}^\infty \bar{\rho}_\nu \sin(\nu  \omega_{z} t + k_{z,l} z_0) \label{eqn:rhoc_exp}\\
    % \omega_c(t) &= \bar{\omega}_c + \sum_{\nu = 1}^\infty \bar{\omega}_\nu \sin(\nu \omega_{z} t + k_{z,l} z_0 ) , \label{eqn:omegac_exp} \\
    \mathrm{P}^M_{nm,u}(t) &= \sum\limits^{\infty}_{\nu = 0} \mathrm{P}^{M,\nu}_{nm,u} \cos(\nu \omega_{z} t) \label{eqn:rc_exp}\\
    \Phi_{n,u}(t) &=  \sum\limits_{\nu = -\infty}^\infty \Phi^\nu_{n,u} \cos\left(u (\bar{\omega}_c + \nu \omega_{z})t \right) \label{eqn:phic_exp} \\
    Z_l(t) &= \sum^{\infty}_{\nu = 1} Z^\nu_l e^{i \nu \omega_{z} t} \label{eqn:sinkz_exp} ,
\end{align} 
\end{subequations}\noindent
The expansions above are written such that their dominant Fourier components are near $\nu=0$: that is, the cyclotron radius and z-position are modulated mostly at the axial frequency $\omega_{z}$. 
% This assumption allows us to characterize the magnitude of $s^M_{\alpha}(\omega)$. 

\begin{figure}[ht]
    \centering
    \resizebox{0.48\textwidth}{!}{\begin{tikzpicture}[every node/.style={font=\Large}]
    % Parameters
    \def\L{5}        % Length L
    \def\A{0.8}      % Amplitude for sharp sinusoid
    \def\offset{0.0} % Offset for sharp sinusoid
    \def\modL{1*\L}
    \def\tuboff{.7}
    \def\phaserphi{90}
    % Axes
    \draw[->] (0, 0) -- (2.1*\L, 0) node[below] {time};
    \draw[->] (0, -0.5) -- (0, 1.5) node[left, rotate=90, yshift=.4cm] {Amplitude};
    % Function 1: f(z) = sin(pi z / L)
    \draw[thick, blue, domain=0:2*\L, samples=100, smooth] 
        plot (\x, {\A*(\tuboff/2 + 0.75 * (1 + abs(sin(180*\x/\L)))*abs(sin(180*\x/\L)))});
    % Function 1b: twice the frequency of the above
    % \draw[thick, blue, domain=0:\L, samples=100, smooth] 
    %     plot (\x, {\A*(0 + 0.75 * (1 + abs(sin(180*\x/\L)))*sin(180*1.8*\x/\L + 2/.2/2))});
    % \draw[thick, blue, domain=\L:2*\L, samples=100, smooth] 
    %     plot (\x, {\A*(0 + 0.75 * (1 + abs(sin(180*\x/\L)))*sin(-180*1.8*\x/\L + 2/.2/2 + 2*180*1.8))});
    \node[blue, right] at (2.0*\L, 0.3) {$Z_l(z_c(t))$};
    % Function 2: g(z) = A(z) sin(w(z) pi z / L)
    \draw[thick, red, domain=\offset:2*\L-\offset, samples=300, smooth] 
        plot (\x, {\A * (\tuboff - 0.2 * (1 + abs(sin(180*\x/\L)))) * sin(180*30*(2 - 2*\x/\L))});
    \node[red, below] at (1.5, -0.7) {$\Phi_{n,u}(\omega_c(t))$};
    % Function 3: Bathtub-like function
    \draw[thick, black, domain=0:2*\L, samples=100, smooth] 
        plot (\x, {\A*(\tuboff + .2 - 0.35 * (1 - abs(sin(180*\x/\L))))});
    \node[black, above, xshift=.2cm] at (2.25*\L, 0.6) {$\mathrm{P}^M_{nm,u}(\rho_c(t))$};
    % Dashed lines for reference at z=0 and z=L
    \draw[dashed] (\L, 0) -- (\L, 1.5) node[above] {$T_a/2$};
    \draw[dashed] (2*\L, 0) -- (2*\L, 1.5) node[above] {$T_a$};
\end{tikzpicture}}
    \caption{Illustration of the variation of the quantities in \cref{tab:coupling_coefficients} with time when the electron is undergoing periodic axial motion with period $T_{a}$. $\Phi_{n,u}$ quickly oscillates due to the cyclotron motion, $\mathrm{P}^M_{nm,u}$ varies with changes in the cyclotron radius, and $Z_l$ varies with the mode amplitude's $\hat{z}$ profile. The periodicity and correlation between these functions dictate cavity CRES signal topology.}
    \label{fig:axial-quants}
\end{figure}
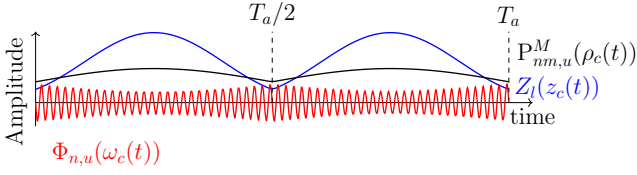

The `main carrier' signal, which is the spectral component not modulated by the axial frequency, corresponds to the average of the cyclotron frequency over the entire periodic motion \cite{ashtariesfahaniElectronRadiatedPower2019}. Thus, regardless of the spatial location of the magnetic field variations that make $\omega_c $ different from its value in the flat trap center, the main carrier observed by the cavity will be at $\bar{\omega}_c $. This can be interpreted as a consequence of the cyclotron motion accumulating a phase within a magnetic field that varies in $\hat{z}$. Secondly, the magnitude of the main carrier is partially limited by the average values of the functions $\mathrm{P},\Phi$, and $Z$. 
These quantities are plotted qualitatively for a period of axial motion in \cref{fig:axial-quants}. 
The correlations between each member of \cref{eqns:axialexpansions} dictate the signal strength of the main carrier and sidebands.
%----------------------------------------------------------------------------------------%
\subsection{Sidebands in a ``box'' magnetic trap} \label{ssec:flatTrap}
The magnetic field lines of the magnetic bottle have a large impact on the amplitude and location of the CRES signal peaks. 
A box trap is an idealized magnetic trap in which the magnetic field amplitude in the trap well is constant and homogeneous, and the magnetic gradients that form the trap's walls arise abruptly at the trap edges.  
As the magnetic field approaches an ideal box trap, the modulation of the cyclotron frequency due to edge magnetic field gradients decreases.
In the limit of an idealized box trap, the cyclotron radius and accompanying $\mathrm{P}$ functions are constant, and the cyclotron frequency is constant, resulting in the spectrum of $\Phi$ given by \cref{eqn:phinu_FT}. 
Then the signal decomposition \cref{eqn:snml-axial} is the convolution of the two time-dependent functions, 
\begin{align}
    j^M_{nml}(\omega)
    &= -e B^M_{nml} \sum^{\infty}_{u=-\infty} \mathrm{P}^M_{nm,u} \mathcal{F} \left\{ \Phi_{n,u}(t) Z_l(t) \right\} \\
    &= -e B^M_{nml} \sum^{\infty}_{u=-\infty} \mathrm{P}^M_{nm,u} \int \Phi_{n,u}(\omega-\omega') Z_l(\omega') d\omega' . \label{eqn:boxtrapspectra-formal}
\end{align}
We next calculate the Fourier transform of $Z_{l}(t)$ in order to characterize the structure of $j^M_{nml}(\omega)$.

In an ideal box trap the electron's axial speed is constant and its axial velocity instantaneously reverses at the trap walls.
The corresponding Fourier transform of $Z_{l}(t)$ (\cref{eqn:sinkz_exp}) can be calculated by summing the two phases of axial motion, each analogous to the throughgoing electron calculation \cref{eqn:streakingspectra}. Assuming the electron starts moving to the right at $t=0$, these contributions to the Fourier transform are
\begin{align}
\begin{cases}
    \frac{1}{T_{a}} \int^{T_{a}/2}_{0} \sin\left( k_{l} (z_{1} + v_{z} t) \right) e^{i \nu \omega_{z} t} dt  & \text{right-going} \\
    \frac{1}{T_{a}} \int_{T_{a}/2}^{T_{a}} \sin\left( k_{l} (z_{2} + L_{a} - v_{z} t) \right) e^{i \nu \omega_{z} t} dt  & \text{left-going}  ,
\end{cases}
\label{eqns:boxtrapFT}
\end{align}
where $z_{1}$ is the location of the trap wall closest to the cavity wall at $z=0$ and ${z_{2}=z_{1} + L_{a}}$ (trap length $L_{a}$) is the location of the other end of the trap%
\footnote{Both $z_1$ and $z_{2}$ are assumed within the cavity, with range $[0,L]$.}.
The resulting decomposition into $Z^\nu_l$ is, using ${k_{l} = \pi l / L}$, ${\omega_{z} = \pi v_{z}/L_{a}}$, ${T_{a}= 2 L_{a}/v_{z}}$,
\begin{align}
    Z^\nu_l &= \frac{ l \frac{L_{a}}{L} \left( \cos\left(\pi  l \frac{ z_1}{L}\right)- (-1)^\nu \cos\left(\pi l \frac{(L_{a} + z_1)}{L}\right)\right)}{\pi  \left( l^2 \frac{L_{a}^2}{L^2} - \nu^2\right)} .
\end{align}

\begin{figure}[ht]
    \centering
    \subfloat{%
    \includegraphics[width=.48\textwidth]{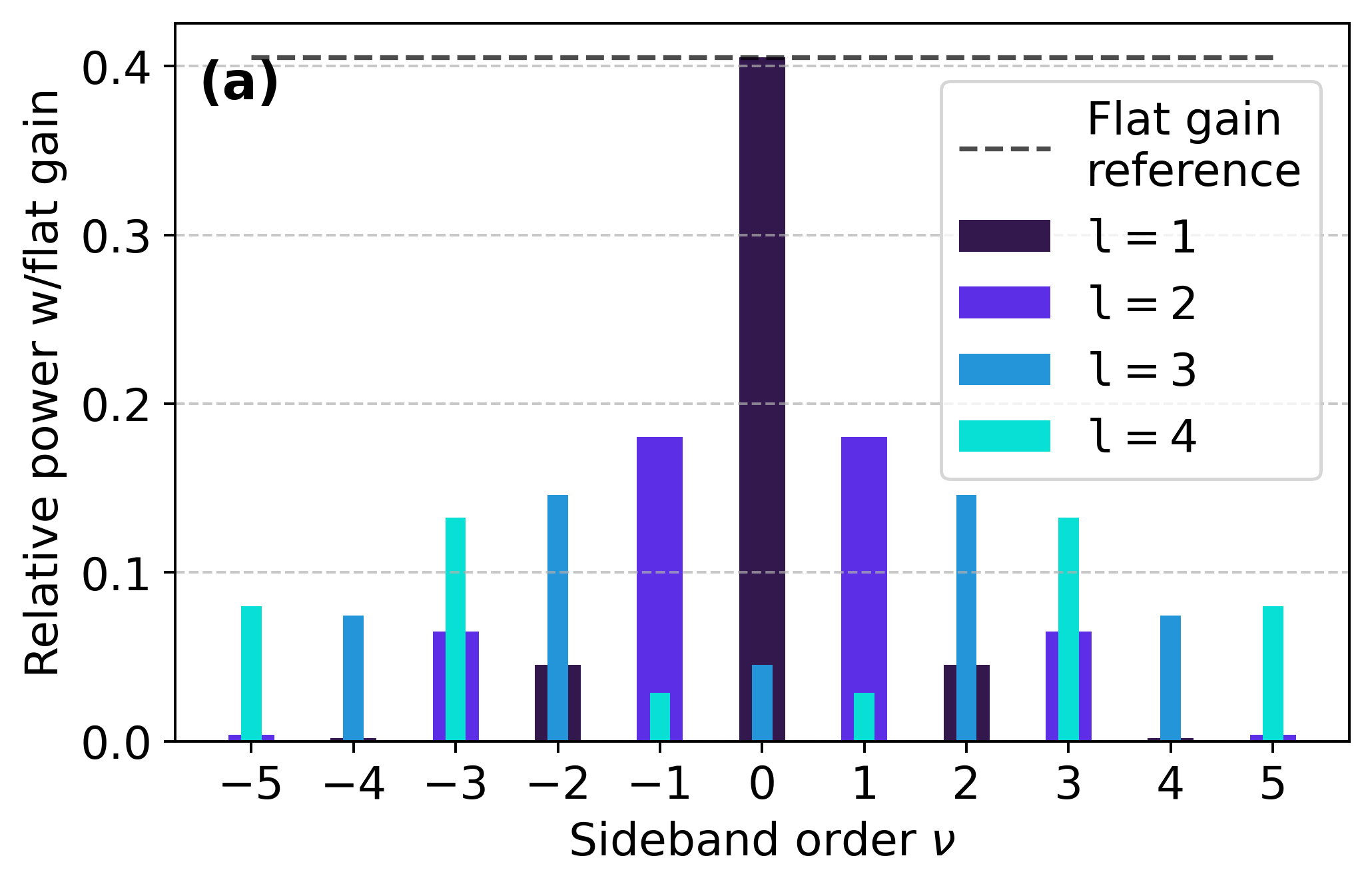}%
    \label{fig:boxtrapsidebands_a}%
    }%
    \\[-1ex]%
    \subfloat{%
    \includegraphics[width=.48\textwidth]{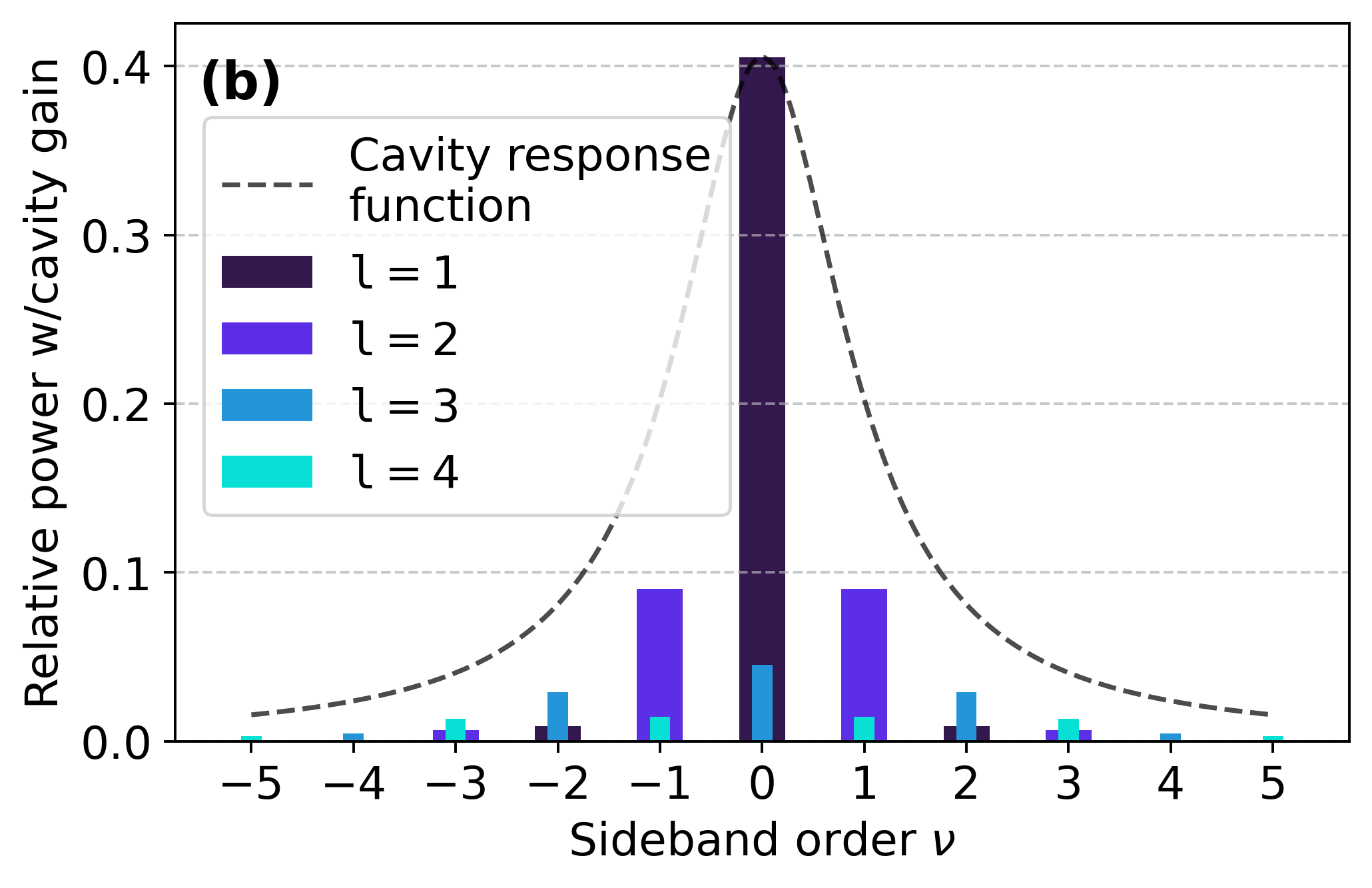}%
    \label{fig:boxtrapsidebands_b}%
    }
    \caption{Power emitted into the sidebands of the CRES signal of \cref{eqn:Zl_boxtrap_spectrum} for several $\hat{z}$ index $l$ (all resonating cavity modes with the same $l$ index will have the same relative sideband spectra). The sideband power of the signal peak at $\omega_{\alpha} + \nu \omega_{z}$ as a fraction of the maximum single-mode signal power in the 90$^\circ$ case (with no axial motion) is plotted \textbf{(a)} omitting the gain of the cavity mode and \textbf{(b)} multiplying by the cavity mode gain. In the latter case, the sidebands off-resonance are greatly suppressed.}
    \label{fig:boxtrapsidebands}
\end{figure}

When the trap length $L_a = z_2 - z_1$ is equal to the cavity length $L$,
\begin{align}
    Z_l(\omega) &= \mathcal{F}\{Z_{l}(t)\} = \sum_\nu Z^\nu_l \delta(\omega - \nu \omega_{z}) \\
     &= \sum_{|\nu| \neq l} \frac{ l \left( 1 - (-1)^{\nu+l} \right)}{\pi  \left( l^2 - \nu^2\right)} \delta(\omega - \nu \omega_{z}) . \label{eqn:Zl_boxtrap_spectrum}
\end{align}
The radiation spectrum of the electron $j^M_{nml}(\omega)$ is different for even and odd $l$. %
If the cavity mode's $\hat{z}$ index $l$ is odd, then there are even sidebands only; and if $l$ is even then there are odd sidebands only. %
As a consequence, CRES signals from even $l$ cavity modes lack a main carrier. %
The amplitude of the sidebands decreases quickly as $1/\nu^2$. 
The coefficients $j^M_{nml}(\omega)$ set the maximum signal strength for a given location within the cavity frequency response.
After using \cref{eqn:Zl_boxtrap_spectrum} in the convolution of \cref{eqn:boxtrapspectra-formal}, the amplitude of the CRES sideband signals may be calculated; the results are shown in \cref{fig:boxtrapsidebands}.

As a qualitative comparison, we calculate the power in the sideband relative to the maximum power emitted in the 90$^\circ$ pitch angle case, without modulation by the cavity response function, when the trap length is equal to the cavity length in \cref{fig:boxtrapsidebands_a}.
In this case, the maximum possible signal strength of a peak in the CRES spectrum with axial motion is always less than that of a 90$^\circ$ electron.
However, it is important to note that CRES events are dominated by electrons with sub-90$^\circ$ pitch angles, and that the box trap calculation (or a similar approximation) may be a more faithful estimate of the maximum power emitted in a typical CRES event.

More central to our study is \cref{fig:boxtrapsidebands_b}, where we show the sideband power with modulation by the cavity response function of \cref{eqn:powercondensed} when the cavity quality factor is large enough to ignore $Q_\alpha$-induced cavity mode frequency shifts.
We assume the electron's cyclotron frequency is exactly on resonance with the cavity mode, and that the axial frequency is one half of the cavity's bandwidth, $\omega_{z} = \frac{1}{2} \text{BW}_{\alpha}$.
The gain of the cavity response function is normalized artificially to one, making the gain at $\omega_{\alpha} \pm \omega_{z}$ one-half.
These conditions are a proxy for a plausible experimental CRES configuration.
Detectable sidebands are instrumental in measuring electron energies with CRES, and in this context \cref{fig:boxtrapsidebands_b} represents an optimization problem that couples the choice of operating mode, cavity mode's bandwidth, and electron energy range of interest.
The optimization of these parameters in signal reconstruction is left to future work.

The above analysis allows for the possibility that a sideband of a cyclotron harmonic contributes near the frequency of a different harmonic: for example a high-order sideband $\kappa$ of the second harmonic at $2 \omega_{c}$ may have frequency $2\omega_c - \kappa \omega_{z}$ near the fundamental harmonic at $\omega_{c}$. 
We remark that in Project 8 CRES experiments, the cyclotron frequency is an order of magnitude (or more) larger than the axial frequency, and the sideband amplitude decreases fast enough to make such contributions to the power spectrum at the fundamental cyclotron harmonic negligible.

The impact of non-negligible grad-B motion is a further splitting of the CRES signal.
We have seen that, in this case, the main carrier signal is split (\cref{eqn:phinu_GradB_FT}). 
The power spectrum is further complicated by any axial motion, whereby both of the shifted grad-B peaks will be accompanied by sidebands with amplitudes given by \cref{eqn:Zl_boxtrap_spectrum} or similar coefficients. 
The calculation of the accompanying spectrum is beyond the scope of this work.

We have used the ideal box trap to approximate the general features of the cavity CRES signal of cyclotron electrons with periodic axial motion. Other idealized traps, such as the `bathtub' trap~\cite{ashtariesfahaniElectronRadiatedPower2019}, or the `harmonic' trap~\cite{thomasDetectionEstimationLimits} may give more precise insight into the signal structure of cavity CRES signals and are a topic for future exploration.
%----------------------------------------------------------------------------------------%

\section{Noise Model \& Microwave Readout} \label{sec:noiseModel}
The ability to use CRES signals to reconstruct the electron's energy is highly dependent on the signal-to-noise ratio (SNR) of the event. Following the analysis of CRES signal power and topology, we now address the noise power coming from the cavity. Cavity noise has been studied in detail, most notably in the context of equivalent circuits \cite{kimRevisitingDetectionRate2020,dahmBlackbodyRadiationSinglemode1975,moskowitzAnalysisMicrowaveCavity1988} for one or more ports. Here, we employ the cavity model derived in \cref{sec:responseModel} in our derivation of the cavity noise. This is made possible through the use of generalized electromagnetic power flow in steady-state. We first cover thermal noise and signal power coming from the cavity through the transmission line and into an amplifier, which will be considered the final RF component in the chain. We then explicitly model noise propagation along a realistic lossy RF signal chain. Finally, we derive the physical SNR and explore the impact of experimental design in simplified parametric studies.

\subsection{Thermal cavity noise}
A cavity with distinct ports (with the cavity walls considered a port) can be modeled as an out-of-equilibrium, steady state system. Here, we model the cavity as the medium through which thermal radiation from the ports is transported. We can characterize the steady state behavior of this system by assuming that the effective noise temperature from the cavity's walls, interior, and port are constant. The gas temperature is assumed to be negligible compared to the walls and ports; thus, its contribution to RF noise is ignored.%
\footnote{This assumption is backed by preliminary atomic tritium operation concepts, in which the gas temperature will be approximately two orders of magnitude lower than the cavity walls.}
The energy flux through the boundaries must satisfy
\begin{align}
    \sum\limits_i P^N_{\alpha, i}(\omega) &= 0 , \label{eqn:poyntingnoise}
\end{align}
where $P^N_{\alpha, i}(\omega)$ is the noise power through the boundary indexed $i$ mediated by mode indexed $\alpha$.
% \begin{align}
%     P_{i} = \RE\left[\int_{S_{i}} \mathbf{S}(\mathbf{r}, \mathbf{T}_{\alpha}) \cdot\mathbf{\hat{n}}(\mathbf{r})dS\right]
% \end{align}
% where $\Theta_{h}= \frac{\hbar\omega}{2}\coth(\frac{\hbar\omega}{2 k_{B} T_h})$ is the thermal photon population at physical temperature $T_h$, 
The power through the boundary is the sum of two incoherent sources, coming from both sides of the boundary:
% , $S_0$ is the cavity wall boundary, and $S_p$ is the cavity port boundary. The second term is the quantity of interest, the noise power coming from the cavity. The Poynting vectors, $\mathbf{S}(\mathds{T}) = \left(\mathbf{E}_{\alpha} \times \mathbf{H}^*_{\alpha}\right)|_{\mathds{T}}$, at the boundaries are defined in our study through the impedance boundary approximation. On surface $l=0,p$:
% \RE\left[\int_{S_{l}} (\mathbf{E}_{\alpha} \times \mathbf{H}^*_{\alpha})\cdot\mathbf{\hat{n}}(\mathbf{r})dS\right] 
\begin{align}
	P^N_{\alpha, i} &= \RE\left[\int_{S_{i}} \mathbf{S}_{\alpha}(\mathbf{r}, \mathbf{T}_{\alpha}) \cdot\mathbf{\hat{n}}(\mathbf{r})dS\right] \nonumber\\
    & \myspace{5} + \RE\left[\int_{S_{i}} \mathbf{S}_{\alpha}( \mathbf{r}, \mathds{T}_{i})\cdot\mathbf{\hat{n}}(\mathbf{r})dS\right] ,
\end{align}
where $\mathbf{S}_{\alpha}(\mathbf{r}, \mathds{T}) = \frac{1}{4\pi^2} \langle\mathbf{E}_{\alpha}(\mathbf{r}) \times \mathbf{H}^*_{\alpha}(\mathbf{r})\rangle|_{\mathds{T}}$ is the time averaged Poynting vector for mode $\alpha$ (frequency dependence implied), $\mathbf{\hat{n}}(\mathbf{r})$ points outward from the cavity interior parallel to the boundary normal. 
We define the noise equivalent temperature (NET) $\mathds{T}_i$ of the boundary indexed $i$ by the blackbody equivalent noise power~\cite{nyquistThermalAgitationElectric1928}. For a general RF component at temperature $T$, the corresponding noise temperature takes into account the quantum mechanical photon population $\eta(\omega)$:
\begin{align}
\eta(\omega) &= \frac{\hbar\omega}{k_B T} \left( \frac{1}{\exp(\hbar\omega/k_B T)-1} + \frac{1}{2} \right) \\
\mathds{T}&\equiv T \eta(\omega) \\
P_{RF} &= k_B \mathds{T} .
\end{align}
In this section, $\mathbf{T}_{\alpha}$ is the NET of the cavity mode $\alpha$ that we must derive, and $\mathds{T}_{i}$ denotes the NET of the corresponding part of the cavity boundary $S_i$. 

Since the fields at the boundaries are dictated by impedance boundaries in our model (\cref{eqn:impedanceBdry}), we can write
\begin{align}
    P^N_{\alpha, i} &= \frac{\mu}{4\pi^2} \left(|d^-_{\alpha}(\omega)|^{2} - |d^+_{\alpha}(\omega)|^{2}\right) \frac{\RE(\omega_{\alpha})}{Q_{\alpha,i}} , \label{eqn:heattransport1}
\end{align}
where $d^\pm_{\alpha}(\omega)$ is the magnetic field amplitude of the mode $m$ due to thermal fluctuations on the inside/outside of the boundary.

The magnetic field coefficients $|d^\pm_{\alpha}|^{2}$ can be found from the fluctuation dissipation theorem and by invoking the reciprocity of thermal sources at the port boundary. Specifically, a thermal source at a port will excite the same thermal mode amplitude at the boundary as a cavity mode with the same temperature (\cref{eqn:Qdefn}). From the magnetic Green's dyadic $\bar{\mathbf{G}}_m$, one can apply the fluctuation dissipation theorem to find the average stored magnetic energy $W^{h}_{\alpha}$ \cite{novotnyPrinciplesNanooptics2006,narayanaswamyGreensFunctionFormalism2014}, which we can use to solve for the magnetic field coefficients. The component of the magnetic dyadic Green's function that holds energy may be defined through \cref{eqn:shortdm},
\begin{align}
    \mathbf{H}(\mathbf{r}')&= \int_{V} \left[i \omega \epsilon_{0} \epsilon_h \mathbf{\bar{G}}_{m}(\mathbf{r}, \mathbf{r}')\cdot \mathbf{J}^{m}(\mathbf{r}) \right. \nonumber\\
    & \myspace{5} \left. + \mathbf{\bar{G}}_M(\mathbf{r}, \mathbf{r}')\cdot \mathbf{J}^{e}(\mathbf{r}) \right] d\mathbf{r} \\
    \bar{\mathbf{G}}_{m}(\mathbf{r},\mathbf{r'},\omega) &= - c^2 \sum_{\alpha} L_{\alpha}(\omega) \mathbf{h}_{\alpha}(\mathbf{r})\mathbf{h}_{\alpha}(\mathbf{r'}) ,
\end{align}
where $\mathbf{J}^{e}$ ($\mathbf{J}^m$) is the electric (magnetic) current inside the cavity, and $\epsilon_h =\frac{1}{\epsilon_0}\left( \epsilon + \frac{\sigma}{i\omega}\right)$. Applying the theorem when near the resonant frequency of mode $\alpha$, $\bar{\mathbf{G}}_{m}(\mathbf{r},\mathbf{r'},\omega) \approx - c^2 L_{\alpha}(\omega) \mathbf{h}_{\alpha}(\mathbf{r})\mathbf{h}_{\alpha}(\mathbf{r'})$, and 
\begin{align}
	W^{h}_{\alpha} &= \frac{1}{2} \frac{\mu}{4\pi^2} |d^\pm_{\alpha}(\omega)|^{2} N_{h,\alpha} \\
	&= 2 \frac{\omega}{c^2} k_B \mathds{T}_\pm \int_{V}\text{Tr}\IM \left[\epsilon_h \mathbf{\bar{G}}_{m}(\mathbf{r},\mathbf{r},\omega)\right] d\mathbf{r}\\
	&= 2 k_B \omega \mathds{T}_\pm \IM \left[- \epsilon_h L_{\alpha}(\omega)\right] N_{h,\alpha} ,
\end{align}
leading to 
\begin{align}
    \frac{\mu}{4\pi^2} & |d^\pm_{\alpha}(\omega)|^{2} = 4 k_{B} \omega \mathds{T}_\pm \IM \left[- \epsilon_h L_{\alpha}(\omega)\right] ,
\end{align}
where $\mathds{T}_\pm$ is the effective temperature on the inside (+) or outside ($-$) of the boundary, $\text{Tr}\mathbf{\bar{A}} = \sum_{p=1}^3 A_{pp} $, $\IM[x]$ takes the imaginary component of $x$, and the magnetic field $\mathbf{h}_{\alpha}$ volume integrals are normalized to $N_{h,\alpha}$. For more details on the Green's dyadics used above, see \cref{app:gfheat}. 

We can now return to \cref{eqn:heattransport1}, using these magnetic field amplitudes to write (with $\Omega_{\alpha} = \RE[\omega_{\alpha}]$),
\begin{align}
    P^N_{\alpha, i} &= 4 k_B \omega \frac{\Omega_{\alpha}}{Q_{\alpha,i}} \left(\mathbf{T}_{\alpha} - \mathds{T}_i\right) \IM \left[- \epsilon_h L_{\alpha}(\omega)\right] . \label{eqn:poyntingport}
\end{align}
Since it has been assumed that the cavity has no internal thermal sources, its thermal temperature is determined by the temperature of its inputs. The effective cavity temperature $\mathbf{T}_{\alpha}$ is then dictated by the relation \cref{eqn:poyntingnoise}, and can be evaluated by substituting in \cref{eqn:poyntingport}, grouping the terms associated with $\mathbf{T}_\alpha$ to equal \cref{eqn:impedanceAndQ}, then dividing out the common factors:
\begin{align}
	\mathbf{T}_{\alpha} &= \sum\limits_{i} \frac{Q_{\alpha} }{Q_{\alpha, i}} \mathds{T}_{i} .
\end{align}
Again, $Q_{\alpha}$ is the loaded $Q$ of mode indexed $\alpha$. 

Now that the cavity temperature is known, we can compute the noise power out of the cavity through port $p$ at noise temperature $\mathds{T}_p$:
\begin{align}
	P^N_{\alpha, p} &= 4 k_B \omega \left( \sum\limits_{i} \frac{Q_{\alpha} }{Q_{\alpha, i}} \mathds{T}_{i} - \mathds{T}_{p}\right) \IM \left[- \epsilon_h L_{\alpha}(\omega)\right] .
\end{align}
Recall the cavity wall boundary is denoted $i=0$, and the unloaded Q by $Q_{\alpha, 0}$. For a cavity with a single port, under the same conditions that define the power formula \cref{eqn:powercondensed} (single mode, non-conductive cavity volume), 
\begin{align}
    & \mathbf{T}_{\alpha} - \mathds{T}_{p} = \frac{Q_{\alpha}}{Q_{\alpha,0}} \left(\mathds{T}_0 - \mathds{T}_p \right) \\
    &\IM \left[L_{\alpha}(\omega)\right] = - \frac{\frac{Q_{\alpha}}{\Omega_{\alpha}^{2}} }{1 + \frac{Q_{\alpha}^2}{\Omega_{\alpha}^{4}} \left(\Omega_{\alpha}^2- \omega^2 - \frac{\Omega_{\alpha}^{2}}{(2 Q_{\alpha})^{2}}\right)^2} ,
\end{align}
which gives
\begin{align}
    & P^N_{\alpha, p} = \frac{4 k_B \frac{\omega}{\Omega_{\alpha}} \left(\mathds{T}_0 - \mathds{T}_p \right) \frac{Q^2_{\alpha}}{Q_{\alpha,0} Q_{\alpha, p}} }{1 + \frac{Q_{\alpha}^2}{\Omega_{\alpha}^4} \left(\Omega_{\alpha}^2 - \omega^2 - \frac{\Omega_{\alpha}^{2}}{(2 Q_{\alpha})^{2}}\right)^2} \label{eqn:snrcondensed} . 
\end{align}
%--------------------------------------------------------------------%
\subsection{Signal power extracted} \label{ssec:signalpower}
Similarly, as a consequence of implementing impedance boundaries, the signal power can be written in terms of the magnetic field amplitude $d_{\alpha}(\omega)$ (\cref{eqn:shortdm}),
\begin{align}
	P^{\text{sig}}_{\alpha, i} &= \frac{1}{2\pi^2} \RE\left[\int_{S_{i}} (\mathbf{E}_{\alpha} \times \mathbf{H}^*_{\alpha})\cdot\mathbf{\hat{n}}(\mathbf{r})dS \right] \\
	&= \frac{\mu}{2\pi^2} \frac{\Omega_{\alpha}}{Q_{\alpha,i}} |d_{\alpha}(\omega)|^{2} .
\end{align}
% Again under the same conditions as \cref{eqn:powercondensed}, 
We can write the signal power $P^{\text{sig}}_{\alpha, i}$ explicitly in terms of the total power emitted into mode $\alpha$ $P_{\alpha}$ (as in \cref{eqn:powercondensed}), given by
\begin{align}
    P_{\alpha}(\omega) = \frac{1}{2\pi^2} \RE\left[b_{m}(\omega) s_{\alpha}(\omega)\right] &= \frac{\mu \Omega_{\alpha}}{2 \pi^2 Q_{\alpha}} \left| d_{\alpha}(\omega) \right|^2 ,
\end{align}
to find
\begin{align}
    \rightarrow P^{\text{sig}}_{\alpha, i}
	&= \frac{Q_{\alpha}}{Q_{\alpha,i}} P_{\alpha}(\omega) \label{eqn:signalextracted} .
\end{align}
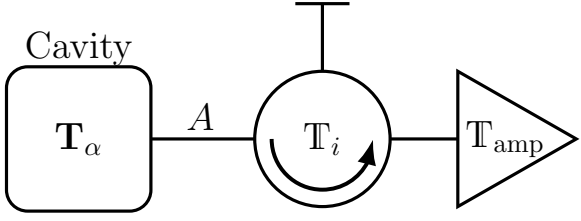
\begin{figure}
    \centering
    \begin{tikzpicture}[scale=1.2,>=Latex,font=\Large]
\def\rectX{-1.5}
\def\rectXlen{.8}
\def\circx{2.0}
\def\circrad{.75}
\def\arcrad{.75*\circrad}
\def\ampx{\circx + 2*\circrad}
\def\ampy{\circrad}
\def\amplen{1.75*\circrad}
\def\lineoutlen{2.0 * \circrad}
\def\Apos{.5*\rectX + .5*\rectXlen + .5*\circx}
% Cavity block (cylinder look)
\draw[rounded corners=8pt, line width=1.2pt] (\rectX,\rectXlen) -- (\rectX,-\rectXlen) -- (\rectX + 2*\rectXlen,-\rectXlen) -- (\rectX + 2*\rectXlen,\rectXlen) -- cycle;
% \draw[line width=1.2pt] (0,0.8) arc (90:-90:0.8);
\node at (\rectX + \rectXlen,1.25*\rectXlen) {Cavity};
\node at (\rectX + \rectXlen,0) {$\mathbf{T}_{\alpha}$};
%
% Line out of cavity with label beta
\draw[line width=1.2pt] (\rectX + 2*\rectXlen,0) -- (\circx-\circrad,0);
\node[above] at (\Apos, 0) {$A$};
% \node[above] at (0.5,0) {$\beta$};
%
% Circle block for circulator
\draw[line width=1.2pt] (\circx,0) circle (\circrad);
\node at (\circx,0) {$\mathds{T}_{i}$};
% \node[below] at (\circx,-1.2) {$A_{\text{cir}}$};
%
% Circular arrow inside circulator
\draw[line width=1.5pt, ->] (\circx-\arcrad, 0) arc (180:360:\arcrad);
%
% Vertical port on circulator
\draw[line width=1.2pt] (\circx,\circrad) -- (\circx,\lineoutlen);
\draw[line width=1.2pt] (\circx-.3,\lineoutlen) -- (\circx+.3,\lineoutlen); % termination bar
%
% Line to amplifier
\draw[line width=1.2pt] (\circx+\circrad,0) -- (\ampx,0);
%
% Amplifier triangle
\draw[line width=1.2pt] (\ampx,\ampy) -- (\ampx+\amplen, 0.0) -- (\ampx,-\ampy) -- cycle;
\node at (\ampx + .8*\amplen/2,0) {$\mathds{T}_{\text{amp}}$};
\end{tikzpicture}
    \caption{Schematic diagram of a hypothetical cavity radiometer composed of 
    a cavity operated at the mode indexed $\alpha$, 
    a readout line with attenuation $A$ connected to the $i^{th}$ cavity port,
    which is fed into an RF component (here, a circulator) with noise equivalent temperature $\mathds{T}_i$.
    The last component in the readout chain is an amplifier with NET $\mathds{T}_{amp}$. In this analysis the third port of the circulator is assumed to be terminated such that the device becomes an isolator. Adapted from~\cite{kimRevisitingDetectionRate2020}.}
    \label{fig:rf_readout}
\end{figure}
\begin{figure*}[htbp] 
    \centering
    % \fbox{%
    % \subfloat[]{
    \includegraphics[width=.99\textwidth]{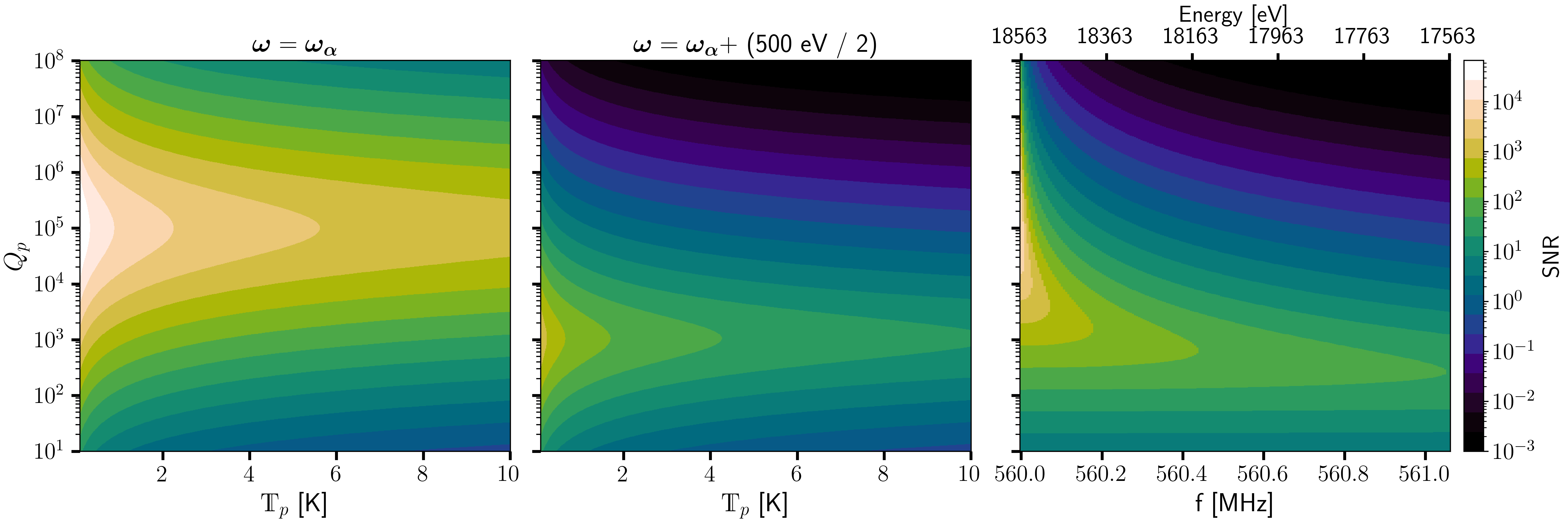}
    \put(-480,140){\color{white}\scriptsize \textbf{(a)}}%
    \put(-332,140){\color{white}\scriptsize \textbf{(b)}}%
    \put(-177,140){\color{white}\scriptsize \textbf{(c)}}%
    \caption{SNR calculated via \cref{eqn:rectangleIdealSNR} in a 1 kHz noise bandwidth for a 560 MHz TE$_{011}$ cylindrical cavity mode with $L/D$=8 and internal quality factor $Q_0=10^5$. \textbf{(a,b)} SNR as a function of external quality factor $Q_p$ and port temperature $\mathds{T}_p$. \textbf{(b)} The SNR for a 90$^\circ$ pitch angle electron 250 eV in frequency above or below the resonant frequency. The choice of 500 eV is intended to set a reference for the frequency region of interest once sidebands due to axial motion are taken into account. \textbf{(c)} SNR as a function of electron energy for $\mathds{T}_p$= 1 K. Each horizontal $Q_p$ slice is a potential experiment configuration. The variation of SNR with frequency presents an optimization problem, where the signal SNR off resonance benefits from larger cavity bandwidth (lower $Q_p$) while the SNR of signals on resonance suffer from it.}
    \label{fig:SNR}
\end{figure*}

\subsection{Noise power generated along a realistic RF readout chain} \label{ssec:total_noise}
While the cavity noise is derived from field theory, the subsequent readout chain is best modeled using cascaded noise temperature analysis. The noise power from a cavity as derived above can now be interpreted in a plausible readout configuration shown in \cref{fig:rf_readout}. We have given an expression for the noise extracted up to the cavity's output port. Past this point along the output line, the components couple more strongly to the spectral continuum of states than the cavity mode (see \cref{ssec:cavityboundaries}). 

The effective temperature of the port $\mathds{T}_p$ has contributions from each lossy RF component along the line, up to the isolator (we assume the isolator back propagation is negligible). A lossy RF component generates noise that is dependent on the noise power at its input and the physical temperature of the component itself. We define the attenuation $A$ as 
\begin{align}
    A = 10^{\text{loss [dB]}/10}.
\end{align}
For example, a coaxial cable or waveguide with a total loss of \SI{3}{\dB} will transmit half of the power, $A = 0.5$. $A=1$ corresponds to a lossless line, and $A=0$ to no transmission at all.
In general, for a passive component with physical temperature $T_i$ at the start and $T_f$ at the end, the noise power at the end is given by~\cite{kimRevisitingDetectionRate2020}
\begin{align}
    P^N_{RF_1} = k_B \Delta\nu (A T_i  + (1-A) T_f) ,
    \label{eq:lossy_element}
\end{align}
 which can be restated as 
 \begin{align}
     P^N_{RF_1} = k_B \mathds{T} \Delta\nu \hspace{1cm} \mathds{T} = T_i A + T_f (1-A).
 \end{align}
 For lossy transmission lines, RF propagation will undergo gradual temperature changes. This can be represented by a continuum of N identical small segments with temperature variation $\delta T$ and attenuation $\delta A$. The noise temperature at the n$^{th}$ segment is
 \begin{align}
     \mathds{T}_n = \mathds{T}_{n-1} \delta A + T_n (1-\delta A).
 \end{align}
 Integrating over all segments gives \cite{kimRevisitingDetectionRate2020}
\begin{align}
   \mathds{T}_f = \mathds{T}_i + (T_f-T_i)\left( 1+\frac{1-A}{\ln A}\right) + (T_i-\mathds{T}_i)(1-A) .
   \label{eq:lossy_line}
\end{align}
% \begin{figure}
%     \centering
%     \includegraphics[width=.48\textwidth]{figs/NoiseModel/fig6.pdf}
%     \caption{Effective noise temperature seen by the amplifier (device under test - DUT) as a function of its physical temperature relative to that of the cavity for different values of the line attenuation. The right panel zooms in to the region where the physical amplifier temperature is negligible compared to the cavity's physical temperature.}
%     \label{fig:thermal_coupling}
% \end{figure}

Continuing away from the cavity along the RF chain, the noise added by the first stage amplifier is:
\begin{equation}
    P^N_{RF_2} = k_B \mathds{T}_{amp} \Delta \nu, \myspace{1} \mathds{T}_{amp} = \frac{1}{\varepsilon} \frac{\hbar\omega}{k_B} ,
\end{equation}
where $\varepsilon$ is the quantum efficiency of the amplifier. Realistic values for $\varepsilon$ are 0.1 for a HEMT and 0.5 for a quantum-based amplifier. It is assumed that the further stages in the RF chain, both amplifiers or attenuation, affect the noise temperature in a manner that we consider negligible. If this assumption is invalid, higher stages can be included by replacing the first stage amplifier with an effective system noise temperature.

\subsection{Physical signal-to-noise ratio} \label{ssec:SNR}
Totaling the noise power contributions from the full RF chain, we can compute the noise power in an angular frequency interval $\Delta\nu$ centered at $\omega$:
\begin{align}
	P^{\Delta\nu}_{\alpha, i}(\omega) &= \int^{\omega + \Delta\nu/2}_{\omega - \Delta\nu/2} \left(P^N_{\alpha, i} + P^N_{RF_1} + P^N_{RF_2}\right) d\omega' . \label{eqn:physicalNoise}
\end{align}
Assuming a noise bandwidth $\Delta\nu$ much less than the cavity bandwidth, the integral of noise over the bin width can be approximated as a rectangular distribution. If we also assume that the insertion losses from the cable and the isolator can be neglected, the SNR of a mode $\alpha$ in port $p$ is
\begin{align}
    P^{\Delta\nu}_{\alpha, i}(\omega) &\approx \left(P^N_{\alpha, i} + P^N_{RF_1} + P^N_{RF_2}\right)  \Delta \nu \\
	\text{SNR}^{\Delta\nu}_{\alpha,i}(\omega) &= \frac{P^{\text{sig}}_{\alpha,i}(\omega)}{P^{\Delta\nu}_{\alpha,i}(\omega)} \label{eqn:snrphys}\\
    &\approx \frac{\frac{1}{\Omega_{\alpha} \epsilon } |s_{\alpha}(\omega)|^{2}}{\Delta\nu k_B } \left[\frac{4}{Q_{\alpha,0}} \left(\mathds{T}_{0} - \mathds{T}_{p} \right) \right. \nonumber\\
    & \myspace{3} + \frac{Q_{\alpha,  p}}{Q_{\alpha}^{2}}\left(\mathds{T}_{p} + \frac{\hbar\omega}{k_B \varepsilon}\right)  \nonumber \\
    & \myspace{2} \left. \cdot \frac{\Omega_{\alpha}}{\omega} \left( 1 + \frac{Q^{2}_{\alpha}}{\Omega^{4}_{\alpha}}\left(\Omega_{\alpha}^{2} - \omega^{2} \right)^{2}\right)\right]^{-1} , \label{eqn:rectangleIdealSNR}
\end{align}
% \begin{align}
%     P^{\Delta\nu}_{\alpha, i}(\omega) &\approx \left(P^N_{\alpha, i} + P^N_{RF_1} + P^N_{RF_2}\right)  \Delta \nu \\
% 	\text{SNR}^{\Delta\nu}_{\alpha,i}(\omega) &= \frac{P^S_{\alpha,i}(\omega)}{P^{\Delta\nu}_{\alpha,i}(\omega)} \label{eqn:snrphys}
% \end{align}
% \begin{align}
%     &= \frac{\frac{1}{\Omega_{\alpha} \epsilon } |s_{\alpha}(\omega)|^{2}}{\Delta\nu k_B \left[\frac{4}{Q_{\alpha,0}} \left(\mathds{T}_{0} - \mathds{T}_{p} \right) + \frac{\Omega_{\alpha} Q_{\alpha,  p}}{\omega Q_{\alpha}^{2}}L_{\alpha}^{-1}(\omega) \left(\mathds{T}_{p} + \frac{\hbar\omega}{k_B \varepsilon}\right) \right]} , \label{eqn:rectangleIdealSNR}
% \end{align}
using \cref{eqn:snrcondensed,,eqn:signalextracted} and ignoring the small shift term in the Lorentzian. This equation is plotted in \cref{fig:SNR}.  In the plot, the cavity wall temperature is 4K, and the electron is emitted with 90$^\circ$ pitch angle in the middle of the cavity at $r_0=a/2$ with 18.6 keV kinetic energy. The readout line is assumed lossless, and the first RF component encountered is at effective temperature $\mathds{T}_p$. The signal is then fed into an amplifier with quantum efficiency $\epsilon=.5$. When the cavity bandwidth (\cref{eqn:BWdefn}) is less than 100 noise bandwidths, the denominator in \cref{eqn:snrphys} is integrated rather than rectangle approximated. In the case of an ideal RF readout chain ($\mathds{T}_{p}=1/\varepsilon=0$), the SNR is independent of frequency and loaded $Q$, and only scales with unloaded $Q$. When non-ideal RF components are implemented, the SNR suffers off resonance due to dwindling signal power and frequency independent noise generated along the RF chain. If the cavity is critically coupled to the port, $Q_{\alpha,0}=Q_{\alpha,p}=2 Q_{\alpha}$, and $\frac{Q_{\alpha,p}}{Q^{2}_{\alpha}} = \frac{4}{Q_{\alpha,0}}$, which leads to a denominator of \cref{eqn:rectangleIdealSNR} that is independent of the port NET on resonance. For larger noise bandwidths (relative to the cavity mode's bandwidth) it is sometimes required to integrate \cref{eqn:physicalNoise} explicitly, in which case the $\mathds{T}_{p}$ independence on resonance is compromised (see \cref{fig:SNR}). In lab conditions, this formula can only provide a first estimate due to the neglect of important cable losses.
 %%% Cavity Noise Model

% Expansion options
% \input{sections/SNR/SNR} %%% Signal To Noise Ratio
% \input{sections/SNR/ReconstructionBackgroundSensitivity} %%% Implications for Reconstruction, Backgrounds and Energy Resolution

%%% Conclusion
\section{Conclusion}
We have developed a comprehensive theoretical framework for understanding the electromagnetic coupling between electrons undergoing cyclotron motion and resonant cavity modes, and in doing so addressed the unique challenges posed by cavity CRES experiments. This formalism expands previous work focused on Penning trap experiments to account for an electron's changing coupling to cavity modes as it moves through the modes' spatial nodes and antinodes.

Our derivation of the power spectrum for electrons in cyclotron motion radiating in general cavities, with specific application to cylindrical TE and TM modes, provides the necessary tools for understanding signal and noise characteristics in large-volume cavity CRES detectors. 
The predicted mapping of cavity configuration space to signal characteristics provides essential design guidance for implementing cavity CRES. 
Our noise model, which details how thermal noise sources couple through the cavity to the readout system, establishes the framework for optimizing detector performance for single and multiple mode operation. 

Beyond its immediate application to neutrino mass measurements, this work establishes a general theoretical framework for cavity-enhanced charged particle spectroscopy that is applicable to arbitrary particle motion and cavity geometries.
The formalism developed here for cavity response, noise modeling, and electromagnetic field coupling may benefit any experiment dependent on cavity SNR and signal structure, including cavity axion searches and other precision physics applications requiring large-volume detectors with enhanced electromagnetic coupling. 

The full validation of cavity CRES as a scalable approach for achieving the statistical sensitivity required for next-generation neutrino mass measurements is left to future work. Topics not explored here yet crucial to cavity CRES--for example energy reconstruction algorithms, detector geometry and readout--will benefit from results developed here.

\section*{Acknowledgments}
This material is based upon work supported by the following sources: the U.S. Department of Energy Office of Science, Office of Nuclear Physics, under Award No.~DE-SC0020433 to Case Western Reserve University (CWRU), under Award No.~DE-SC0011091 to the Massachusetts Institute of Technology (MIT), under Field Work Proposal Number 73006 at the Pacific Northwest National Laboratory (PNNL), a multiprogram national laboratory operated by Battelle for the U.S. Department of Energy under Contract No.~DE-AC05-76RL01830, under Early Career Award No.~DE-SC0019088 to Pennsylvania State University, under Award No.~DE-SC0024434 to the University of Texas at Arlington, under Award No.~DE-FG02-97ER41020 to the University of Washington, and under Award No.~DE-SC0012654 to Yale University; the National Science Foundation under Grant No.~PHY-2209530 to Indiana University, and under Grant No.~PHY-2110569 to MIT; the Cluster of Excellence ``Precision Physics, Fundamental Interactions, and Structure of Matter" (PRISMA+ EXC 2118/1) funded by the German Research Foundation (DFG) within the German Excellence Strategy (Project ID 39083149); the Karlsruhe Institute of Technology (KIT) Center Elementary Particle and Astroparticle Physics (KCETA); the Laboratory Directed Research and Development (LDRD) program at Lawrence Berkeley National Laboratory; LDRD 18-ERD-028 and 20-LW-056 at Lawrence Livermore National Laboratory (LLNL), prepared by LLNL under Contract DE-AC52-07NA27344, LLNL-JRNL-2014244; the LDRD Program at PNNL; University of Pittsburgh; and Yale University.
%-------------------------------------------------------------------------------------------------%
% \let\clearpage\relax
\bibliographystyle{apsrev4-2}
% \bibliography{top/myrefs, top/myrefs2, top/massreview}
\bibliography{top/myrefs2}

\appendix
\section{Helmholtz expansion of the electromagnetic field in a cavity} \label{app:helmholtz}
The solenoidal and irrotational fields \cref{eqn:EfieldExpansion,eqn:HfieldExpansion} within a cavity resonator obey distinct eigenvalue equations:
\begin{align}
    \mathbf{\nabla}\times\left(\frac{1}{\mu_{r}}\mathbf{\nabla}\times \mathbf{e}_{\beta}\right) &= k^2_{\beta} \epsilon_{r} \mathbf{e}_{\beta} \\
    \mathbf{\nabla}\times\left(\frac{1}{\epsilon_{r}}\mathbf{\nabla}\times \mathbf{h}_{\beta}\right) &= k^2_{\beta} \mu_{r} \mathbf{h}_{\beta} \\
    \nabla\left(\nabla\cdot\mathbf{f}_{\alpha}\right) &= k^2_{\alpha} \mathbf{f}_{\alpha}  \\
    \nabla\left(\nabla\cdot\mathbf{g}_{\alpha}\right) &= k^2_{\alpha} \mathbf{g}_{\alpha} ,
\end{align}
where $k_\alpha = \omega_{\alpha}/c$ and $\mu_{r}, \epsilon_{r}$ are the relative permeability and permittivity of the cavity volume (respectively). The resonant wavenumber $k_{\alpha}$ may be complex to account for energy loss. The solenoidal electric and magnetic fields are related by $\nabla\times\mathbf{e}_{\beta} = - i \omega_{\beta} \mu \mathbf{h}_{\beta}$.

If the cavity medium is homogeneous, then the solenoidal fields are solutions to the homogeneous wave equation
\begin{align}
    \nabla^{2} \mathbf{e}_{\beta} - k^2_{\beta} \epsilon_{r}\mu_{r} \mathbf{e}_{\beta} &= 0 \\
    \nabla^{2} \mathbf{h}_{\beta} - k^2_{\beta} \epsilon_{r}\mu_{r} \mathbf{h}_{\beta} &= 0 .
\end{align}
\section{Electric fields of cylindrical cavity eigenmodes} \label{app:efields}
This section lists the $\phi_1$ components of the cylindrical cavity electric fields in the electron's guiding center coordinate system. The calculation is straightforward, substituting \cref{eqn:graf} into the scalar potentials \cref{eqns:scalarpots}. The field definitions \cref{eqns:efieldsbymode} still hold in the guiding center coordinate system: the electric field $\hat{\phi}$ components for each mode type are calculated as
\begin{align}
    e^{\,TE}_{\phi,nml} &= A_{nml}^{TE} \chi'_{nm} \sin(k_{l} z) \sum^{\infty}_{u=-\infty} J_{n+u}(\chi'_{nm} r_0) J'_{u}(\chi'_{nm} \rho) \nonumber \\
    & \myspace{3.5} \cdot (-1)^u  \cos(u(\phi_0 - \phi_c) - n\phi_0) \\
    e^{\,TM}_{\phi,nml} &= \frac{A_{nml}^{TM} k_{l}}{\rho k_{nml}} \sin(k_{l} z) \sum^{\infty}_{u=-\infty} J_{n+u}(\chi_{nm} r_0) J_{u}(\chi_{nm} \rho) \nonumber \\
    & \myspace{3.5} \cdot u (-1)^u \cos(u(\phi_0 - \phi_c) - n\phi_0) \\
    e^{\,EI}_{\phi,nml} &= \frac{A_{nml}^{EI}}{\rho} \sin(k_{l} z) \sum^{\infty}_{u=-\infty} J_{n+u}(\chi_{nm} r_0) J_{u}(\chi_{nm} \rho) \nonumber \\
    & \myspace{3.5} \cdot u (-1)^u \cos(u(\phi_0 - \phi_c) - n\phi_0) .
\end{align}
\section{Convergence of the cavity fields} \label{app:convergence}
Numerical simulations of the electron dynamics within a cavity are greatly simplified by modeling a superposition of finite number of eigenmodes rather than seeking the full-wave solution to Maxwell's equations. It is then relevant to investigate the convergence properties of the field solutions derived in the previous sections by estimating the mode amplitude when the eigenfrequency is much larger than the driving frequency. There is also the related question of where the fields are convergent, as the electron's self-field is divergent at the electron's position.

Theoretical investigations into electromagnetic Green's functions have shown that their divergences arise from their zero-frequency behavior \cite{johnsonIrrotationalComponentElectric1979,yaghjianElectricDyadicGreens1980,bladelSingularElectromagneticFields2000}. Additionally, these studies have revealed that the singularities are features of the irrotational mode amplitudes in the standard eigenfunction expansion, as opposed to the solenoidal (TE/TM) modes. We note that the coefficients presented in \cref{sec:responseModel} can be used to construct the dyadic Green's functions \cite{collinFieldTheoryGuided1991}, and so our resonant mode coefficients should also be convergent.

Prior modeling of this interaction employed the dipole approximation \cite{brownCyclotronMotionMicrowave1985,ashtariesfahaniElectronRadiatedPower2019}, which is valid when the wavelength of the mode is much larger than the oscillation length scale, in this case, the cyclotron radius. When the total electric field is calculated as a superposition of all cavity modes, this approximation has been shown to result in divergent fields for any choice of cyclotron frequency \cite{brownCyclotronMotionMicrowave1985}. On the other hand, when the full classical cyclotron motion is accounted for, which is the case for the results in this section, the resulting non-zero frequency electric field becomes well-defined, and no renormalization methods are necessary to justify the modal superposition--as expected from the classical electromagnetic Green's function theory.  We will show this next for the case of the cylindrical cavity.

The convergence properties of the electric field amplitudes $a_{nml}, b_{nml}$ in \cref{ssec:couplingIntegrals} can be found by deriving their asymptotic behavior in the limit of $n,m,l$ large. We can break up the asymptotic behavior in three components: (1) that of the resonance structure of the fields (\cref{eqn:cavitylorentzian}), the current's expansion normalizations $B^M_{nml}$ (\cref{tab:coupling_coefficients}), and the contribution from the coupling integral (\cref{eqn:sm_formal}).

The first of these is trivial and goes as $ 1/\omega_{nml}^{2} \sim \left[ (m+n/2)^2/a^{2}+ \pi^{2}l^{2}/L^{2}\right]^{-1}$ far below resonance. For large real arguments $z$ and ignoring oscillations in $z$, $J_n(z) \sim \sqrt{\frac{2}{\pi z}} , \ J'_n(z) \sim -\sqrt{\frac{2}{\pi z}} $ \cite{abramowitzHandbookMathematicalFunctions2013}. For fixed n, as $m\rightarrow\infty$, and to leading order in n: $\chi_{nm} \sim (m + \frac{1}{2}n - \frac{1}{4})\pi, \ \chi'_{nm} \sim (m + \frac{1}{2}n - \frac{3}{4})\pi$. The contribution from the mode normalization varies depending on the mode type:
\begin{align}
    s_{nml} &\propto \frac{1}{\left(1 - \frac{n^2}{{\chi'}^2_{nm}}\right) J^2_n(\chi'_{nm})} \qquad (\text{TE}) \\
    &\sim \frac{(m+n/2)}{1 - \frac{n^2}{(m+n/2)^2}} \\
	s_{nml} &\propto \frac{k^2_{l}}{k^2_{nml}} \frac{1}{\chi^2_{nm} {J'}^2(\chi_{nm})} \qquad (\text{TM}) \\
    &\sim \frac{l^2/L^2}{(m+n/2)^2/a^2 + l^2/L^2} \frac{1}{m+n/2} \\
	q_{nml} &\propto \frac{1}{k^2_{nml} {J'_n}^2(\chi_{nm})} \qquad (EI) \\
    &\sim \frac{m+n/2}{(m+n/2)^2/a^2 + \pi^2 l^2/L^2} .
\end{align}
Finally, the contribution from the coupling integral is common to all mode types in \cref{eqn:currentExpansion} since $J_n$ and $J'_n$ have the same asymptotic behavior:
\begin{align}
	s_{nml}, \ q_{nml} &\propto \left(\frac{1}{m+n/2}\right)^2 \frac{a^2}{\rho_c r_0} .
\end{align}
By multiplying these three contributions, we have shown all modes contain at least inverse square dependencies on their mode numbers, and thus converge uniformly.
\section{Axial motion in TM coupling} \label{sec:AxialTM}
From \cref{eqn:tmmode}, we see the $\hat{z}$ component of the TM eigenmodes is
\begin{align}
    e^{TM}_{z,nml} &= \frac{1}{k_{nml}}\left(\partial^2_z + k^2_{nml}\right) \psi^{TM}_{nml} \\
    &= -\frac{-k_{z,l}^2 + k^2_{nml}}{k_{nml}} \psi_{nml}
    = -\frac{\chi^2_{nm}}{k_{nml}} \psi^{TM}_{nml} .
\end{align}
We seek to define the analogy of \cref{tab:coupling_coefficients}:
\begin{align}
    s^{z}_{nml}(\omega) &= -e \mathcal{F} \left\{ v_z(t) \int\delta\left(\mathbf{r} - \mathbf{r}_c(t)\right) e^{TM}_{z,nml} dV \right\} \\
    &= B_{nml}^{z} Z^{z}_l(t) \sum^{\infty}_{u=-\infty} \mathrm{P}^{z}_{nm,u}(t) \Phi^z_{n,u}(t) ,
\end{align}
where $\Phi^z_{n,u}(t) = (-1)^u \sin(u(\phi_0 - \phi_c(t)) - n\phi_0)$ is as before, and 
\begin{align}
    B_{nml}^{z} &= -\frac{\chi^4_{nm}}{k^2_{nml}} A_{nml}^{TM} \\
    \mathrm{P}^{z}_{nm,u}(t) &= J_{n+u}(\chi_{nm} r_0) J_{u}(\chi_{nm} \rho_c(t)) \\
    Z^{z}_l &= v_z(t) \cos(k_{l} z(t)) .
\end{align}
In the Fourier representation, the structure is quite similar to \cref{eqns:axialexpansions}, except for the modulation from the axial velocity and mode amplitude modulation $Z_l$.
% \begin{subequations}
% \begin{align}
%     Z^{z}_l(t) &= \sum_{\nu = 1}^\infty Z^{z}_\nu \cos( (\bar{v}_a k_{l} + \nu \omega_{z} ) t ) \label{eqn:sinkz_exp_TMz} ,
% \end{align}
% \end{subequations}
% which is now modulated by the axial velocity.
 Compared to the signal generated from $\phi$ motion, there is a change in radiated power proportional to $ \left(\bar{v}_z / \bar{v}_\perp\right)^2$ (where $v_\perp$ is the component of the electron's velocity perpendicular to the magnetic field).

\section{Green's dyadics conventions} 
\label{app:gfheat}
We provide additional context for the use of Green's dyadics to derive the cavity noise. The cavity fields are defined through dyadic Green's functions
\begin{align}
	\mathbf{E}(\mathbf{r}') &= \int_{V_{l}} \left[i \omega \mu_{0}\mu(\mathbf{r}) \mathbf{\bar{G}}_{e}(\mathbf{r}, \mathbf{r}')\cdot \mathbf{J}^{e}(\mathbf{r}) \right. \nonumber \\
	& \myspace{3} \left. - \mathbf{\bar{G}}_E(\mathbf{r}, \mathbf{r}')\cdot \mathbf{J}^{m}(\mathbf{r}) \right] d\mathbf{r} \\
	\mathbf{H}(\mathbf{r}')&= \int_{V_{l}} \left[i \omega \epsilon_{0} \epsilon(\mathbf{r}) \mathbf{\bar{G}}_{m}(\mathbf{r}, \mathbf{r}')\cdot \mathbf{J}^{m}(\mathbf{r}) \right. \nonumber\\
	& \myspace{3} \left. + \mathbf{\bar{G}}_M(\mathbf{r}, \mathbf{r}')\cdot \mathbf{J}^{e}(\mathbf{r}) \right] d\mathbf{r} ,
\end{align}
with $\epsilon(\mathbf{r}) = \epsilon_h =  \frac{1}{\epsilon_0}\left( \epsilon + \frac{\sigma}{i\omega}\right)$, $\mu(\mathbf{r}) = \mu_h = 1 $, and 
\begin{align}
	\bar{\mathbf{G}}_{e}(\mathbf{r},\mathbf{r'},\omega) &= -c^2 \sum\limits_{\alpha} L_{\alpha}(\omega) \mathbf{e}_{\alpha}(\mathbf{r})\mathbf{e}_{\alpha}(\mathbf{r'}) \nonumber\\
	& \myspace{3} + \frac{-\mu^{-1}(\mathbf{r})}{\omega\mu_{0}\sigma + i \omega^{2}/c^{2}} \sum\limits_{\beta} \mathbf{f}_{\beta}(\mathbf{r})\mathbf{f}_{\beta}(\mathbf{r'}) \\
	%--------
	\bar{\mathbf{G}}_{m}(\mathbf{r},\mathbf{r'},\omega) &= -c^2 \sum\limits_{\alpha} L_{\alpha}(\omega) \mathbf{h}_{\alpha}(\mathbf{r})\mathbf{h}_{\alpha}(\mathbf{r'}) \nonumber\\
	& \myspace{3} + \frac{-\epsilon^{-1}(\mathbf{r})}{\mu\omega^{2}} \sum\limits_{\beta} \mathbf{g}_{\beta}(\mathbf{r})\mathbf{g}_{\beta}(\mathbf{r'})\\
	%--------
	\bar{\mathbf{G}}_{M}(\mathbf{r},\mathbf{r'},\omega) &= c \sum\limits_{\alpha} \omega_{\alpha} L_{\alpha}(\omega) \mathbf{e}_{\alpha}(\mathbf{r})\mathbf{h}_{\alpha}(\mathbf{r'}) \\
	&= \bar{\mathbf{G}}_{E}(\mathbf{r'},\mathbf{r},\omega) .
\end{align}
\section{Glossary of symbols in the main text}
\begin{description}[font=\normalfont]
    \item[$F_{\alpha}$] Purcell factor
    \item[$c$] Speed of light
    \item[$Q_{\alpha}$] Cavity loaded quality factor of mode $\alpha$
    \item[$\omega$] Angular frequency of emission
    \item[$V$] Cavity volume
    \item[$\omega_c$] Relativistic cyclotron frequency
    \item[$e$] Elementary electric charge
    \item[$\mathbf{B}$] Magnetic field (vector)
    \item[$B$] Magnetic field magnitude ($B=|\mathbf{B}|$)
    \item[$\gamma$] Lorentz factor
    \item[$m_e$] Electron mass
    \item[$E_{kin}$] Electron kinetic energy
    \item[$\mathbf{v}$] Electron velocity (vector)
    \item[$\rho_c$] Cyclotron radius
    \item[$v_\perp$] Magnitude of velocity perpendicular to $\mathbf{B}$
    \item[$\mathbf{v}_\perp$] Velocity component perpendicular to $\mathbf{B}$
    \item[$\theta_p$] Pitch angle (between momentum and local magnetic field)
    \item[$B_0$] Magnetic field strength at the trap center
    \item[$B_{\text{max}}$] Maximum magnetic field of the trap
    \item[$\mathbf{v}_{grad-B}$] grad-B drift velocity
    \item[$\nabla_{\perp} B$] Spatial gradient of the magnetic field magnitude in the plane orthogonal to the dominant component of $\mathbf{B}$
    \item[$P(\gamma,\theta_p)$] Larmor power (radiated power of the electron in free space)
    \item[$\epsilon_0$] Permittivity of free space
    \item[TE, TM] Cavity transverse electric and transverse magnetic mode labels (example: TE$_{011}$)
    \item[$\mathbfcal{E}(\mathbf{r},t)$] Electric field (time domain)
    \item[$\mathbf{f}_{\alpha}(\mathbf{r})$] Electric field irrotational (EI) mode basis function
    \item[$a_{\alpha}(t)$] Expansion coefficient for electric irrotational mode $\mathbf{f}_{\alpha}(\mathbf{r})$
    \item[$\mathbf{e}_\beta(\mathbf{r})$] Solenoidal electric mode basis function
    \item[$b_\beta(t)$] Expansion coefficient for solenoidal electric modes $\mathbf{e}_\beta(\mathbf{r})$
    \item[$\mathbfcal{H}(\mathbf{r},t)$] Magnetic field (time domain)
    \item[$\mathbf{g}_{\alpha}(\mathbf{r})$] Magnetic irrotational (MI) mode basis function $\mathbf{g}_{\alpha}(\mathbf{r})$
    \item[$c_{\alpha}(t)$] Expansion coefficient for solenoidal magnetic mode $\mathbf{g}_\alpha(\mathbf{r})$
    \item[$\mathbf{h}_\beta(\mathbf{r})$] Solenoidal magnetic mode basis function
    \item[$d_\beta(t)$] Expansion coefficient for solenoidal magnetic mode $\mathbf{h}_\beta(\mathbf{r})$
    \item[$\omega_{\alpha}$] Complex eigenfrequency of mode $\alpha$
    \item[$\epsilon$] Real part of permittivity of the cavity volume
    \item[$\mu$] Permeability of the cavity volume
    \item[$\sigma(\mathbf{r})$] Conductivity of the cavity volume
    \item[$\mathbf{\hat{n}}(\mathbf{r})$] Unit normal vector pointing out from the boundary
    \item[$\mathbfcal{J}_s (\mathbf{r},t)$] Surface current
    \item[$Z(\mathbf{r})$] Complex impedance on the cavity volume boundary
    \item[$\mathbfcal{H}_{tan}(\mathbf{r},t)$] Tangential magnetic field at the boundary
    \item[$S$] Cavity volume boundary surface
    \item[$S_i$] Boundary segment $i$
    \item[$S_0$] Surface of the cavity walls
    \item[$P_{\alpha}^{S}$] Total power radiated through the boundary for mode $\alpha$
    \item[$W_{\alpha}$] Total stored energy in mode $\alpha$
    \item[BW$_{\alpha}$] Bandwidth of mode $\alpha$
    \item[$Q_{\alpha,i}$] Quality factor of mode $\alpha$ due to boundary $S_i$
    \item[$Q_{\alpha,0}$] Unloaded quality factor (due to walls $S_0$)
    \item[$\mathcal{F}^{-1}\{ x \}$] Inverse Fourier transform of x
    \item[$\mathbf{E}(\mathbf{r},\omega)$] Electric field (frequency domain)
    \item[$\mathbf{H}(\mathbf{r},\omega)$] Magnetic field (frequency domain)
    \item[$\mathbf{J}(\mathbf{r},\omega)$] Current density (frequency domain)
    \item[$a_{\alpha}(\omega)$] Expansion coefficient for electric irrotational mode $\mathbf{f}_{\alpha}(\mathbf{r})$ in the frequency domain
    \item[$b_{\alpha}(\omega)$] Expansion coefficient for solenoidal electric modes $\mathbf{e}_\beta(\mathbf{r})$ in the frequency domain
    \item[$c_{\alpha}(\omega)$] Expansion coefficient for magnetic irrotational mode $\mathbf{g}_{\alpha}(\mathbf{r})$ in the frequency domain
    \item[$d_{\alpha}(\omega)$] Expansion coefficient for solenoidal magnetic mode $\mathbf{h}_\beta(\mathbf{r})$ in the frequency domain
    \item[$k_{\alpha}$] Wavevector of mode $\alpha$
    \item[$L_{\alpha}(\omega)$] Cavity Lorentzian response function
    % \item[$\bar{P}(t)$] Time-averaged electromagnetic power
    % \item[$T$] Period of oscillation
    % \item[$\omega_0$] Single oscillation frequency
    \item[$q_{\alpha}(\omega)$, $q^{EI}_{\alpha}(\omega)$] Spectral coefficient of electric current for irrotational electric modes
    \item[$s_{\alpha}(\omega)$, $s^M_{\alpha}(\omega)$] Spectral coefficient of electric current for solenoidal electric modes
    \item[$\mathcal{F}\{ x \}$] Fourier transform of x
    \item[$P_{\alpha}(\omega)$] Time-averaged electromagnetic power emitted into cavity mode $\alpha$
    \item[$\RE {[x]} $] Real part of $x$
    \item[$\Omega_{\alpha}$] Real part of eigenfrequency $\omega_{\alpha}$
    \item[$a$] Cylindrical cavity radius
    \item[$L$] Cylindrical cavity length
    \item[$\psi^M_{nml}$] Scalar potential for mode $M$ ($M \in \{\text{TE, TM, EI}\}$)
    \item[$n$] Azimuthal mode index
    \item[$m$] Radial mode index
    \item[$l$] Axial mode index
    \item[$k_l$] Axial wavenumber ($k_l = l\pi/L$)
    \item[$J_n(x)$] Cylindrical Bessel function of order $n$
    \item[$k^M_{nm}$] Radial wavenumber
    \item[$\chi_{nm}$] $m$-th zero of $J_n(x)$
    \item[$\chi'_{nm}$] $m$-th zero of $J'_n(x)$
    \item[$A^M_{nml}$] Mode normalization factor
    \item[$\delta_{ij}$] Kronecker delta function
    \item[$f^M_{nml}$] Resonant frequency of mode indexed by $n,\ m,\ \text{and } l$
    \item[$\mathbf{e}^{M}_{nml}$] Electric field for mode $M$ ($M \in \{\text{TE, TM}\}$)
    \item[$\mathbf{f}^{EI}_{nml}$] Electric field for the irrotational electric mode $nml$
    \item[$j^M_{\alpha}(\omega)$] Generalized modal expansion coefficient for mode type $M$ 
    \item[$\mathbf{r}_c$] Electron position vector
    \item[$z_c$] Electron axial position
    \item[$\phi_c$] Electron azimuthal cyclotron phase
    \item[$\mathbf{r}_0$] Electron guiding center position
    \item[$\phi_0$] Electron guiding center azimuthal angle
    \item[$\mathbf{r}_1$] Position vector in guiding center coordinates
    \item[$\Delta \phi$] Phase difference between cavity and guiding center frames
    \item[$\omega_g$] Grad-B drift frequency
    \item[$E^M_{\phi_1,\alpha}(\mathbf{r}_1)$] Phi component of the electric field in guiding center coordinates for mode type $M$
    \item[$\Phi_{n,u}$] Phase coupling function
    \item[$Z_l$] Axial coupling function
    \item[$B^M_{nml}$] Coupling pre-factor for mode type $M$
    \item[$\mathrm{P}^M_{nm,u}$] Radial coupling function for mode type $M$
    \item[$j^M_{nml}(p \omega_c)$] Generalized modal expansion coefficient, representing $s^M_{nml}$ ($M$ is TE or TM) or $q^{EI}_{nml}$ at the $p$-th harmonic of cyclotron frequency
    \item[$j^{M\pm}_{nml}(p \omega_c)$] Mode orientation summed modal expansion coefficient
    \item[$\left| j^{M\pm}_{nml}(p \omega_c) \right|^2$] Squared magnitude of mode orientation summed modal expansion coefficient
    \item[$\left| s^{TE\pm}_{nml}(p \omega_c) \right|^2$] Squared magnitude for TE modes
    \item[$\left| s^{TM\pm}_{nml}(p \omega_c) \right|^2$] Squared magnitude for TM modes
    \item[$\left| q^{EI\pm}_{nml}(p \omega_c) \right|^2$] Squared magnitude for EI modes
    \item[$v_z$] Electron axial velocity
    \item[$T_a$] Time spent in the cavity (for throughgoing electron), or period of axial motion
    \item[$\text{rect}(x)$] Rectangle function
    \item[$\text{sinc}(x)$] Sinc function, $\text{sinc}(x) = \frac{\sin x}{x}$
    \item[$z_1$] Initial axial position (or trap wall position)
    \item[$L_a$] Trap length ($L_a = z_2 - z_1$)
    \item[$z_2$] Second trap wall position ($z_2 = z_1 + L_a$)
    \item[$\omega_z$] Axial oscillation frequency
    \item[$\nu$] Sideband order relative to main carrier for harmonics of axial motion
    \item[$\bar{\omega}_c$] Average cyclotron frequency
    \item[$Z^\nu_l$] Fourier coefficient of axial coupling function at sideband order $\nu$
    \item[NET] Noise equivalent temperature 
    \item[$\mathds{T}_i$] Noise equivalent temperature (NET) of boundary segment $i$
    \item[$k_B$] Boltzmann constant
    \item[$\eta(\omega)$] Quantum mechanical photon population
    \item[$P_{RF}$] Noise power from an RF component
    \item[$\mathbf{T}_{\alpha}$] NET of cavity mode $\alpha$
    \item[$d^\pm_{\alpha}(\omega)$] Magnetic field amplitude of mode $\alpha$ due to thermal fluctuations on inside ($+$) or outside ($-$) of boundary
    \item[$\mathbf{J}^e$] Electric current density inside the cavity
    \item[$\mathbf{J}^m$] Magnetic current density inside the cavity
    \item[$\bar{\mathbf{G}}_m$] Magnetic dyadic Green's function
    \item[$W^{h}_{\alpha}$] Average stored magnetic energy for mode $\alpha$
    % \item[$N_{h,\alpha}$] Normalization factor for magnetic field volume integrals of mode $\alpha$
    \item[$\epsilon_h$] Effective permittivity including conductivity, ${\epsilon_h = \frac{1}{\epsilon_0}\left( \epsilon + \frac{\sigma}{i\omega}\right)}$
    \item[$\IM {[x]} $] Imaginary part of $x$
    \item[$P^N_{\alpha, i}$] Noise power from cavity at port $i$
    \item[$P^{\text{sig}}_{\alpha, i}$] Signal power extracted at port $i$
    \item[$\text{Tr}\mathbf{\bar{A}}$] Trace of a dyadic $\mathbf{\bar{A}}$
    \item[$A$] Attenuation factor of the RF transmission line ($A = 10^{\text{loss [dB]}/10}$)
    \item[$T_i$] Physical temperature at the input of the transmission line
    \item[$T_f$] Physical temperature at the output of the transmission line
    \item[$\Delta\nu$] Noise bandwidth
    \item[$\delta T$] Temperature variation per segment along a transmission line
    \item[$\delta A$] Attenuation per segment along a transmission line
    \item[$\mathds{T}_n$] Noise temperature at the $n$th segment
    \item[$\mathds{T}_f$] Final noise temperature at end of transmission line
    \item[$\mathds{T}_{amp}$] Noise equivalent temperature of amplifier
    \item[$\varepsilon$] Quantum efficiency of amplifier
    \item[$P^N_{\alpha, i}$] Noise power from cavity at port $i$
    \item[$P^N_{RF_1}$] Noise power from first RF component (e.g., cable, isolator)
    \item[$P^N_{RF_2}$] Noise power from the amplifier
    \item[$P^{\Delta\nu}_{\alpha, i}(\omega)$] Total noise power in bandwidth $\Delta\nu$ at frequency $\omega$ for mode $\alpha$ at port $i$
    \item[$\text{SNR}^{\Delta\nu}_{\alpha,i}(\omega)$] Signal-to-noise ratio within bandwidth $\Delta\nu$ for mode $\alpha$ at port $i$
\end{description}

\end{document}